\documentstyle[12pt]{article}
\input math_macros
\font\twebf=cmbx12
\font\tenbf=cmbx10
\font\tenrm=cmr10
\font\tenit=cmti10
 1
 1
 1

\begin{document}
\thispagestyle{empty}

\centerline{\normalsize hep-ph/9508351 \hfill OITS-556}
\centerline{\normalsize \hfill September 1994}
\centerline{\normalsize \hfill {\it revised,} August 1995}

\vspace{39mm}

\renewcommand{\theequation}{\thesection.\arabic{equation}}
\renewcommand{\thesection}{\Roman{section}.}
\hspace{-5mm}{\bf Perturbative QCD calculations of total cross sections and
                  decay widths}

\hspace{43mm}{\bf in hard inclusive processes}

\begin{quotation}
\noindent
{\tenrm Levan R. Surguladze\\}
{\tenit Institute of Theoretical Science, University of Oregon, Eugene,
        Oregon 97403}

\vspace{3mm}

\noindent
{\tenrm Mark A. Samuel\\}
{\tenit Department of Physics, Oklahoma State University, Stillwater,
        Oklahoma 74078}

\vspace{9mm}

\noindent
A summary of the current understanding of methods of analytical higher
order perturbative  computations of total cross sections and decay widths
in Quantum Chromodynamics is presented. As examples, the quantities
\linebreak
$\sigma_{\mbox{\scriptsize{tot}}}(e^{+}e^{-}\rightarrow \mbox{hadrons})$,
$\Gamma(\tau^{-}\rightarrow\nu_{\tau}+\mbox{hadrons})$ and
$\Gamma(H\rightarrow \mbox{hadrons})$ up to  $O(\alpha^{2}_{s})$ and
$O(\alpha^{3}_{s})$ are considered. The evaluation of the
four-loop QED $\beta$ - function at an intermediate step of the
calculation is briefly described.
The problem of renormalization group ambiguity of perturbative results is
considered and some of the existing prescriptions are discussed.
The problem of estimation of theoretical uncertainty in perturbative
calculations is briefly discussed.
\end{quotation}

\vspace{39mm}

\begin{center}
{\em To be published in} {\it Reviews of Modern Physics}
\end{center}

\newpage
%
%
\renewcommand{\thesection}{\Roman{section}}
\renewcommand{\thesubsection}{\Alph{subsection}}
\tableofcontents
\newpage
\baselineskip=14pt

\renewcommand{\thesection}{\Roman{section}}
\section{\twebf Introduction}
\renewcommand{\thesection}{\arabic{section}}
\setcounter{equation}{0}

The Standard Model (SM) (Weinberg, 1967; Salam, 1969;
Glashow {\it et al.}, 1970)
of strong and electroweak interactions has been tested
with precise experiments at the present colliders (for
reviews see, e.g., Altarelli, 1989; Marciano, 1991, 1993a,b; Bethke, 1992;
Bethke and Pilcher, 1992; R.\ K.\ Ellis, 1992;
Langacker, Luo and Mann, 1992;  Brodsky, 1993; \linebreak
Sirlin, 1993a,b; Kniehl, 1994a; Soper, 1995).
However, the decisive confirmation of the SM or its modification is still
ahead. One awaits with great interest precise measurements from
LEP, SLC, HERA, Fermilab Tevatron, etc.

The progress in a very precise experiment requires adequate
progress in the developing of calculational methods and performing the
theoretical computations of various observables. This can be consistently
done within perturbation theory for processes with large momentum transfer.
Nowadays, the standard way to evaluate experimentally measurable
quantities from first principles of the theory is to use perturbation
methods. Lattice calculations provide an alternative method.

The main goal of this paper is to review some of the recent achievements in
methods of high order analytical perturbative calculations of a wide class
of observable quantities. These quantities are
total cross sections, decay widths and
structure functions in deep inelastic processes, several key
theoretical quantities, such as renormalization group functions,
renormalization constants, Wilson coefficient functions, etc.
We will present a simplified description of some of the recent calculations.

A decisive role in the construction of the SM has been played by
experimental studies of so called inclusive processes, in particular deep
inelastic lepton-hadron processes like
$e^{+}e^{-} \rightarrow \mbox{hadrons}$, deep inelastic $e$, $\mu$ and $\nu$
-scattering, etc. The discovery of scaling of
deep inelastic structure functions
(Bjorken, 1968, 1969; Yang, 1969) led to the parton
model (Feynman, 1969, 1972; see also Drell and Yan, 1971).
The explanation of the observed scaling properties has been given by
Matveev, Muradyan and Tavkhelidze (1970, 1972), using the universal
principle of automodelity and dimensional analysis.
The quark counting formulae, allowing one to obtain the high energy
asymptotic behavior for cross sections and hadron form factors at large
momentum transfers, have been derived by Brodsky and Farrar (1973),
and Matveev, Muradian and Tavkhelidze (1973).
The discovery of
asymptotic freedom in nonabelian gauge field models (Gross and Wilczek, 1973;
Politzer, 1973) together with the conception of
spin half, fractionally electric charged fundamental constituents of hadrons
-  quarks (Gell-Mann, 1964; Zweig, 1964) with an additional quantum number
color (Bogolyubov, Struminsky and Tavkhelidze, 1965; Tavkhelidze, 1965;
Han and Nambu, 1965; also Miyamoto, 1965; Greenberg, 1964),
 interacting via eight massless, non-abelian,
spin 1, self interacting gauge fields - gluons, led to the creation of
Quantum Chromodynamics (QCD) (Fritzsch, Gell-Mann and Leutwyler, 1973) - the
present theory of strong
interactions. For an introductory review of QCD see, e.g., Marciano and Pagels
(1978), and a short historical review has been given recently by
Tavkhelidze (1994).
QCD is based on a local SU$_{c}$(3) symmetry group, which
implies the minimal locally gauge invariant Lagrangian density of the model.
QCD is a renormalizable quantum field model. There exists well defined rules
for removing of ultraviolet divergences from
S-matrix amplitudes at each order of the interaction coupling constant.
After the renormalization, the calculated physical quantities are free of
ultraviolet regularization parameters. The problem of renormalizability
of non-abelian gauge theories has been considered since the early 60's
(Feynman, 1963; deWitt, 1967; Mandelstam, 1968, etc.). After the Lorentz
covariant quantization of gauge fields, based on the path integral approach
(Faddeev and Popov, 1967; for a textbook, see also Faddeev and Slavnov, 1980),
the proof of renormalizability was given ('t~Hooft, 1971; Lee and
Zinn-Justin, 1972, 1973).
Besides the short distance effects, in QCD one has to deal with infrared
divergences associated with long distance infinities. In other words,
in addition to the large parameter
(large momentum transferred - $Q^{2}$), there are small parameters such as,
for instance, hadron mass - $m$
or momenta of some of the participating particles, and in the calculation
one faces senseless large logarithmic contributions
$\sim \log m^{2}/Q^{2}$.  The infrared divergence problem was considered
long ago for QED (Bloch and Nordsieck, 1937; Yennie, Frautschi and Suura,
1961).
A modern treatment of this problem in the SM is based on the
operator product expansion technique (Wilson, 1969) and factorization
theorems. This, in some cases, in particular, for deep inelastic processes,
allows one to factorize the large and small distance
contributions (Libby and Sterman, 1978; Mueller, 1978;
Efremov and Radyushkin, 1980a,b; Radyushkin, 1983; Collins and Soper, 1987;
Collins, Soper and Sterman, 1983, 1984, 1985, 1989 and references therein).
The concrete prescriptions of dealing with infrared divergent Feynman
integrals have been given by Vladimirov (1978, 1980),
Pivovarov and Tkachov (1988),
Tkachov (1991, 1993),
Chetyrkin and Smirnov (1985). For earlier references, see the work
by Tkachov (1993). For a textbook, see Collins (1984).
The interference
of long and short distance effects is still problematic in actual
higher order calculations.

The group character of renormalizing transformations in quantum field
theory was first discovered by Stueckelberg and Peterman (1953), and
Gell-Mann and Low (1954) have applied it
to study the ultraviolet asymptotics of Green's functions in spinor
electrodynamics.  The mathematical formalism of the
renormalization group has been worked out by Bogolyubov and Shirkov
(1955, 1956a,b), and Bogolyubov and Parasyuk (1955a,b, 1956, 1957)
have introduced the $R$-operation for subtracting ultraviolet
divergent contributions recursively
on the level of loop Feynman diagrams. The renormalization group and
$R$-operation techniques are the crucial tools in any perturbative
calculation within the Standard Model (see the textbook by
Bogolyubov and Shirkov, 1980). For a historical review on renormalization
group see, for example, Shirkov (1992; also Peterman, 1979)
and references therein.
For a textbook on the modern renormalization
theory and the references, see Collins (1984).

The property of asymptotic freedom, the method of renormalization group and
factorization theorems are the basis of the present perturbative QCD. In
order to relate perturbative QCD to measurable quantities,
it was necessary along with the renormalization group, dispersion relations,
operator product expansion techniques and factorization, to develop
a technique for evaluation of loop Feynman diagrams. Indeed, in each order
of perturbation theory, contributions to the physical observables come from
a finite set of divergent Feynman integrals with the same number of
internal momentum integrations (number of loops). Thus, one has to deal
with very singular (ultraviolet and infrared) Feynman integrals, and a
correct mathematical apparatus suitable for calculational purposes
is necessary.  The dimensional regularization technique
('t~Hooft and Veltman, 1972, 1973;  Bollini and Giambiagi, 1972;
Ashmore, 1972; Cicuta
and Montaldi, 1972) for ultraviolet divergent Feynman integrals is based on the
idea of integration over the space-time of noninteger dimension less than 4.
In this case, the Feynman integrals
become well defined, and divergences appear as poles in terms of the deviation
from the physical space-time dimension of 4. The important property of
dimensional regularization is that it preserves explicit gauge invariance
and is very convenient for practical calculations. In fact, almost all recent
progress in higher order analytical perturbative calculations has been made
within the dimensional regularization, using 't Hooft's (1973) minimal
subtraction
prescription.

    The systematic study of strong interaction effects for the various
observables in processes with large momentum transfer requires
one to evaluate at least the first few coefficients in the perturbative
expansion in terms of the strong coupling. Here
the problem of calculating multiloop Feynman diagrams arises. The recursive
type algorithm for analytical evaluation of one-, two- and three-loop massless,
propagator type, dimensionally regularized Feynman diagrams has been given by
Chetyrkin and Tkachov (1981), and Tkachov (1981, 1983a). This algorithm,
together
with the so called Gegenbauer x-space technique
(Chetyrkin and Tkachov, 1979) allows one to evaluate an expansion in the
Laurent series in $\varepsilon=(4-D)/2$
of all massless propagator type Feynman diagrams up to the
three-loop level, where $D$ is the
noninteger dimension of the space-time.
The above algorithm is applicable to
a wide class of problems up to the four-loop level. These are, for instance,
calculation of renormalization constants, renormalization group functions,
some of the cross sections and decay widths.
We note that this algorithm
deals only with propagator type massless diagrams. Nevertheless, due to the
remarkable properties of dimensional regularization ('t~Hooft and Veltman,
1972)
 and the minimal subtraction prescription ('t~Hooft, 1973), namely, that the
counterterms are polynomials in dimensional parameters within minimal
subtraction (Collins, 1974; Speer, 1974; see also the
textbook by Collins, 1984), a wide class of problems can be reduced to the
evaluation of
propagator type diagrams (Vladimirov, 1978, 1980).
At high energies, in some cases, it is possible to neglect the masses of
participating particles and
consider the massless diagrams. The
mass corrections of the type $m^{2n}/s^{n}$, where $s$ is the center-of-mass
energy squared, can also be evaluated through the calculation of massless
diagrams (see, e.g., Gorishny, Kataev and Larin, 1986;
Surguladze, 1989a, 1994a,b,c). Feynman graphs can also contain virtual
heavy particle
propagators regardless of the energy scale of the particular process.
If the masses of the virtual particles are much larger than the energy scale,
one can neglect them, since their effcts are supressed by powers of large
mass, according to the decoupling theorem (Appelquist and Carazzone, 1975).
However, in some cases, such effects may not be entirely negligible
(Soper and Surguladze, 1994).
The prescriptions to study asymptotic expansions of Feynman integrals
in powers of $m^{2}/s$ can be obtained from Chetyrkin and
Tkachov (1982), Tkachov (1983b,c, 1991, 1993), and Chetyrkin (1991; see also
Smirnov 1990, 1991 and references therein).
An exact general expression for one-loop, N-point, massive Feynman integrals
has been obtained by Davydichev (1991), and  Boos and Davydychev (1992).
This expression contains the generalized hypergeometric function
and is complicated, except for some particular cases. An alternative method
for massive Feynman integrals has been suggested by Kotikov (1991).

   In practice, the calculation of physical quantities within perturbation
theory is very cumbersome and tedious already beyond the one-loop level,
especially in realistic quntum field theory models, like QCD. However, the
recursive type algorithms by Chetyrkin and Tkachov (1981)
allow convenient implementation within algebraic programming systems like
{\small REDUCE} (Hearn, 1973),
{\small SCHOONSCHIP} (Veltman, 1967; Strubbe, 1978; Veltman, 1989)
and {\small FORM} (Vermaseren, 1989). Several computer programs
 were written in
the last decade for analytical computation of multiloop Feynman diagrams.
Among them we mention the programs which fully implement the above
mentioned recursive algorithms. The program {\small LOOPS}
 (Surguladze and Tkachov, 1989a),
written on the {\small REDUCE} system, calculates one- and
 two-loop massless, propagator type Feynman diagrams for
 arbitrary structure in the numerator
of the integrand and for an arbitrary space-time dimension.
The program {\small MINCER} (Gorishny, Larin,
Surguladze and Tkachov, 1989), written on the {\small SCHOONSCHIP}
 system, and the program {\small HEPL}oops (Surguladze, 1992),
 written on the {\small FORM} system, calculates one-,
 two- and three-loop massless, propagator type diagrams.
The status of the existing program packages has
 been discussed recently in
Surguladze (1994d).
The above methods, algorithms and computer programs
allow one to make significant
progress in high order analytical perturbative calculations of
several important physical observables.

The other outstanding problem in perturbative calculations is
the renormalization group ambiguity of perturbation theory
predictions. Indeed, starting from a
certain order, the perturbative coefficients become scheme-scale
dependent, while it is obvious that the calculated observable cannot
 depend on any subjective choice of nonphysical parameters.
Several approaches have been
suggested to deal with the scheme-scale ambiguity problem. Among them
we consider the so called {\it fastest apparent convergence}
approach (Grunberg, 1980),
suggesting one absorb the leading QCD corrections in the definition
of the ``effective'' running coupling. We will consider an
approach based on the {\it principle of minimal sensitivity} of the
physical observables to nonphysical parameters (Stevenson, 1981a,b), and
Brodsky, Lepage and Mackenzie (1983) (BLM) method,
suggesting one should fix the scale according to the size of the quark
vacuum polarization effects.
The commensurate scale relations by Brodsky and Lu (1994, 1995)
allow one to make scale-fixed perturbative predictions without
referring to the particular renormalization prescription.

  In the recent works, some authors try
to predict the perturbative coefficients without
calculating the relevant Feynman graphs. First, we mention
the method by West (1991) which is based on the renormalizability, analyticity
arguments, and the saddle point technique. For comments
on this work see Barclay and Maxwell (1992a),
Brown and Yaffe (1992), Surguladze and Samuel (1992), and
Duncan {\it et al.} (1993).
The method of Samuel {\it et al.}
(Samuel and Li, 1994a,b,c; Samuel, Li and Steinfelds, 1994a,b,c),
based on Pad\`{e} approximants, work surprisingly well for the large
number of cases considered. However, a theoretical basis of this method is
necessary. Recent developments have put Pad\`{e} approximant method
on a more rigorous basis, which justifies its application
to perturbation series in QED, QCD, Atomic physics, etc.
This is discussed in recent papers (see, e.g., Ellis, Karliner and Samuel,
1995).
An alternative method for estimation of higher order perturbative
contributions can be obtained based on Stevenson's (1981a,b) approach
(Surguladze and Samuel, 1993; Kataev and Starshenko, 1994).
The important problem of large-order behavior of perturbation
theory has been considered  by Barclay and Maxwell (1992b),
and Brown and Yaffe (1992). The same problem has been discussed
during the past twenty years. The part of papers on the subject have been
collected in the book edited by Le Guillou and Zinn-Justin (1990).
The application of renormalon calculus in the study of the behaviour of
perturbative QCD series is a subject of intensive discussions in the recent
literature (see, e.g., Zakharov, 1992; Mueller, 1992; Lovett-Turner
and Maxwell, 1994; Vainshtein and Zakharov, 1994, Soper and Surguladze, 1995).

   After a brief historical review,
we turn to the main subject of the present article. Namely, we discuss
the analytical high order perturbative calculations of several physical
observables which have been completed recently with the help
of the above mentioned methods, algorithms and computer programs.
First, we consider the analytical calculation of
$R(s)$ in electron - positron annihilation at the four-loop level
of perturbative QCD (Surguladze and Samuel, 1991a,b; Gorishny,
Kataev and Larin, 1991), which turned out to be the most difficult
among the problems of this type.
This is the first and so far the only
four-loop calculation of a physical quantity in QCD.
\footnote{
This calculation was attempted earlier by Gorishny, Kataev and Larin (1988)
but, unfortunately, errors were found.
}
As a byproduct, the four-loop $R_{\tau}$ in $\tau$ decay
(Gorishny, Kataev and Larin, 1991; Samuel and Surguladze, 1991) and four-loop
QED $\beta$ function (Surguladze, 1990; Gorishny, Kataev and Larin, 1990)
have been evaluated.
\footnote{ For the joint publication of the results of the two
independent calculations of the four-loop
QED $\beta$-function, see Gorishny, Kataev, Larin and Surguladze (1991a).
}
For earlier works, we mention, for instance, the calculation of the
three-loop correction to $R(s)$ in electron - positron annihilation
(Chetyrkin, Kataev and Tkachov, 1979;
Dine and Sapirstein, 1979;
Celmaster and Gonsalves, 1980),
the calculation of the three-loop
QCD ${\beta}$ function (Tarasov, Vladimirov and Zharkov, 1980) and the
calculation of the three-loop anomalous dimensions of quark masses
(Tarasov, 1982).
We would also like to list some other three- and two-loop calculations.
These are:the calculation of the total decay width of the neutral Higgs boson
into hadrons at the three-loop level
(Gorishny, Kataev, Larin and Surguladze, 1990, 1991b;
Surguladze, 1994a,b),
the calculation of the two- and three-loop Wilson coefficients in QCD
sum rules (Surguladze and Tkachov, 1986, 1988, 1989b, 1990),
the calculation of the two-loop anomalous dimensions of the proton
current (Pivovarov and Surguladze, 1991). So far only one
five-loop calculation exists. This is the calculation of the five-loop
renormalization group functions in $\phi^4$-theory (Kleinert {\it et al.},
1991).

The scope of the present paper is limited and
we are not planning to review  perturbative
QCD. This has already been done and excellent reviews exist. We
recommend, for instance, the recent work {\it Handbook of
Perturbative QCD} by CTEQ collaboration (Brock, {\it et al.}), edited by
G.~Sterman (1993). Here we focus on a somewhat simplified description of the
key methods which allow one to perform analytical high order
perturbative calculations up to and including the four-loop level.
As an example, we will demonstrate the main points of the calculation of
$\sigma_{\mbox{\scriptsize{tot}}}(e^{+}e^{-} \rightarrow \mbox{hadrons})$,
$\Gamma(\tau^{-} \rightarrow \nu_{\tau} + \mbox{hadrons})$,
$\Gamma(H \rightarrow \mbox{hadrons})$ and the QED $\beta$ function.
We also outline the calculation
of the Wilson coefficient functions of higher twist operators
in the operator product expansion and discuss various
approaches to resolve the renormalization group ambiguity of
perturbation theory predictions.

   The paper is organised as follows. In the 2nd section we introduce
our notation and present some general relations. In this section we
discuss the relevant methods and tools of perturbative QCD.
We briefly consider the necessary dispersion
relation, the operator product expansion (OPE), the renormalization
relations and the method for evaluation of the renormalization constants.
We also discuss the main ideas of the method of projectors for calculating
Wilson coefficients in OPE. In the 3rd section we evaluate the quantity
$\Gamma(H \rightarrow \mbox{hadrons})$ at the three-loop level. In the 4th
section we calculate the corrections to the correlation functions due to the
nonvanishing quark masses. In the 5th section we describe the calculation of
the Wilson coefficient functions of the dim=4 operators in the OPE of the
two-point correlation function of quark currents.
In the 6th section we describe the four-loop calculation of
$\sigma_{\mbox{\scriptsize{tot}}}(e^{+}e^{-} \rightarrow \mbox{hadrons})$.
Sections 7 and 8 are dedicated to the evaluation of
$\Gamma(\tau^{-} \rightarrow \nu_{\tau} + \mbox{hadrons})$ and the QED
$\beta$ function respectively.
In section 9 we discuss the problem of the renormalization
group ambiguity of perturbative QCD results. As an example, we consider
calculated quantities and use the known approaches to try to fix the
scheme-scale parameter within the one parametric family of MS-type
schemes. Next, we outline the original method of scheme-invariant
analysis and optimization procedure by Stevenson (1981a,b).
The paper ends with summarizing notes.

\renewcommand{\thesection}{\Roman{section}}
\section{\bf Calculational methods}
\renewcommand{\thesection}{\arabic{section}}
\setcounter{equation}{0}
\subsection{\tenbf Notation and general relations of perturbative QCD}

    Throughout this paper we work within the standard
 model of strong interactions - QCD. For a review on QCD see, for
 example, Marciano and Pagels (1978), Mueller (1981), Reya (1981), and
Altarelli (1982). For a textbook see, e.g., Yndurain (1983), Quigg (1986),
Muta (1987), and Ellis and Stirling (1990).
 For the most recent source see, e.g., {\it Handbook of
 Perturbative QCD} by CTEQ collaboration (Brock {\it et al.}),
edited by G.~Sterman (1993).
The four-loop QED calculations will be discussed in section 8.

   The Lagrangian density of standard QCD is
\begin{eqnarray}
L(x)=-1/4(G_{\mu\nu}^{a})^{2}-\frac{1}{2\alpha_G}
(\partial^{\mu}A_{\mu}^{a})^{2}
+\sum_{f}\overline{q}_{f}(i\hat{\partial}-m_{f})q_{f}
+g\sum_{f}\overline{q}_{f}T^{a}\hat{A}^{a}q_{f}        \nonumber\\
+\partial^{\mu}c^{a^{\dag}}(\partial_{\mu}\delta^{ac}
+gf^{abc}A_{\mu}^{b})c^{c},
\label{eq:lagrangian}
\end{eqnarray}
where
$G_{\mu\nu}^{a}=\partial_{\mu}A_{\nu}^{a}-\partial_{\nu}A_{\mu}^{a}
+gf^{abc}A_{\mu}^{b}A_{\nu}^{c}$ \ ($a=1,2,...,8$) are the
 Yang-Mills field (Yang and Mills, 1954) strengths, \ $A^{a}$ and $q_{f}$ are
gluon and
quark fields, $m_{f}$ are the quark masses, \ $c^{a}$ are the
Faddeev-Popov ghosts and $\alpha_{\mbox{\tiny G}}$ is the gauge parameter.
 We use the standard notation
$\hat{\partial}=\gamma^{\mu}\partial_{\mu}$ and
$\hat{A}^{a}=\gamma^{\mu}A_{\mu}^{a}$.
The index $f$ enumerates the
quark flavors, total number of which is $N$. The generators $T^{a}$
of the SU$_{\mbox{\scriptsize c}}$(N) gauge group,
the structure constants $f^{abc}$ and $d^{abc}$ obey the following relations
\begin{eqnarray}
[T^a,T^b]=if^{abc}T^c,\hspace{2mm}
\{T^a,T^b\}=\frac{1}{N}\delta^{ab}+d^{abc}T^c, \nonumber\\
f^{acd}f^{bcd}=C_A\delta^{ab},\hspace{2mm}
T^aT^a=C_F\hat{\bf 1}, \hspace{2mm}
\mbox{tr}T^aT^b=T\delta^{ab}.
\label{eq:casimirs}
\end{eqnarray}
The eigenvalues of the Casimir operators for the adjoint ($N_A=8$) and the
fundamental ($N_F=3$) representations of SU$_{\mbox{\scriptsize c}}$(3) are
\begin{equation}
C_{A}=3,\hspace{2mm}C_{F}=4/3, \hspace{2mm} \mbox{and} \hspace{2mm}
 T=1/2,\hspace{2mm} d^{abc}d^{abc}=40/3.
\label{eq:casimirsnum}
\end{equation}

  We use the standard QCD Feynman rules (see, e.g., Abers and Lee, 1973;
Muta, 1987).

\vspace{3mm}

{\bf Propagators}

\vspace{4mm}

\hspace{7mm} quark \hspace{43mm} $= \frac{1}{i}\frac{m+\hat{P}}
                                           {m^2-P^2}\delta_{ij}$

\vspace{13mm}

\hspace{7mm} gluon \hspace{43mm} $= \frac{1}{i}\frac{\delta_{ab}}{P^2}
                                       \left[g^{\mu\nu}
                                      -(1-\alpha_{\mbox{\tiny G}})
                                      \frac{P_{\mu}P_{\nu}}{P^2}\right]$
\vspace{13mm}

\hspace{7mm} ghost \hspace{43mm} $= \frac{1}{i}\frac{\delta_{ab}}{P^2}$

\vspace{10mm}

{\bf Vertices}

\vspace{7mm}

\hspace{7mm} quark-quark-gluon \hspace{43mm} $= ig\gamma^{\mu}T^{a}_{ij}$

\vspace{29mm}

\hspace{7mm} ghost-ghost-gluon \hspace{43mm} $= igf^{abc}P_{\mu}$

\vspace{26mm}

\hspace{7mm} 3-gluon \hspace{43mm}
                        $=gf^{abc}[g_{\mu\nu}(q-p)_{\lambda}
                          +g_{\nu\lambda}(k-q)_{\mu}
                                +g_{\mu\lambda}(p-k)_{\nu}]$
\vspace{29mm}

\hspace{7mm} 4-gluon \hspace{43mm} $=ig^2[f^{abe}f^{cde}(g^{\mu\lambda}
                                              g^{\nu\rho}
                                          -g^{\mu\rho}g^{\nu\lambda})$

\hspace{74mm}      $+f^{ace}f^{bde}(g^{\mu\nu}
                             g^{\lambda\rho}-g^{\mu\rho}g^{\nu\lambda})$

\hspace{74mm}      $+f^{ade}f^{cbe}(g^{\mu\lambda}g^{\nu\rho}
                                   -g^{\mu\nu}g^{\lambda\rho})]$

\vspace{11mm}

The sum of all momenta coming in each vertex of the
               Feynman diagram is zero (momentum conservation).

\vspace{4mm}

{\bf Factors}

\vspace{3mm}

(-1) for each closed fermion or ghost loop

\vspace{3mm}

Statistical factors (for derivations see, e.g., 't~Hooft and Veltman, 1973):

\vspace{9mm}

$\frac{1}{2}$ for each graph (subgraph)

\vspace{14mm}

$\frac{1}{6}$ for each graph (subgraph) \hspace{5cm} etc.

\vspace{9mm}

{\bf Integration}

\vspace{3mm}

Each loop corresponds to the integration \hspace{7mm}
                                  $\int\frac{d^4P}{(2\pi)^4}$.

\vspace{7mm}

    In general, the Feynman integral constructed according to the above rules
is divergent.
There are two kind of divergences. One, the so called ultraviolet (UV)
 divergence is due to large integration
momenta and the other one - the so called infrared divergence is associated
 with the small integration momenta in the massless limit. The most convenient
 regularization of Feynman integrals is dimensional regularization
('t~Hooft and Veltman, 1972;  Bollini and Giambiagi, 1972; Ashmore, 1972;
 Cicuta and Montaldi, 1972),
where the space-time dimension is analytically continued from the physical
 value, 4, to a complex value $D=4-2\varepsilon$.
In the limit $\varepsilon \rightarrow 0$, the divergences appear as poles
$1/\varepsilon$, defining the counterterms. One of the remarkable properties
of dimensional regularization is that the Ward identities implied by gauge
invariance are maintained for arbitrary space-time dimension D, in
contrast with the old Pauli-Villars regularization (Pauli and Villars, 1949).
Another useful property is a convenience in practical multiloop calculations.
Thus, in dimensional regularization we formally replace
$\int\frac{d^4P}{(2\pi)^4}$ $\rightarrow$ $\int\frac{d^{D}P}{(2\pi)^D}$.
It is straightforward to extend the all necessary tensor algebra into
$D$-dimensions.
For example, $g^{\mu\nu}g_{\mu\nu}=D$, $\mbox{Tr}{\gamma_{\mu}\gamma_{\nu}}
=2^{D/2}g_{\mu\nu}$,
etc. For the complete list of formulae see, e.g., Collins (1984) and also
Narison (1982).
Note, however, that the extension of the usual definition of the
matrix $\gamma_{5}$\\
\begin{center}
$\gamma_{5} = \frac{1}{4!}\varepsilon_{\alpha\beta\mu\nu}\gamma_{\alpha}
\gamma_{\beta}\gamma_{\mu}\gamma_{\nu}$\\
\end{center}
is not straightforward. The totally antisymmetric tensor
$\varepsilon_{\alpha\beta\mu\nu}$ is defined only in the
 four-dimensional space.
In some cases the calculation of the quantities involving
$\gamma_{5}$ is still possible within dimensional regularization.
 For a discussion of the problem of
$\gamma_{5}$ in dimensional regularization see Delbourgo and Akyeampong
 (1974), Trueman (1979), Bonneau (1980), Narison (1982), Collins (1984),
and Larin (1993).
For a calculation involving $\gamma_{5}$ within
dimensional regularization see, e.g., Pivovarov and Surguladze (1991).

In order to get finite physical quantities,
the divergences in dimensionally regularized Feynman
integrals, appearing as poles in $1/\varepsilon$, need to be subtracted
by adopting of some specific rule.
This rule is usually called a renormalization scheme.
Throughout this paper we use 't~Hooft's minimal subtraction (MS) type
 scheme ('t~Hooft, 1973). The subtraction of divergences is equivalent to
 the redefinition (renormalization) of the parameters (coupling,
 mass and gauge fixing parameter)
and fields in the original ``bare'' lagrangian
\begin{center}
$\alpha_s^{\mbox{\tiny B}}=\mu^{2\varepsilon}Z_{\alpha_s}\alpha_s$,
 \hspace{1cm} $(g^{2}/4\pi\equiv\alpha_{s})$
\end{center}
\begin{equation}
m^{\mbox{\tiny B}} = mZ_{m},
\label{Zdefinitions}
\end{equation}
\begin{center}
$\alpha_{\mbox{\tiny G}}^{\mbox{\tiny B}} = \alpha_{\mbox{\tiny G}}
                                                  Z_{\mbox{\tiny G}}$.
\end{center}
$\mu$ is a quantity of dimension of mass which is introduced
within dimensional regularization in order to make an action
dimensionless. Superscript ``B''  denote the unrenormalized quantity.
We renormalize the gluon, quark and ghost fields analogously.
Within the MS scheme the N-point Green function
is renormalized in the following way
\begin{equation}
\Gamma(p_1,...,p_N,g,m,\alpha_{\mbox{\tiny G}},\mu) =
 Z_{\Gamma}\Gamma^{\mbox{\scriptsize B}}
                     (p_1,...,p_N,g,m,\alpha_{\mbox{\tiny G}}),
\label{Zgreen}
\end{equation}
where $Z_{\Gamma}$ is a polynomial in $1/\varepsilon$, and thus
multiplying by $Z_{\Gamma}$,
we subtract
only pole parts from the divergent $\Gamma^{\mbox{\scriptsize B}}$.
The evaluation of the renormalization constants $Z$ will be
discussed in the next subsections.

It is easy to see that the $\mu$ parameter entered through the
renormalization and hence the unrenormalized Green's function is
independent of $\mu$
\begin{displaymath}
\mu\frac{d}{d\mu}\Gamma^{\mbox{\scriptsize B}}(p_1,...,p_N,g,m,
                                   \alpha_{\mbox{\tiny G}}) = 0.
\end{displaymath}
Using eq.\ (\ref{Zgreen}) and expanding the full derivative we get
 the renormalization group equation in the following form
\begin{equation}
\left[\mu^2\frac{\partial}{\partial\mu^2}
+\beta(\alpha_s)\alpha_s\frac{\partial}{\partial\alpha_s}
-\gamma_{m}(\alpha_s)m\frac{\partial}{\partial m}
+\beta_{\mbox{\tiny G}}(\alpha_s)\frac{\partial}
{\partial\alpha_{\mbox{\tiny G}}}
-\gamma_{\Gamma}\right]
\Gamma(p_1,...,p_N,m,\alpha_s,\alpha_{\mbox{\tiny G}},\mu) = 0.
\label{RGE}
\end{equation}
The QCD renormalization group functions
 - the $\beta$-function and the anomalous dimension functions
 - $\gamma$ are defined in the following way
\begin{displaymath}
\alpha_s\beta(\alpha_s)=\mu^2\frac{d\alpha_s}{d\mu^2},
\end{displaymath}
\begin{displaymath}
\beta_{\mbox{\tiny G}}(\alpha_s)
=\mu^2\frac{d\alpha_{\mbox{\tiny G}}}{d\mu^2},
\end {displaymath}
\begin{equation}
\gamma_{m}(\alpha_s)=-\frac{\mu^2}{m}\frac{dm}{d\mu^2},
\label{eq:RGfunctions}
\end{equation}
\begin{displaymath}
\gamma_{\Gamma}(\alpha_s)=\frac{\mu^2}{Z_{\Gamma}}\frac{dZ_{\Gamma}}{d\mu^2},
\end{displaymath}
with bare coupling and mass fixed. In the present paper
 we use the renormalization group equation in the above form.
 The other forms are also known in the literature.
 The group properties
 of the renormalization was first discovered by
 Stueckelberg and Peterman (1953). The ultraviolet asymptotics
 of the  Green function
was studied by Gell-Mann and Low (1954) in quantum electrodynamics using
 the group of multiplicative renormalizations.
The renormalization group formalism was further
 developed in the original works by Bogolyubov and Shirkov
 (1955, 1956a,b).
 For the detailed monograph see
 Bogolyubov and Shirkov (1980). The renormalization group
 equation was studied by Ovsyannikov (1956),
 Callan (1970), and Symanzik (1970). For a recent historical
 review see Shirkov (1992) and references therein.

The renormalization group $\beta$-function and anomalous
 dimensions of quark masses are calculated up to the three-loop level
 (Tarasov, Vladimirov and Zharkov, 1980; Tarasov, 1982).
The QCD $\beta$-function up to and including the three-loop
 level in MS type schemes is
\begin{equation}
\beta(\alpha_{s})=-\beta_{0}\frac{\alpha_s}{\pi}
-\beta_1(\frac{\alpha_s}{\pi})^2
-\beta_2(\frac{\alpha_s}{\pi})^3+O(\alpha_s^4),
\label{eq:beta}
\end{equation}
where (Tarasov, Vladimirov and Zharkov, 1980)

\vspace{3mm}

\noindent
$\beta_0=\frac{1}{4}\biggl(\frac{11}{3}C_A
-\frac{4}{3}TN\biggr)$,

\noindent
$\beta_1=\frac{1}{16}\biggl(\frac{34}{3}C_A^2-\frac{20}{3}C_ATN
-4C_FTN\biggr)$,

\noindent
$\beta_2=\frac{1}{64}\biggl(
\frac{2857}{54}C_A^3-\frac{1415}{27}C_A^2TN
+\frac{158}{27}C_AT^2N^2-\frac{205}{9}C_AC_FTN
+\frac{44}{9}C_FT^2N^2+2C_F^2TN\biggr)$.

\vspace{3mm}

\noindent
The quark mass anomalous dimension up to and including three-loop level is
\begin{equation}
\gamma_{m}(\alpha_{s})=\gamma_{0}\frac{\alpha_s}{\pi}
+\gamma_1(\frac{\alpha_s}{\pi})^2
+\gamma_2(\frac{\alpha_s}{\pi})^3+O(\alpha_s^4),
\label{eq:gamma}
\end{equation}
where (Tarasov, 1982)

\vspace{3mm}

\noindent
$\gamma_0=\frac{3}{4}C_F$,

\noindent
$\gamma_1=\frac{1}{16}\biggl(\frac{3}{2}C_{F}^2+\frac{97}{6}C_{F}C_A
-\frac{10}{3}C_FTN\biggr)$,

\noindent
$\gamma_2=\frac{1}{64}\biggl[\frac{129}{2}C_{F}^3-\frac{129}{4}C_{F}^{2}C_{A}
+\frac{11413}{108}C_{F}C_{A}^2-(46-48\zeta(3))C_{F}^{2}TN$

\hspace{7cm} $-\biggl(\frac{556}{27}+48\zeta(3)\biggr)C_{F}C_{A}TN
             -\frac{140}{27}C_{F}T^2N^2\biggr]$.

\vspace{3mm}

As it was shown by Caswell and Wilczek (1974) and Banyai, Marculescu
 and Vescan (1974), the above renormalization group functions
 are gauge independent, which greatly simplifies their evaluation.
 In fact, the QCD $\beta$-function and the quark mass anomalous dimension
have been evaluated in the Feynman gauge $\alpha_{\mbox{\tiny G}}=1$.
We note that the perturbative coefficients of the renormalization
 group functions are the same within the one parametric family
 of the MS type schemes.
Note also the independence of these perturbative coefficients
on the quark masses by their definition within the MS type schemes.

\renewcommand{\thesection}{\arabic{section}}
\subsection{\tenbf Vacuum polarization function and Dispersion relation}

\indent
The vacuum polarization functions for various types of quark
 currents are crucial in the theoretical evaluation of total
 cross sections and decay widths. Indeed, for example, the quantity
$\sigma_{\mbox{\scriptsize{tot}}}(e^{+}e^{-}\rightarrow \mbox{hadrons})$,
according
 to the well known optical theorem (see, e.g., the textbook by
 Bogolyubov and Shirkov, 1980),
is proportional to the imaginary part of the function
 $\Pi(-q^2+i0)$, defined from the hadronic vacuum polarization
 function
\begin{equation}
\Pi_{\mu\nu}(q)=i\int e^{iqx}<Tj_\mu(x)j_\nu(0)>_0d^4x
=(g_{\mu\nu}Q^2-Q_{\mu}Q_{\nu})\Pi(Q^2)\frac{1}{(4\pi)^2}.
\label{eq:pifunction}
\end{equation}
Here, $j_{\mu}(x)=Q_f\overline{q}_f\gamma_{\mu}q_f$, $Q_f$ is the electric
charge of the quark of flavor $f$ and $Q^2=-q^2$ is the
Euclidean momentum squared. The sum over all participating quark
 flavors is assumed in $\Pi$.
 The transverse form in the r.h.s.
 is conditioned by the conservation of electromagnetic currents.
 In this paper we also consider the two-point function of quark
 axial vector currents associated with the quantity
 $\Gamma(Z \rightarrow \mbox{hadrons})$ and two-point function of
 quark scalar currents associated with the quantity
$\Gamma(H \rightarrow \mbox{hadrons})$ - the
 total decay width of the neutral Standard Model Higgs boson into hadrons.

   The renormalized vacuum polarization function obeys the
 dispersion relation
\begin{equation}
\Pi(Q^2) = \frac{4}{3}\int_{s_{0}}^{\infty}\frac{R(s)}{s+Q^2}ds
 - \mbox{subtractions},
\label{eq:disprelat}
\end{equation}
where
\begin{equation}
R(s)=\frac{\sigma_{\mbox{\scriptsize{tot}}}(e^{+}e^{-}\rightarrow
\mbox{hadrons})}
                 {\sigma(e^{+}e^{-}\rightarrow \mu^{+}\mu^{-})}
=\frac{3}{4\pi} \mbox{Im}\Pi(s+i0).
\label{eq:Rsdefin}
\end{equation}
Recall also that the muon pair production cross-section
$\sigma(e^{+}e^{-}\rightarrow \mu^{+}\mu^{-})=4\pi\alpha^2/3s$,
where $\alpha=e^2/4\pi$ is the electromagnetic fine structure constant.
The above dispersion relation allows one to connect the
 experimentally measurable quantity $R(s)$ to the $\Pi(Q^2)$
 calculable perturbatively in the deep Euclidean region
 ($Q^2$ is large compared to the typical hadron mass).
For the discussion on theoretical calculability of R(s)
see earlier references: Adler (1974), Appelquist and Politzer (1975),
De~R\'{u}jula and Georgi (1976), Poggio, Quinn and Weinberg (1976),
Shankar (1977), and Barnett, Dine and McLerran (1980).
The combination of the idea of local duality in the dispersion
 relations (Logunov, Soloviov and Tavkhelidze, 1967) and the
 Operator Product Expansion technique (Wilson, 1969) became a basis
 of various versions of QCD sum rules (Shifman, Vainshtein
 and Zakharov, 1979; Novikov {\it et al.}, 1978, 1985;
 Krasnikov, Pivovarov and Tavkhelidze, 1983;
 Shifman, 1992 and references therein).
 The methods of QCD sum rules  are widely used to obtain
 quantitative information on the observed hadron spectrum
and to extract the fundamental theoretical parameters.

In practice, sometimes it is more convenient to introduce the Adler
function (Adler, 1974)
\begin{equation}
D(Q^2)=-\frac{3}{4}\frac{\partial}{\partial \log Q^2}\Pi(Q^2)
=Q^2\int_{s_0}^{\infty}\frac{R(s)}{(s+Q^2)^2}ds.
\label{eq:Ddefin}
\end{equation}
Derivative here avoids an inconvenient extra subtraction in the r.h.s.

The leading (parton) approximation of $D(Q^2)$ in the zero quark mass limit
coincides with $R(s)$
\begin{equation}
D(Q^2) = 3\sum_{f}Q_{f}^{2},
\label{eq:D0}
\end{equation}
where the sum runs over all participating  quark charges at the given
 energy.
 3 stands for the number of different colors. The leading ``non-QCD''
 contribution is completely free of ultraviolet divergences,
 while the $\Pi(Q^2)$ needs an additive renormalization
  even at the leading order.
 At higher orders of perturbative expansion of the $D$-function
 the ultraviolet divergences appear
 and one should employ a procedure (usually called
 renormalization scheme) for their subtraction order-by-order.
 Because of ambiguity in the choice of
 subtraction scheme, the amplitude calculated within the perturbation theory
 depends on nonphysical parameters. Within the one-parametric family
of the MS type schemes (t~'Hooft, 1973) such a parameter is usually
called $\mu$. Thus, up to
power corrections, the $D$-amplitude will be a function of
 $\log(\mu^2/Q^2)$
and the strong coupling $\alpha_s$. On the other hand, since $D$
 is connected to the observable $R(s)$, it can not depend on our
 subjective choice of nonphysical parameter $\mu$. This can be
 achieved if the strong coupling becomes a function of $\mu$,
providing independence of observables on the choice of parameter $\mu$.
 Here, it is asumed that all orders of perturbation theory are summed up.
 Otherwise, if one considers a truncated series, the $\mu$ dependence
 remains. The problem of scheme-scale dependence and some possible
solutions will be discussed later in this review. The set of
 transformations which leave observables independent of renormalization
 parameters has a group character and forms the renormalization group.
 The renormalization group in renormalizable theories (like QCD)
 fixes the dependence of the coupling on the $\mu$-parameter.

The function $D(Q^2)$ calculated in perturbative QCD within the MS
type schemes obeys the  renormalization group equation
\begin{equation}
\left(\mu^2\frac{\partial}{\partial\mu^2}
+\beta(\alpha_s)\alpha_s\frac{\partial}{\partial\alpha_s}
-\gamma_m(\alpha_s)m\frac{\partial}
{\partial m}\right)D(\mu^2/Q^2,m,\alpha_s)=0.
\label{eq:RGD}
\end{equation}
Below we consider the limit of the massless light quarks and
the infinitely large top mass which decouples (Appelquist and Carazzone, 1975).
The solution of eq.\ (\ref{eq:RGD}) at $\mu^2=Q^2$ is
\begin{equation}
D(\mu^2/Q^2,\alpha_s(\mu))=D(1,\alpha_{s}(Q))=\sum_{i\geq0}
R_i(\alpha_{s}(Q)/\pi)^i,
\label{eq:RGDsolut}
\end{equation}
where the $\alpha_{s}(\mu^2)$ is the running coupling, usually
parametrized up to the three-loop level as follows
\begin{equation}
\frac{\alpha_s(\mu^2)}{\pi}=\frac{1}{\beta_0 L}-\frac{\beta_1 \log L}
{\beta_0^3 L^2}+\frac{1}{\beta_0^5 L^3}(\beta_1^2 \log^2 L-\beta_1^2 \log L
+\beta_2 \beta_0-\beta_{1}^{2})+O(L^{-4}),
\label{eq:Asparametr}
\end{equation}
where $L=\log (\mu^2/\Lambda^2)$. Parametrization (\ref{eq:Asparametr})
has the same form and
the QCD $\beta$-function coefficients are the same within the MS type
schemes. The scale parameter $\Lambda$ depends on the particular
modification of the MS prescription. In fact, $\Lambda$ is used
to parametrize other versions of renormalization prescription as well.
It is shown by Celmaster and Gonsalves (1979)
that the transformation relations valid to all orders between $\Lambda$'s
defined by any two renormalization prescription can be deduced from a
one-loop calculation. Comparing the bare coupling constants
within different renormalization prescriptions and using the results
for the one-loop renormalization constants and the property of asymptotic
freedom, one obtains for, e.g.,
momentum subtraction (MOM) and MS schemes (Celmaster and Gonsalves, 1979)
\begin{equation}
\Lambda_{\mbox{\scriptsize MS}}=\Lambda_{\mbox{\scriptsize MOM}}
\mbox{exp}\biggl[\frac{A(\alpha_{\mbox{\tiny G}},N)}
{4\beta_0}\biggr],
\label{eq:MOMMS}
\end{equation}
where
\begin{eqnarray}
\lefteqn{\hspace{-75mm} A(\alpha_{\mbox{\tiny G}},N)=C_A\biggl[
-\frac{11}{6}(\gamma_{\mbox{\tiny E}}-\ln 4\pi)
+\frac{11}{3}+\frac{23}{72}I
+\frac{3}{8}\alpha_{\mbox{\tiny G}}(1-I)
-\frac{1}{12}\alpha_{\mbox{\tiny G}}^2(3-I)
+\frac{1}{24}\alpha_{\mbox{\tiny G}}^3\biggr]} \nonumber\\
 && \hspace{-64mm}
+TN\biggl[\frac{2}{3}(\gamma_{\mbox{\tiny E}}-\ln 4\pi)
-\frac{4}{3}-\frac{8}{9}I\biggr]
\label{eq:A}
\end{eqnarray}
and the integral
\begin{equation}
I=-2\int_0^1 \frac{\ln x}{x^2-x+1}dx=2.3439072...
\label{eq:I}
\end{equation}
One note due to Stevenson (1981b, 1994) is in order.
Despite its convenient form, the parametrization
(\ref{eq:Asparametr}) produces an additional ambiguity due to the
freedom with a particular definition of $\Lambda$ parameter, even when the
renormalization prescription is already specified. This problem
was discussed by Abbot (1980), Shirkov (1980),
Stevenson (1981b), Monsay and Rosenzweig (1981)
and Radyushkin (1983). In fact, one can take advantage of this
freedom in the choice of $\Lambda$ and try to optimize the expansion
in $1/L$. Indeed, as was shown by Radyushkin (1983),
if one takes $0.6\Lambda$ in eq.\ (\ref{eq:Asparametr})
instead of standard (Buras, Floratos, Ross and Sachrajda, 1977)
$\Lambda$ then the $1/L^2$ and $1/L^3$ terms contribute only a few
percent for a reasonably wide range of $\mu$. On the other hand,
Stevenson (1981b, 1994) has suggested to avoid the entire problem of
ambiguity in the definition of $\Lambda$ by abandoning the $1/L$ expansion
and solving the renormalization group equation (\ref{eq:RGfunctions})
for $\alpha_s$ and resulting transcendental equation numerically,
using the truncated $\beta$ function.

According to the operator product expansion technique (Wilson, 1969), one can
separate perturbative and nonperturbative contributions to the
function $\Pi(Q^2)$. As shown by Shifman, Vainshtein and Zakharov (1979),
this function can be represented in the following form
\begin{equation}
\Pi(Q^2)=\mbox{perturbation theory}
+\sum_{n\geq2}\frac{C_{n}(Q)<O_n>_0}{Q^{2n}}
+\mbox{instanton contributions,}
\label{eq:PIOPE}
\end{equation}
where $<O_n>_{0}$ denote vacuum condensates parametrizing the
nonperturbative contributions and $C_n(Q)$ are their coefficient
functions. The last term in the above equation describes the instanton
contributions, which, in the case of electromagnetic currents, was estimated
to be small (Krasnikov and Tavkhelidze, 1982;
Kartvelishvili and Margvelashvili, 1995). The coefficient functions
of the condensates can be calculated within perturbation theory.
High order perturbative corrections to the coefficient functions of
dimension 4 and 6 power terms  have been
calculated in Loladze, Surguladze and Tkachov (1984, 1985), Surguladze
 and Tkachov (1989b, 1990), Chetyrkin, Gorishny and Spiridonov (1985), and
 Lanin, Chetyrkin and Spiridonov (1986). In subsection E we discuss
the method for evaluation of Wilson coefficient functions. Examples
will be outlined in section 4.
Note, that we consider the  region of very high energies where, in fact,
only perturbation theory contributions survive in $\Pi(Q^2)$.
The nonperturbative corrections could have some (small) effect in the case of,
for instance, $\tau$ lepton decay (see section 7). Note also that, in fact, the
effects of neglected light quark masses are not entirely negligible in some
phenomenological applications (see section 4).

\vspace{2cm}

\renewcommand{\thesection}{\arabic{section}}
\subsection{\tenbf Renormalization relations}

\indent
There are several approaches for the ultraviolet renormalization
 of Green's functions known in the literature. Throughout this
 paper we use 't Hooft's minimal subtruction method ('t Hooft, 1971, 1973).
 For alternative prescriptions we refer to the works by
 Weinberg (1967), Gell-Mann and Low (1954), Callan (1970), Symanzik (1970),
and Collins, Wilczek and Zee (1978).
 For an analysis of various renormalization methods see Collins and
 Macfarlane (1974). For a review see, e.g., Narison (1982),
 the textbook by Collins (1984) and references therein.
We focus on the renormalization relations for the
 two point correlation function of quark currents relevant
for the further evaluation of total cross sections and decay widths.

It is known that the vacuum polarization function is renormalized additively
\begin{equation}
\Pi(\mu^2/Q^2,\alpha_s)
=\Pi^{\mbox{\scriptsize B}}
(\mu^2/Q^2,\alpha_s^{\mbox{\tiny B}})+Z_{\Pi}\equiv \mbox{finite}.
\label{eq:PiR}
\end{equation}
The bare coupling $\alpha_s^{\mbox{\tiny B}}$ is related to the
renormalized one by the
relation (\ref{Zdefinitions}).
The perturbative expansion for $Z_{\alpha_s}$ can be found
based on eqs.\ (\ref{eq:RGfunctions}) and (\ref{eq:beta}), the MS definition
of $Z_{\alpha_s}$ and the renormalization group equation
\begin{equation}
\mu^2\frac{d}{d\mu^2}\alpha_s^{\mbox{\tiny B}}=0.
\label{eq:AsR1}
\end{equation}
We obtain
\begin{equation}
Z_{\alpha_s}=1-\frac{\alpha_s}{\pi}\frac{\beta_0}{\varepsilon}
+\left(\frac{\alpha_s}{\pi}\right)^2 \left(\frac{\beta_0^2}{\varepsilon^2}
-\frac{\beta_1}{2\varepsilon}\right)
-\left(\frac{\alpha_s}{\pi}\right)^3
\biggl(\frac{\beta_0^3}{\varepsilon^3}
-\frac{7}{6}\frac{\beta_0\beta_1}{\varepsilon^2}
+\frac{\beta_2}{3\varepsilon}\biggr)
+O(\alpha_s^4).
\label{eq:AsR}
\end{equation}
In general, the polarization function depends on quark masses and
we will need the relation between ``bare'' and renormalized masses
up to $O(\alpha_s^2)$ (Tarasov, 1982)
\begin{eqnarray}
\lefteqn{(m_{f}^{\mbox{\tiny B}})^2 =
    m_{f}^{2} \biggl\{ 1
       -\left(\frac{\alpha_{s}}{4\pi}\right)\frac{6C_F}{\varepsilon}
     +\left(\frac{\alpha_{s}}{4\pi}\right)^2C_{F} \biggl[
          \biggl(11C_{A}+18C_{F}-4TN\biggr)\frac{1}{\varepsilon^2}}
                                                               \nonumber\\
 && \quad \hspace{43mm}
         -\biggl(\frac{97}{6}C_{A}+\frac{3}{2}C_{F}-\frac{10}{3}TN\biggr)
               \frac{1}{\varepsilon} \biggr]+O(\alpha_{s}^{3}) \biggr\}.
\label{eq:mrenorm}
\end{eqnarray}

Within the minimal subtraction prescription ('t Hooft, 1973)
 the renormalization constant $Z_{\Pi}$ can be expressed
 as the following double sum
\begin{equation}
Z_{\Pi}=\sum_{\stackrel{-l\leq k<0}{l>0}}
        \left(\frac{\alpha_s}{\pi}\right)^{l-1}Z_{l,k}
\varepsilon^k,
\label{eq:Zexpans}
\end{equation}
where $Z_{lk}$ are numbers. Furthermore, for the ``bare'' vacuum polarization
function one has the following expansion in a perturbation series
\begin{equation}
\Pi^{\mbox{\scriptsize B}}\left(\frac{\mu^2}{Q^2},
\alpha_s^{\mbox{\tiny B}}\right)=
\sum_{\stackrel{-l\leq k}{l>0}}
\left(\frac{\alpha_s^{\mbox{\tiny B}}}{\pi}\right)^{l-1}
\left(\frac{\mu^2}{Q^2}\right)^{l\varepsilon}\Pi_{l,k}\varepsilon^k,
\label{eq:Piexpans}
\end{equation}
where the first index denotes the number of loops of the corresponding
Feynman diagrams at the given order of $\alpha_s$.
Substituting eqs.\ (\ref{eq:Piexpans}) and (\ref{eq:PiR})
 into the definition (\ref{eq:Ddefin}),
after the renormalization of the coupling via (\ref{eq:AsR}) we obtain
at $\mu^2=Q^2$
\begin{eqnarray}
\lefteqn{\hspace{-7mm}D(\alpha_{s})=\frac{3}{4} \biggl\{ \Pi_{1,-1}
                         +\frac{\alpha_{s}}{\pi}
     \biggl[2\Pi_{2,-2}\frac{1}{\varepsilon}+2\Pi_{2,-1}\biggr]
    +\left(\frac{\alpha_{s}}{\pi}\right)^2
 \biggl[
   \frac{1}{\varepsilon^2}\left(3\Pi_{3,-3}-2\beta_0\Pi_{2,-2}\right)}
                                                                \nonumber\\
 && \quad \hspace{51mm}
  +\frac{1}{\varepsilon}  \left(3\Pi_{3,-2}-2\beta_0\Pi_{2,-1}\right)
          +\left(3\Pi_{3,-1}-2\beta_0\Pi_{2,0}\right)\biggr]     \nonumber\\
 && \quad +\left(\frac{\alpha_{s}}{\pi}\right)^3
\biggl[
   \frac{1}{\varepsilon^3}\left(4\Pi_{4,-4}-6\beta_0\Pi_{3,-3}
                                    +2\beta_0^2\Pi_{2,-2}\right) \nonumber\\
 && \quad \hspace{16mm}
     +\frac{1}{\varepsilon^2}\left(4\Pi_{4,-3}-6\beta_0\Pi_{3,-2}
                  -\beta_1\Pi_{2,-2}+2\beta_0^2\Pi_{2,-1}\right) \nonumber\\
 && \quad \hspace{16mm}
     +\frac{1}{\varepsilon}\left(4\Pi_{4,-2}-6\beta_0\Pi_{3,-1}
                   -\beta_1\Pi_{2,-1}+2\beta_0^2\Pi_{2,0}\right) \nonumber\\
 && \quad \hspace{16mm}
     +\left(4\Pi_{4,-1}-6\beta_0\Pi_{3,0}
            -\beta_1\Pi_{2,0}+2\beta_0^2\Pi_{2,1}\right) \biggr]
              +O(\alpha_{s}^{4}) \biggr\}.
\label{eq:DDexpans}
\end{eqnarray}
Because of the renormalization group invariance of $D(\mu^2/Q^2,\alpha_s)$,
in the above equation we take $\mu^2=Q^2$ to avoid unnecessary logarithms.
The renormalized expression for the $D$-function must be finite in the
 limit $\varepsilon \rightarrow 0$. Thus the coefficients of pole
 terms must vanish identically. This implies relations between the
 perturbative coefficients of
$\Pi$ and the QCD $\beta$-function. First, we note that prior to any
renormalization
the leading poles must cancel at each order of $\alpha_s$ in the
 sum of all relevant Feynman diagrams. As shown by the
 actual calculation, this happens in each gauge invariant set of diagrams.
\begin{equation}
\Pi_{4,-4} = \Pi_{3,-3} = \Pi_{2,-2} = 0.
\label{eq:Pirelat0}
\end{equation}
Moreover, from the cancellation of nonleading poles we get
\begin{displaymath}
3\Pi_{3,-2}-2\beta_{0}\Pi_{2,-1} = 0,
\end{displaymath}
\begin{equation}
4\Pi_{4,-3}-6\beta_{0}\Pi_{3,-2}+2\beta_{0}^2\Pi_{2,-1} = 0,
\label{eq:Pirelat1}
\end{equation}
\begin{displaymath}
4\Pi_{4,-2}-6\beta_{0}\Pi_{3,-1}-\beta_{1}\Pi_{2,-1}
                                +2\beta_{0}^2\Pi_{2,0} = 0.
\end{displaymath}
The above relations provide powerful tests of the calculation
 at its intermediate stages and are crucial.

{}From eq.\ (\ref{eq:PiR}) we see that
fully renormalized $\Pi(Q^2,\alpha_s)$ must be finite. Thus,
 substituting eqs.\ (\ref{eq:AsR}) -
(\ref{eq:Piexpans}) and (\ref{eq:Pirelat0}) in eq.\ (\ref{eq:PiR})
 we obtain the following expression for the divergent part of
 $\Pi(\mu^2/Q^2,\alpha_s)$ at $\mu^2=Q^2$
\begin{eqnarray}
\lefteqn{\mbox{div}\Pi(\alpha_{s}) =
         \frac{1}{\varepsilon}(\Pi_{1,-1}+Z_{1,-1})
    + \frac{\alpha_{s}}{\pi} \biggl[
      \frac{1}{\varepsilon}(\Pi_{2,-1}+Z_{2,-1})\biggr]}
                                                           \nonumber \\
 && \quad
    +\left(\frac{\alpha_{s}}{\pi}\right)^2 \biggl[
          \frac{1}{\varepsilon^2}(\Pi_{3,-2}
                      -\beta_0\Pi_{2,-1}+Z_{3,-2})
         + \frac{1}{\varepsilon}(\Pi_{3,-1}
                      - \beta_0\Pi_{2,0}+Z_{3,-1}) \biggr]
                                                            \nonumber \\
 && \quad
       +\left(\frac{\alpha_{s}}{\pi}\right)^3 \biggl[
                 \frac{1}{\varepsilon^3}(\Pi_{4,-3}
                      -2\beta_0\Pi_{3,-2}
                      +\beta_0^2\Pi_{2,-1}
                      +Z_{4,-3})
                                                            \nonumber \\
 && \quad \hspace{16mm}
   +\frac{1}{\varepsilon^2}(\Pi_{4,-2}-2\beta_0\Pi_{3,-1}
                        +\beta_0^2\Pi_{2,0}
                        - \beta_1\Pi_{2,-1}/2+Z_{4,-2})
      \nonumber \\
 && \quad \hspace{16mm}
   +\frac{1}{\varepsilon}(\Pi_{4,-1}-2\beta_0\Pi_{3,0}
             +\beta_0^2\Pi_{2,1}
             -\beta_1\Pi_{2,0}/2+Z_{4,-1})\biggr]\hspace{2mm} \equiv 0.
\label{eq:PiPiexpans}
\end{eqnarray}
The leading poles in $Z_{\Pi}$ are absent at each order of
$\alpha_s$ ($Z_{2,-2}=Z_{3,-3}=Z_{4,-4}=0$) except the zeroth order.
Taking into account eq.\ (\ref{eq:Pirelat1}), we obtain the other
set of relations between the perturbative coefficients of $\Pi$, $Z$
 and QCD $\beta$-function
\begin{displaymath}
3Z_{3,-2}+\beta_0Z_{2,-1}=0,
\end{displaymath}
\begin{equation}
2Z_{4,-3}+\beta_0Z_{3,-2}=0,
\label{eq:Pirelat2}
\end{equation}
\begin{displaymath}
4Z_{4,-2}+2\beta_0Z_{3,-1}+\beta_1Z_{2,-1}=0.
\end{displaymath}

\vspace{3mm}

\begin{displaymath}
\Pi_{1,-1}=-Z_{1,-1},
\end{displaymath}
\begin{displaymath}
\Pi_{2,-1}=-Z_{2,-1},
\end{displaymath}
\begin{displaymath}
\Pi_{3,-2}=-Z_{3,-2}-\beta_0Z_{2,-1},
\end{displaymath}
\begin{equation}
\Pi_{3,-1}=-Z_{3,-1}+\beta_0\Pi_{2,0},
\label{eq:Pirelat3}
\end{equation}
\begin{displaymath}
\Pi_{4,-1}=-Z_{4,-1}+2\beta_0\Pi_{3,0}+\beta_1\Pi_{2,0}/2-\beta_0^2\Pi_{2,1},
\end{displaymath}
\begin{displaymath}
\Pi_{4,-2}=-Z_{4,-2}-2\beta_0Z_{3,-1}-\beta_1Z_{2,-1}/2+\beta_0^2\Pi_{2,0},
\end{displaymath}
\begin{displaymath}
\Pi_{4,-3}=-Z_{4,-3}-2\beta_0Z_{3,-2}-\beta_0^2Z_{2,-1}.
\end{displaymath}
In section 6, the above relations will be used in the
calculations of the four-loop total cross-section in electron-positron
annihilation.

\vspace{2cm}

\renewcommand{\thesection}{\arabic{section}}
\subsection{\tenbf Method for evaluation of renormalization constants}

We now discuss the evaluation of renormalization
constants within 't~Hooft's MS scheme ('t~Hooft, 1973), using
 Vladimirov's method (Vladimirov, 1978)  and the so-called
 infrared rearrangement procedure
(Vladimirov, 1980; Chetyrkin and Tkachov, 1982).

To calculate the renormalizaton constant $Z_{\Gamma}$ for the
one-particle-irreducible
Green's function $\Gamma$, it is convenient to use the following
representation (Vladimirov, 1978)
\begin{equation}
Z_{\Gamma}=1-{\cal K}R'\Gamma.
\label{eq:Kdef}
\end{equation}
The operator $\cal K$ picks out all singular terms from
the Laurent series in $\varepsilon$
\begin{displaymath}
{\cal K}\sum_{i}c_{i}\varepsilon^i=\sum_{i<0}c_{i}\varepsilon^i.
\end{displaymath}
$R'$ is defined by the recursive relation
\begin{equation}
R'G=G-\sum_{G_i} {\cal K}R'G_{1}...{\cal K}R'G_{n}\times
 G_{/(G_{1}\cup...\cup G_{n})},
\label{eq:KR'}
\end{equation}
where the sum runs over all sets of one-particle-irreducible
divergent subgraphs $G_i$ of the diagram $G$. $G_{/(G_{1}+...+G_{n})}$ is
 the diagram G with the subgraphs $G_{1},...,G_{n}$ shrunk to a point.
In fact, $R'$ is the ordinary Bogolyubov-Parasyuk $R$-operation (Bogolyubov
 and Parasyuk, 1955a,b, 1956, 1957; for a textbook see
 Bogolyubov and Shirkov, 1980)
without the last (overall) subtraction. Thus, $R'$ subtracts
 all ``internal'' divergences
only and is connected to the ordinary $R$-operation in the following way
\begin{displaymath}
R=(1-{\cal K})R'.
\end{displaymath}
To calculate the renormalization constant $Z$  in
eq.\ (\ref{eq:PiR}),  one should
write a diagram representation of $\Pi$ and apply ${\cal K}R'$
to the corresponding graphs ( eq.\ (\ref{eq:Kdef}) )
or, in other words, one should evaluate the
counterterms for each graph. The benefit
 of using relation (\ref{eq:Kdef}) is based
on the fact that the ${\cal K}R'$ for each diagram is a
polynomial in dimensional parameters (Collins, 1974; Speer 1974).
 This fundamental property of the
't~Hooft's minimal subtraction prescription is the basic idea of the
various versions of the infrared rearrangement technique
(Vladimirov, 1980; Chetyrkin and Tkachov, 1982).

As an example, we demonstrate the application of the ${\cal K}R'$ operation
 to the three-loop QCD diagram contributing to the $O(\alpha_{s}^2)$
 total cross section for the process $e^{+}e^{-} \rightarrow \mbox{hadrons}$.

\newpage

\begin{displaymath}
R'\biggl\{ \hspace{2cm} \biggr\} = \hspace{2cm}
  -2{\cal K}R'\biggl\{ \hspace{1cm} \biggr\}\hspace{1.5cm}
  -2{\cal K}R'\biggl\{ \hspace{1cm} \biggr\}\hspace{1.5cm}
\end{displaymath}
\begin{displaymath}
  +\biggl({\cal K}R'\biggl\{ \hspace{1cm} \biggr\}\biggr)^2
\end{displaymath}
\vspace{3mm}
\begin{displaymath}
{\cal K}R'\biggl\{ \hspace{1cm} \biggr\}={\cal K}\biggl( \hspace{1cm}
-{\cal K}R'\biggl\{ \hspace{1cm} \biggr\}\hspace{1.5cm} \biggr),
\end{displaymath}
\vspace{3mm}
\begin{displaymath}
{\cal K}R'\biggl\{ \hspace{1cm} \biggr\}={\cal K}\biggl\{ \hspace{1cm}
\biggr\}.
\end{displaymath}

\vspace{3mm}

The benefit of using the ${\cal K}R'$ operation besides its convenience
in actual calculations is as follows.
Using the fact that the result of ${\cal K}R'$ operation is
a polynomial in masses and external momenta of the diagram, one can remove
the  dependence on the external momenta by differentiating (usually twice is
sufficient) with respect to the external momentum and then
 setting the external
momentum to zero. However, in this case infrared divergences appear. In order
to prevent this, one can introduce a new fictitious external momentum  as
 an infrared regulator flowing along some of the lines of the diagram
 (Chetyrkin and Tkachov, 1982). Alternatively, one can introduce
a fictitious mass in one of the lines of the diagram as an infrared
regulator (Vladimirov, 1980). An appropriate choice of the
 fictitious momentum can drastically simplify the topology of the given
diagram. Both versions
of the so called infrared rearrangement procedure simplify the calculation
 and make it possible to evaluate counterterms to
 four- and five-loop diagrams.
The main result of the application of the infrared rearrangement technique
can be formulated as follows.
The problem of calculating  the counterterms of an arbitrary
$l$-loop diagram with an arbitrary number of masses and external momenta
within the MS prescription can be reduced
to the problem of calculating some $l-1$ -loop massless integrals
to $O(\varepsilon^0)$ with only one external momentum. In the later sections,
the full calculational procedure will be demonstrated for a typical four-loop
diagram contributing to the photon renormalization constant.

\vspace{2cm}

\renewcommand{\thesection}{\arabic{section}}
\subsection{\tenbf Evaluation of Wilson coefficient functions in operator
product expansion}

\indent
In this subsection we briefly discuss the problem of evaluation of
 higher twist operator contributions to the hadronic vacuum polarization
 function. Those contributions are relevant in the analysis of
 nonperturbative contributions in some processes
(e.g., hadronic decay of the $\tau$-lepton).
We use the Wilson operator product expansion technique
(Wilson, 1969) - mathematical apparatus allowing a factorization of the
short distance contributions, which are calculable perturbatively and large
 distance effects which can be parametrized with the
 vacuum condensates
( Shifman, Vainshtein and Zakharov, 1979; Novikov {\it et al.}, 1985).
In the perturbative evaluation of Wilson coefficient functions, we
rely on the so called method of projectors (Gorishny, Larin and Tkachov, 1983;
 Gorishny and Larin, 1987; see also Pivovarov and Tkachov 1988, 1993
and references therein).
An actual calculation for the coefficient functions of the operators of
 $\mbox{dim}=4$ has been done in the work by Loladze, Surguladze and Tkachov
 (1984, 1985), and Surguladze and Tkachov (1989b, 1990). The present
 discussion is based mainly on those works.
Below we demonstrate
the above technique in the  case of the coefficient functions of gluon
and quark condensates.

Consider the operator product expansion of the T-product of two quark
currents in the deep Euclidean region, $-q^2=Q^2 \rightarrow \infty$
\begin{equation}
{\cal T}(Q)=i\int d^4x e^{iqx} T J(x)J(0) = \sum_{i} C_{i}(Q)O_{i}(0),
\label{eq:OPE}
\end{equation}
where $J$ are quark currents. $C_{i}(Q)$ are c-number coefficient functions
containing all dependence on $Q$. $O_{i}$ are local operators forming in
 general  a complete basis.
If the currents $J$ are gauge invariant then,
 after averaging over the vacuum, only gauge invariant operators
 contribute to the r.h.s. of eq.\ (\ref{eq:OPE}).
However, the renormalization procedure
 mixes gauge invariant operators with non-invariant ones and one has
 to consider the complete basis of operators of the given dimension.
The following set of operators of the dimension 4

\begin{displaymath}
O_{1} = (G_{\mu\nu}^{a})^2, \hspace{3mm}
O_{2}^{f}=m_{f}\overline{q}_{f}q_{f}, \hspace{3mm}
O_{3}^f = \overline{q}_{f}(i\hat{\partial}-m_f+gT^{a}\hat{A^a})q_f,
\end{displaymath}
\begin{equation}
O_{4}=(\partial_{\mu}\overline{c}^a)(\partial_{\mu}c^a)
  +(\partial_{\mu}\delta^{ab}+gf^{abc}A_{\mu}^{c})A_{\nu}^{b}G_{\mu\nu}^{a}
  -g\sum_{f}\overline{q}_{f}T^{a}\hat{A}^{a}q_f,
\label{eq:OPEbasis}
\end{equation}
\begin{displaymath}
O_{5} =\partial_{\mu}\overline{c}^a((\partial_{\mu}\delta^{ab}
     +gf^{abc}A_{\mu}^{c})c^{b}
\end{displaymath}
is closed under renormalization together with the ``operator'' $\sim m^4$
(Spiridonov, 1984; Loladze, Surguladze and Tkachov, 1984, 1985).
Our aim is to calculate coefficient functions of gauge invariant operators
 $O_1$ and $O_{2}^{f}$. Note that $\sim m^4$
 operators can be ignored because of the
 special structure of the renormalization matrix for the basis
 (\ref{eq:OPEbasis}).
The Feynman rules for the operators (\ref{eq:OPEbasis}) are
(Surguladze and Tkachov, 1990)

\newpage

\begin{displaymath}
\lefteqn{O_1} \hspace{6cm} 4\delta^{ab}(p^2g^{\mu\nu}-p^{\mu}p^{\nu})
\end{displaymath}

\vspace{2mm}

\begin{displaymath}
\lefteqn{O_{2}^{f}} \hspace{8cm} \delta_{ff'}m_f
\end{displaymath}

\vspace{2mm}

\begin{displaymath}
O_{3}^{f} \hspace{65mm} \delta_{ff'}(\hat{p}-m_f)
\end{displaymath}

\vspace{2mm}

\begin{equation}
O_4 \hspace{55mm} 2\delta^{ab}(p^2g^{\mu\nu}-p^{\mu}p^{\nu})
\label{eq:OPEFrules}
\end{equation}

\vspace{2mm}

\begin{displaymath}
O_4 \hspace{73mm} \delta^{ab}p^2g^{\mu\nu}
\end{displaymath}

\vspace{2mm}

\begin{displaymath}
O_5 \hspace{73mm} \delta^{ab}p^2g^{\mu\nu}
\end{displaymath}

\vspace{2mm}

\begin{displaymath}
O_5 \hspace{76mm} if^{abc}p^{\mu}
\end{displaymath}

\vspace{11mm}

The operators of the basis (\ref{eq:OPEbasis}) are renormalized as follows
\begin{equation}
O_{i} = (Z_O)_{ij}O_{j}^{\mbox{\scriptsize B}},
\label{eq:OPEbasisren}
\end{equation}
where the superscript B marks the same operators as in (\ref{eq:OPEbasis})
but built from the ``bare'' fields, masses and couplings. The structure of
 the renormalization matrix $Z_O$ has been studied by Spiridonov (1984).
 In the MS type schemes $Z_O$ has the following form
(Surguladze and Tkachov, 1990)

\vspace{3cm}

\begin{equation}
\mbox{}
\label{eq:OPEZ}
\end{equation}

\vspace{23mm}

\noindent
where only the matrix elements $A$ and $B$ are relevant.
\begin{displaymath}
A = \biggl(1-\frac{\beta(\alpha_s)}{\varepsilon}\biggr)^{-1},
\end{displaymath}
\begin{equation}
B = \frac{4\gamma_{m}(\alpha_s)}{\varepsilon}
    \biggl(1-\frac{\beta(\alpha_s)}{\varepsilon}\biggr)^{-1}.
\label{eq:OPEAB}
\end{equation}
Inserting eq.\ (\ref{eq:OPEbasisren}) into the expansion (\ref{eq:OPE})
we get
\begin{equation}
{\cal T}(Q) = \sum_{i,j}C_{i}(Q)O_{i}^{\mbox{\scriptsize B}}(Z_O)_{ij}.
\label{eq:PiRexpans}
\end{equation}
Following the method of projectors (Gorishny, Larin and Tkachov, 1983),
 we define the
 projectors $\pi_{i}$ satisfying the orthogonality condition and vanishing
on higher spin operators
\begin{equation}
\pi_{i}[O_{j}^{\mbox{\scriptsize B}}]=\delta_{ij},
\label{eq:piproject}
\end{equation}
\begin{displaymath}
\pi_{i}[\mbox{higher spin operators}]=0.
\end{displaymath}
Projectors $\pi_{i}$ applied on  the l.h.s.\ of eq.\
(\ref{eq:PiRexpans}) separate in the r.h.s.\ the
coefficient functions we are interested in
\begin{equation}
\pi_{j}[{\cal T}(Q)] = \sum_{i}C_{i}(Q)(Z_O)_{ij}.
\label{eq:pionPi}
\end{equation}
We find the coefficient functions
\begin{equation}
C_{i}(Q)=\sum_{j}\pi_{j}[{\cal T}(Q)](Z_O^{-1})_{ji}.
\label{eq:CF}
\end{equation}
Our aim is to find the coefficient functions of gauge
 invariant operators $O_1=(G_{\mu\nu})^2$ and
 $O_{2}^{f}=m_{f}\overline{q}_{f}q_f$.
 So, we need to construct the corresponding projectors $\pi_1$ and
 $\pi_{2}^{f}$.
Let us represent $\pi_i$ as a linear combinations of
 some ``elementary'' projectors ${\cal P}_j$ defined in the following way.
\begin{displaymath}
{\cal P}_{1}[O] = \frac{1}{N_{A}}\frac{\partial^2}{\partial p^2}
   \delta^{ab}g^{\mu\nu} \biggl\{ \hspace{45mm} \biggr\}_{p=m_{f}=0}
\end{displaymath}

\vspace{1mm}

\begin{displaymath}
{\cal P}_{2}^{f}[O] = \frac{1}{4N_{F}}\frac{\partial}{\partial m_f}
   Tr\biggl\{ \hspace{45mm} \biggr\}_{p=m_{f}=0}
\end{displaymath}

\vspace{2mm}

\begin{equation}
\hspace{-2mm}
{\cal P}_{3}^{f}[O] = \frac{1}{4N_{F}}\frac{\partial}{\partial p^{\sigma}}
   Tr \gamma^{\sigma}\biggl\{ \hspace{45mm} \biggr\}_{p=m_{f}=0}
\label{eq:ElProj}
\end{equation}

\vspace{2mm}

\begin{displaymath}
\hspace{4mm}
{\cal P}_{4}[O] = \frac{1}{N_{A}}\frac{\partial^2}{\partial p^2}
   \delta^{ab} \biggl\{ \hspace{45mm} \biggr\}_{p=m_{f}=0}
\end{displaymath}

\vspace{2mm}

\begin{displaymath}
\hspace{2mm}
{\cal P}_{5}[O] = \frac{if^{abc}}{gN_{A}C_{A}}\frac{\partial}
                                          {\partial p^{\mu}}
    \biggl\{ \hspace{45mm} \biggr\}_{p=m_{f}=0}
\end{displaymath}

\vspace{1mm}

\noindent
where the parentheses contain the one-particle-irreducible Green
function with one operator insertion. In the case of ${\cal P}_{2}^{f}$
 and ${\cal P}_{3}^{f}$ the traces are calculated over Lorentz
 spinor and color indices.

Acting by the projectors ${\cal P}_{j}$ on the operators (\ref{eq:OPEbasis})
we obtain
\begin{displaymath}
{\cal P}_{1}[O_{1}]=8D(D-1), \hspace{6mm} {\cal P}_{1}[O_{4}]=4D(D-1),
 \hspace{6mm} {\cal P}_{2}^{f}[O_{2}^{f'}]=\delta_{ff'},
\end{displaymath}
\begin{equation}
{\cal P}_{2}^{f}[O_{3}^{f'}]=-\delta_{ff'}, \hspace{6mm}
{\cal P}_{3}^{f}[O_{3}^{f'}]=D\delta_{ff'}, \hspace{6mm}
{\cal P}_{4}[O_{4}]={\cal P}_{4}[O_{5}]=2D, \hspace{6mm}
{\cal P}_{5}[O_{5}]=D.
\label{eq:P(o)}
\end{equation}
The results which are not shown in the above list are identicaly zero.
{}From the definition (\ref{eq:piproject}) and eq.\ (\ref{eq:P(o)})
 we obtain the explicit form for the projectors $\pi_{1}$ and $\pi_{2}^{f}$
\begin{displaymath}
\pi_{1} = \frac{1}{8D(D-1)}[{\cal P}_{1}-2(D-1){\cal P}_{4}+4(D-1)
                                                       {\cal P}_{5}],
\end{displaymath}
\begin{equation}
\pi_{2}^{f}={\cal P}_{2}^{f}+\frac{1}{D}{\cal P}_{3}^{f}.
\label{eq:piexplicit}
\end{equation}

    Combining eqs.\ (\ref{eq:OPEZ}) and (\ref{eq:OPEAB}) with
eq.\ (\ref{eq:CF}) we get our final expressions for the coefficient
functions $C_{1}(Q)$ and $C_{2}^{f}(Q)$
(Surguladze and Tkachov, 1989,1990)
\begin{displaymath}
C_{1}(Q)=\pi_{1}[{\cal
T}(Q)]\biggl(1-\frac{\beta(\alpha_s)}{\varepsilon}\biggr),
\end{displaymath}
\begin{equation}
C_{2}^{f}(Q)
=\pi_{2}^{f}[{\cal T}(Q)]-\pi_{1}[{\cal T}(Q)]\frac{4\gamma_{m}(\alpha_s)}
  {\varepsilon}.
\label{eq:CFexplicit}
\end{equation}
The above expressions have a closed form and are valid at any order
 of perturbation theory. We note that ${\cal T}$ must be
constructed with unrenormalized couplings and fields before one
 applies the projectors $\pi_{i}$.

The general theory of Euclidean asymptotic expansions of Feynman integrals
and the methods applicable to high order perturbative calculations
 have been developed in the works of Tkachov (1983b, 1983c, 1991, 1993),
Chetyrkin and Tkachov (1982), and Chetyrkin (1991) (see also
Smirnov 1990, 1991 and references therein).
The technique developed in these works  allows one to derive operator
 product expansions in the MS-scheme for any Feynman integral.
 For more general discussion and further details we refer to the
above works and also to the original calculations
(Surguladze and Tkachov, 1989, 1990). In section 5 we
 present a short description of the calculation of the coefficient
 functions of gluon and quark condensates up to $O(\alpha_{s}^2)$.

\vspace{2cm}

\renewcommand{\thesection}{\Roman{section}}
\section{\bf \ $\Gamma(H\rightarrow hadrons)$
 to $O(\alpha^2_s)$}
\renewcommand{\thesection}{\arabic{section}}
\setcounter{equation}{0}

\renewcommand{\thesection}{\arabic{section}}
\subsection{\tenbf The decay rate in terms of running parameters}

\indent
   In this subsection, using the above methods we calculate the
$O(\alpha^{2}_{s})$ corrections to the total
hadronic decay width of the Standard Model Higgs boson in the massless quark
limit (Gorishny, Kataev, Larin and Surguladze, 1990, 1991b; Surguladze, 1994b).

\vspace{6cm}
\begin{center}
FIG.\ 1.\hspace{2mm} The process $H \rightarrow$ hadrons
\end{center}

\vspace{5mm}

   The standard SU(2)$\times$U(1) Lagrangian density of
fermion-Higgs interaction is
\begin{equation}
L = -g_{Y}\overline{q}_fq_fH
    =-(\sqrt{2}G_F)^{1/2}m_f\overline{q}_fq_fH
     =-(\sqrt{2}G_F)^{1/2}j_fH.
\label{eq:Lagr}
\end{equation}
The decay width of a scalar Higgs boson to the quark-antiquark pair
is determined by the imaginary part of the two-point correlation
function
\begin{equation}
\Pi(Q^2=-s,m_f)=i\int e^{iqx}<Tj_{f}(x)j_{f}(0)>_0d^4x
\label{eq:pifunctions}
\end{equation}
of the quark scalar currents $j_f=m_f\overline{q}_fq_f$ in the following way
\begin{equation}
\Gamma_{H\rightarrow q_f\overline{q}_f}
         =\frac{\sqrt{2}G_F}{M_H} \mbox{Im}\Pi(s+i0,m_f)\biggr|_{s=M_H^2}.
\label{eq:ImPi}
\end{equation}
$M_H$ is the Higgs mass.
The total decay width will be the sum over all participating
(depending on $M_H$) quark flavors
\begin{equation}
\Gamma(H \rightarrow \mbox{\small hadrons}) = \sum_{f=u,d,s,...}
             \Gamma_{H\rightarrow q_f\overline{q}_f}.
\label{eq:total}
\end{equation}

    We follow the work by Gorishny, Kataev, Larin and Surguladze (1990)
 and in analogy to the vector channel introduce the Adler function
(Adler, 1974)
\begin{equation}
D(Q^2,m_f) = Q^2\frac{d}{dQ^2}\frac{\Pi(Q^2,m_f)}{Q^2}.
\label{eq:Dfunctscalar}
\end{equation}
The derivative avoids the additive renormalization of $\Pi$.
In fact, it is possible to proceed without the introduction
 of $D$-function and deal
directly with the correlation function $\Pi$ (Surguladze, 1994b).
Indeed, we are interested in $\mbox{Im}\Pi(s+i0,m_f)$. Since the overall
MS renormalization constant has no terms like
$(\log\mu^2/Q^2)^n/\varepsilon^k$, its imaginary part vanishes identically.
The abscence of the pole logarithms in renormalization constants
is a general feature of MS type schemes.

   The $D$-function obeys the homogeneous renormalization group equation
\begin{equation}
\left(\mu^2\frac{\partial}{\partial\mu^2}
+\beta(\alpha_s)\alpha_s\frac{\partial}{\partial\alpha_s}
-\gamma_m(\alpha_s)\frac{\partial}
{\partial \log m_f}\right)D(\mu^2/Q^2,m_f,\alpha_s)=0.
\label{eq:RGEDscalar}
\end{equation}
The QCD $\beta$-function and the mass anomalous dimension
 $\gamma_m$ are known up to the three loop approximation and
 have been given in the previous section.
The plan for evaluation of
$\Gamma_{H\rightarrow q_f\overline{q}_f}$ is as follows.
First, we write the diagram representation for $\Pi(Q^2,m_f)$
according to the standard Feynman rules
up to the desired loop-level. Second, we evaluate the Feynman diagrams
 using the dimensional regularization
 and renormalize the coupling and quark masses within the MS renormalization
 prescription. Finally, to get the decay rate,
 we analytically continue the result
 for the $D$-function obtained from eq.\ (\ref{eq:Dfunctscalar})
 from Euclidean to Minkowski space.
 Following the above plan, we now demonstrate the
 calculation of $\Gamma_{H\rightarrow q_f\overline{q}_f}$
 up to the 3-loop level.
   First of all, note that the correlation function $\Pi$ and the
 related $D$-function depend on quark masses. The
algorithms for evaluation of the 3-loop Feynman diagrams constructed with
 the propagators of massive particles has not yet been developed.
  However, in the deep Euclidean region ($Q^2\rightarrow \infty$)
it is possible to simplify the
 calculation using the expansion in terms of the small parameter
$m_f^2/Q^2$
\begin{equation}
\frac{1}{m_f^2Q^2}\Pi(Q^2,m_f)
   =\Pi(Q^2)+O\biggl(\frac{m_f^2}{Q^2}\biggr).
\label{eq:Piexpa}
\end{equation}
Such an expansion is legitimate since we consider a Higgs
boson much heavier than the typical hadronic mass scale.
In this section we calculate the first term in the above expansion
and the related decay rate. This is equivalent to the assumption
that all five quarks are massless and the top quark decouples
($m_t \rightarrow \infty$).

  The diagrammatic representation for $\Pi$ in somewhat symbolic
form looks like

\vspace{7mm}

\begin{displaymath}
\Pi(Q^2) \sim \hspace{2cm}
     + \frac{\alpha_s}{\pi}\biggl[\hspace{23mm}+
                      \hspace{3mm} 2\hspace{23mm} \biggr]
\end{displaymath}

\vspace{3mm}

\begin{equation}
     +\biggl(\frac{\alpha_s}{\pi}\biggr)^2\biggl[\hspace{23mm}
     +\cdots +(\mbox{total of 16 three-loop diagrams})\biggr]
     +O(\alpha^{3}_{s}).
\label{eq:pidiagram}
\end{equation}

\vspace{7mm}

\noindent
Next, we evaluate one-, two- and three-loop massless Feynman diagrams. By
simple power counting, it is easy to find that in general the above
diagrams are UV divergent.
The unrenormalized contribution from a typical three-loop diagram
in the $\overline{\mbox{MS}}$
renormalization scheme (Bardeen, Buras, Duke and Muta, 1978) reads

\vspace{3cm}
\begin{displaymath}
\sim \frac{1}{(4\pi)^2}
   \left(\frac{\alpha_{s}^{\mbox{\tiny B}}}{4\pi}\right)^{2}N_{F}
\frac{C_{F}C_{A}}{2}
(m_{f}^{\mbox{\tiny B}})^{2}Q^2
\left(\frac{\mu_{\overline{\mbox{\tiny MS}}}^2}{Q^2}\right)^{3\varepsilon}
\biggl[ \frac{16}{\varepsilon^3}
       +\frac{400}{3\varepsilon^2}
       +\frac{2344}{3\varepsilon}
       -\frac{160}{\varepsilon}\zeta(3)
       +\frac{11800}{3}
\end{displaymath}
\begin{displaymath}
\hspace{9cm} -1312\zeta(3)-240\zeta(4)+320\zeta(5) \biggr],
\end{displaymath}
where $m_{f}^{\mbox{\tiny B}}$ is the $f$-flavor quark mass
 originating from the
quark mass dependence of the Yukawa coupling.
 $\zeta(3)$, $\zeta(4)$ and $\zeta(5)$  are ordinary Riemann
 $\zeta$-functions.
The number 2 in front of the diagram stands for the symmetry factor.
The algorithms for the evaluation of propagator type one-, two- and
 three-loop
massless Feynman diagrams have been given by Tkachov (1981, 1983a) and
 Chetyrkin and Tkachov (1981). For the description of the
algorithms see also Gorishny, Larin, Surguladze and Tkachov, 1989.
The results given in this section were reobtained with the help
 of the program {\small HEPL}oops (Surguladze, 1992) and  the previous
 results (Gorishny, Kataev, Larin and Surguladze, 1990, 1991)
 were independently confirmed (Surguladze, 1994b).

As one can see, each three-loop diagram in general may contain
 a pole with power
$\leq 3$. In the vector channel, after summing the results for
all diagrams with an appropriate symmetry and SU(N) group factor,
 the leading pole cancels.
This is the consequence of the conservation of electromagnetic currents.
In the scalar channel, the leading poles remain in $\Pi$.
This is related to the quark mass dependence of the coupling.

Evaluating the unrenormalized correlation function (\ref{eq:pifunctions})
and using the definition (\ref{eq:Dfunctscalar}), we obtain the
unrenormalized $D$-function in the massless limit.
\begin{eqnarray}
\lefteqn{\hspace{-9mm}
 D\left(
  \frac{\mu_{\overline{\mbox{\tiny MS}}}^{2}}{Q^2},\alpha_s\right) =
          \frac{1}{(4\pi)^2}N_{F}(m_{f}^{\mbox{\tiny B}})^2 \biggl\{
    \left(\frac{\mu_{\overline{\mbox{\tiny MS}}}^2}{Q^2}\right)^{\varepsilon}
    (2+4\varepsilon+8\varepsilon^2) } \nonumber\\
 && \quad \hspace{7mm} +\left(\frac{\alpha_{s}^{\mbox{\tiny B}}}{4\pi}\right)
     \left(\frac{\mu_{\overline{\mbox{\tiny MS}}}^2}{Q^2}\right)^{2\varepsilon}
      C_F\biggl[\frac{12}{\varepsilon}+58+\varepsilon(227-48\zeta(3))\biggr]
                                                                  \nonumber\\
 && \quad \hspace{6mm} +\left(\frac{\alpha_{s}^{\mbox{\tiny B}}}{4\pi}\right)^2
     \left(\frac{\mu_{\overline{\mbox{\tiny MS}}}^2}{Q^2}\right)^{3\varepsilon}
          C_F\biggl[ \hspace{5mm}
            C_{F}\biggl( \frac{36}{\varepsilon^2}+\frac{279}{\varepsilon}
                        +\frac{3139}{2}-360\zeta(3) \biggr) \nonumber\\
 && \quad  \hspace{48mm}
           +C_{A}\biggl( \frac{22}{\varepsilon^2}+\frac{201}{\varepsilon}
                        +\frac{2511}{2}-300\zeta(3) \biggr) \nonumber\\
 && \quad  \hspace{48mm}
           -TN \biggl( \frac{8}{\varepsilon^2}+\frac{68}{\varepsilon}
                         +414-96\zeta(3) \biggr) \biggr]+O(\alpha_{s}^{3})
                        \biggr\}.
\label{eq:Dbare}
\end{eqnarray}
    The above expression requires the renormalization of the
 strong coupling ( eq.\ (\ref{eq:AsR}) ) and the multiplicative
 renormalization ( eq.\ (\ref{eq:mrenorm}) ) originating from the
quark mass dependence of the Yukawa coupling.

  Expanding the factors
 $(\mu_{\overline{\mbox{\tiny MS}}}^{2}/Q^2)^{l\varepsilon}$
in terms of $\varepsilon$ and performing the renormalizations of the
coupling and the quark mass, we get a finite analytical expression
 for the $D$-function in the $\overline{\mbox{MS}}$ scheme
\begin{eqnarray}
\lefteqn{\hspace{-5mm}
  D\left(
  \frac{\mu_{\overline{\mbox{\tiny MS}}}^{2}}{Q^2},\alpha_s\right)=
      \frac{N_F}{8\pi^2}m_{f}^{2}\biggl\{1
       +\biggl(\frac{\alpha_s}{4\pi}\biggr)C_F \biggl[17
          +6\log
   \biggl(\frac{\mu_{\overline{\mbox{\tiny MS}}}^{2}}{Q^2}\biggr) \biggr]}
                                                                   \nonumber\\
 && \quad
       +\biggl(\frac{\alpha_s}{4\pi}\biggr)^2C_F
            \biggl[  C_F\biggl(\frac{691}{4}-36\zeta(3)\biggr)
               +C_A\biggl(\frac{893}{4}-62\zeta(3)\biggr)
                       -TN(65-16\zeta(3))
                                                                   \nonumber\\
 && \quad \hspace{4cm}
           +\log\biggl(\frac{\mu_{\overline{\mbox{\tiny MS}}}^{2}}{Q^2}\biggr)
             \biggl( 105C_F+\frac{284}{3}C_A-\frac{88}{3}TN \biggr)
                                                                   \nonumber\\
 && \quad \hspace{4cm}
     +\log^2\biggl(\frac{\mu_{\overline{\mbox{\tiny MS}}}^{2}}{Q^2}\biggr)
                               ( 18C_F+11C_A-4TN ) \biggr]  \biggr\}.
\label{eq:Danalyt0}
\end{eqnarray}
For standard QCD with the color SU$_{\mbox{\scriptsize c}}$(3) symmetry group,
 the analytical result for the $D$-function reads (Surguladze, 1989d)
\begin{eqnarray}
\lefteqn{D\left(
  \frac{\mu_{\overline{\mbox{\tiny MS}}}^{2}}{Q^2},\alpha_s\right)=}
                                                                  \nonumber\\
 && \frac{3}{8\pi^2}m_{f}^{2}\biggl\{1
       +\biggl(\frac{\alpha_s}{\pi}\biggr)\biggl[\frac{17}{3}
   +2\log\biggl(\frac{\mu_{\overline{\mbox{\tiny MS}}}^{2}}{Q^2}\biggr) \biggr]
                                                                   \nonumber\\
 && \quad \hspace{13mm} +\biggl(\frac{\alpha_s}{\pi}\biggr)^2
       \biggl[ \frac{10801}{144}-\frac{39}{2}\zeta(3)
              -\biggl(\frac{65}{24}-\frac{2}{3}\zeta(3)\biggr)N \\
 && \quad \hspace{24mm}
    +\log\biggl(\frac{\mu_{\overline{\mbox{\tiny MS}}}^{2}}{Q^2}\biggr)
             \biggl(\frac{106}{3}-\frac{11}{9}N \biggr)
  +\log^2\biggl(\frac{\mu_{\overline{\mbox{\tiny MS}}}^{2}}{Q^2}\biggr)
                \biggl(\frac{19}{4}-\frac{1}{6}N\biggr) \biggr] \biggr\}.
                                 \nonumber
\label{eq:Danalyt}
\end{eqnarray}
This completes the evaluation of the correlation function of the two scalar
 quark currents in the massless limit at the three-loop
 approximation.

There is one crucial test of this calculation based on the renormalization
group constraints. The solution of the renormalization group equation
(\ref{eq:RGEDscalar})
can be conveniently rewritten as follows
\begin{equation}
D\biggl(\frac{\mu^2}{Q^2},m_{f}(\mu),\alpha_s(\mu)\biggr)=
       \frac{3}{8\pi^2}
       m_f^2(\mu)\sum_{0\leq j \leq i}
    \biggl(\frac{\alpha_s(\mu)}{\pi}\biggr)^i
          a_{ij}
        \log^j\frac{\mu^2}{Q^2}.
\label{eq:gamma0gen}
\end{equation}
Applying the differential operator $\mu^2 d/d\mu^2$ to
both sides of eq.\ (\ref{eq:gamma0gen}), taking into account
the renormalization group invariance of the $D$-function and
eqs.\ (\ref{eq:beta}) and (\ref{eq:gamma}), we obtain to $O(\alpha_s)$
\begin{equation}
a_{11}=2\gamma_0a_{00},
\label{eq:relations1}
\end{equation}
to $O(\alpha_s^2)$
\begin{displaymath}
a_{21}=2\gamma_1a_{00}+(\beta_0+2\gamma_0)a_{10},
\end{displaymath}
\begin{equation}
a_{22}=(\beta_0+2\gamma_0)\frac{a_{11}}{2}=(\beta_0+2\gamma_0)\gamma_0a_{00},
\label{eq:relations2}
\end{equation}
and to $O(\alpha_s^3)$
\begin{displaymath}
a_{31}=2(\beta_0+\gamma_0)a_{20}+(\beta_1+2\gamma_1)a_{10}+2\gamma_2a_{00},
\end{displaymath}
\begin{displaymath}
a_{32}=(\beta_0+\gamma_0)a_{21}+(\beta_1+2\gamma_1)\frac{a_{11}}{2}
      =(\beta_0+\gamma_0)[2\gamma_1a_{00}+(\beta_0+2\gamma_0)a_{10}]
       +(\beta_1+2\gamma_1)\gamma_0a_{00},
\end{displaymath}
\begin{equation}
a_{33}=\frac{2}{3}(\beta_0+\gamma_0)a_{22}=\frac{2}{3}\gamma_0
         (\beta_0+\gamma_0)(\beta_0+2\gamma_0)a_{00}.
\label{eq:relations3}
\end{equation}
   The relations (\ref{eq:relations1}) and (\ref{eq:relations2}) provide a
powerful check of our calculation, while the relations
(\ref{eq:relations3}) allow one to evaluate the $\log$ terms to
 $O(\alpha_s^3)$,
without explicit calculations of the corresponding four-loop diagrams.
With those relations, the information available at present, namely the
QCD $\beta$-function, mass anomalous dimension and the
two-point correlation function up to the three-loop level is fully exploited.
In fact, similar relations can be derived for the correlation function
$\Pi$. However, the renormalization group equation for $\Pi$ is not
 a homogeneous one and the
anomalous dimension function up to the corresponding order of $\alpha_s$
is necessary.

 We evaluate the decay rate of the neutral Higgs boson
into a quark antiquark pair by analytical continuation of
 $D\left(\mu^{2}/Q^2,m_{f}(\mu),\alpha_s(\mu)\right)$
  from Euclidean to Minkowski space. The total decay rate can be obtained by
summing up over all participating quark flavors.
\begin{eqnarray}
\lefteqn{\hspace{-7mm}\Gamma(H\rightarrow \mbox{\small hadrons})
                  =\frac{3\sqrt{2}G_FM_H}{8\pi}
 \sum_{f=u,d,s,...} m_f^2\biggl\{
  1+\frac{\alpha_s}{\pi}\biggl(\frac{17}{3}
+2\log\frac{\mu_{\overline{\mbox{\tiny MS}}}^2}{M_H^2}\biggr)}\nonumber\\
 && \quad
   +\biggl(\frac{\alpha_s}{\pi}\biggr)^2
      \biggl[\frac{10801}{144}-\frac{19}{2}\zeta(2)-\frac{39}{2}\zeta(3)
         +\frac{106}{3}\log\frac{\mu_{\overline{\mbox{\tiny MS}}}^2}{M_H^2}
         +\frac{19}{4}\log^2
           \frac{\mu_{\overline{\mbox{\tiny MS}}}^2}{M_H^2}\nonumber\\
 && \quad \hspace{1cm}
  -N\biggl(\frac{65}{24}-\frac{1}{3}\zeta(2)-\frac{2}{3}\zeta(3)
         +\frac{11}{9}\log\frac{\mu_{\overline{\mbox{\tiny MS}}}^2}{M_H^2}
         +\frac{1}{6}\log^2
             \frac{\mu_{\overline{\mbox{\tiny MS}}}^2}{M_H^2}\biggr)\biggr]
               \biggr\}.
\label{eq:GHtot}
\end{eqnarray}
The Riemann function $\zeta(2)=\pi^2/6$ arose
from the analytical continuation of the
$\log^2\mu_{\overline{\mbox{\tiny MS}}}^2/Q^2$
term and $\zeta(3)=1.202056903$.
The procedure of analytical continuation and the appearance of invariant
 additional contributions have been discussed in several earlier works
 (Krasnikov and Pivovarov, 1982; Pennington and Ross, 1982;
Radyushkin, 1982; Pivovarov, 1992a). Note that in some cases those
additional corrections are large and affect the result significantly.
This is especially true for the total cross section in the process
$e^{+}e^{-} \rightarrow \mbox{hadrons}$.
To minimize such corrections it was proposed, for instance,
to redefine the expansion
parameter (Pennington and Ross, 1982; Radyushkin, 1982).

\renewcommand{\thesection}{\arabic{section}}
\subsection{\tenbf The decay rate in terms of pole quark mass}

For the heavy flavor decay mode of the Higgs, it is relevant
to parametrize the decay rate in terms of quark
pole mass (see, e.g., Kniehl, 1994a). Let us rewrite the result
for $\Gamma_{H\rightarrow q_f\overline{q}_f}$ in terms of pole quark mass,
assuming that heavy quark is not exactly on-shell.
This subsection is based mainly on recent findings
(Surguladze, 1994a,b).

   Solving the renormalization group equation for the quark mass -
eq.\ (\ref{eq:RGfunctions}), we obtain the following scaling law for the
running quark mass:
\begin{equation}
\frac{m_f(\mu_1)}{m_f(\mu_2)}=\frac{\phi(\alpha_s(\mu_1))}
                                   {\phi(\alpha_s(\mu_2))},
\label{eq:mrun}
\end{equation}
where
\begin{eqnarray}
\lefteqn{\phi(\alpha_s(\mu))=\biggl(2\beta_0\frac{\alpha_s(\mu)}{\pi}\biggr)
                              ^{\frac{\gamma_0}{\beta_0}}
       \biggl\{1
     +\biggl(\frac{\gamma_1}{\beta_0}
         -\frac{\beta_1\gamma_0}{\beta_0^2}\biggr)\frac{\alpha_s(\mu)}{\pi}}
                                                             \nonumber\\
 && \quad \hspace{-3mm}
     +\frac{1}{2}\biggl[\biggl(\frac{\gamma_1}{\beta_0}
         -\frac{\beta_1\gamma_0}{\beta_0^2}\biggr)^2
       +\frac{\gamma_2}{\beta_0}
       -\frac{\beta_1\gamma_1}{\beta_0^2}-\frac{\beta_2\gamma_0}{\beta_0^2}
       +\frac{\beta_1^2\gamma_0}{\beta_0^3}
         \biggr]\biggl(\frac{\alpha_s(\mu)}{\pi}\biggr)^2\biggr\}.
\label{eq:f}
\end{eqnarray}
In the above equation all appropriate quantities
are evaluated for $N$ active quark flavors. $N$ can be determined
according to the scale of $M_H$. At present one usually considers $N=5$.

For the running coupling we obtain the following evolution equation
to $O(\alpha_s^3)$ (Surguladze, 1994b)
\begin{eqnarray}
\lefteqn{\hspace{-1cm}\frac{\alpha_s^{(n)}(\mu_1)}{\pi}
                         =\frac{\alpha_s^{(N)}(\mu_2)}{\pi}}
                                                             \nonumber\\
 &&
      +\biggl(\frac{\alpha_s^{(N)}(\mu_2)}{\pi}\biggr)^2
      \biggl(\beta_0^{(N)}\log\frac{\mu_2^2}{\mu_1^2}
      +\frac{1}{6}\sum_{l}\log\frac{m_l^2}{\mu_1^2} \biggr)
                                                             \nonumber\\
 &&   +\biggl(\frac{\alpha_s^{(N)}(\mu_2)}{\pi}\biggr)^3
      \biggl[\beta_1^{(N)}\log\frac{\mu_2^2}{\mu_1^2}
      +\frac{19}{24}\sum_{l}\log\frac{m_l^2}{\mu_1^2}
                                                             \nonumber\\
 && \quad \hspace{23mm}
      +\biggl(\beta_0^{(N)}\log\frac{\mu_2^2}{\mu_1^2}
      +\frac{1}{6}\sum_{l}\log\frac{m_l^2}{\mu_1^2} \biggr)^2
                                                             \nonumber\\
&& \quad \hspace{23mm}
                           -\frac{25}{72}(N-n)\biggr],
\label{eq:Astransform}
\end{eqnarray}
where the superscript $N$ ($n$) indicates that the corresponding quantity
is evaluated for $N$ ($n$) numbers of
participating quark flavors. Conventionally (see, e.g., Marciano, 1984)
$N$ ($n$) is specified to be the number
of quark flavors with mass $\leq \mu_2$ ($\leq \mu_1$). However,
eq.\ (\ref{eq:Astransform}) is relevant for any $n\leq N$ and arbitrary
$\mu_1$ and $\mu_2$, regardless of the conventional specification of the number
of quark flavors.
The $\log m_l/\mu_1$
terms are due to the ``quark treshold'' crossing effects and the constant
coefficients $1/6=\beta_0^{(k-1)}-\beta_0^{(k)}$,
$19/24=\beta_1^{(k-1)}-\beta_1^{(k)}$ represent
the contributions of the quark loop in the $\beta$-function.
The sum runs over $N-n$ quark flavors (e.g., $l=b$ if $n=4$ and $N=5$).
Note that $m_l$ is the pole mass of the quark with flavor $l$.
For the on-shell definition of the quark masses eq.\ (\ref{eq:Astransform})
changes - the constant $-25/72$ should be substituted by $+7/72$.
The above equation is derived based on eq.\ (\ref{eq:Asparametr}),
the QCD matching conditions for $\alpha_s$ at ``quark thresholds''
(Bernreuter and Wetzel, 1982; Marciano, 1984;
Barnett, Haber and Soper, 1988; Rodrigo and Santamaria, 1993)
and the one-loop relation between on-shell and pole quark masses.
Eq.(\ref{eq:Astransform}) is consistent with the QCD matching relation
at $m_f(m_f)$ (Bernreuter and Wetzel, 1982)
\begin{equation}
\alpha_s^{(N_f-1)}(m_f(m_f))=\alpha_s^{(N_f)}(m_f(m_f))
               +(\alpha_s^{(N_f)}(m_f(m_f)))^3(C_A/9-17C_F/96)/\pi^2
\label{eq:2lmatch}
\end{equation}
Here and below $N_f$ is the number of quark flavors $u,d,...,f$.
Note that the nonlogarithmic constant at $O(\alpha_s^3)$ in
eq.\ (\ref{eq:Astransform}) will not contribute in further analysis.

    Next, using the scaling properties of the MS running mass
and eq.\ (\ref{eq:Astransform}), one obtains
the following matching condition
\begin{eqnarray}
\lefteqn{\hspace{-9mm}m_f^{(N-1)}(\mu)=m_f^{(N)}(\mu)\biggl\{1+
   \biggl(\frac{\alpha_s^{(N)}(\mu)}{\pi}\biggr)^2
   \biggl[\delta(m_f,m_{f'})-\frac{5}{36}\log\frac{\mu^2}{m_f^2}
   -\frac{1}{12}\log^2\frac{\mu^2}{m_f^2}}
                                                               \nonumber\\
 && \quad \hspace{63mm}
    +\frac{1}{6}\log\frac{\mu^2}{m_f^2}\log\frac{\mu^2}{m_{f'}^2}
    -\frac{2}{9}\log\frac{m_{f'}^2}{m_f^2}\biggr]\biggr\}
\label{eq:massmatch},
\end{eqnarray}
where the constant terms are:
$1/12=\gamma_0(\beta_0^{(k-1)}-\beta_0^{(k)})/2$,
$5/36=\gamma_1^{(k-1)}-\gamma_1^{(k)}$ and
$2/9=C_F(\beta_0^{(k-1)}-\beta_0^{(k)})$.
In general, the $\delta(m_f,m_{f'})$ is the
finite contribution of the single virtual heavier
quark with mass $m_{f'}$, entering when one increases the
number of flavors
from $N-1$ to $N$ (one can also consider the particular case
$m_{f'}=m_f$).

{}From the two-loop on-shell quark mass
renormalization one has (Broadhurst, Gray and Schilcher, 1991)
\begin{equation}
\delta(m_f,m_{f'})=-\zeta(2)/3-71/144
    +(4/3)\Delta(m_{f'}/m_f),
\label{eq:deltaM}
\end{equation}
where
\begin{equation}
\Delta(r)=\frac{1}{4}\biggl[\log^2r+\zeta(2)-\biggl(\log r
        +\frac{3}{2}\biggr)r^2
        -(1+r)(1+r^3)L_{+}(r)-(1-r)(1-r^3)L_{-}(r)\biggr],
\label{eq:delta}
\end{equation}
\begin{displaymath}
L_{\pm}(r) = \int_{0}^{1/r}dx\frac{\log x}{x \pm 1}.
\end{displaymath}
$L_{\pm}(r)$ can be evaluated for different quark mass ratios
$r$ numerically.
      We relate the $\overline{\mbox{MS}}$ quark mass $m_f(m_f)$
to the pole mass $m_f$ using the $O(\alpha_s^2)$ on-shell results of
Broadhurst, Gray and Schilcher (1991)
\begin{equation}
m_f^{(N_f)}(m_f)=m_f\biggl[1-\frac{4}{3}\frac{\alpha_s^{(N_f)}(m_f)}{\pi}
     +\biggl(\frac{16}{9}-K_{f}\biggr)
         \biggl(\frac{\alpha_s^{(N_f)}(m_f)}{\pi}\biggr)^2\biggr],
\label{eq:mtopole}
\end{equation}
where
\begin{equation}
K_f= \frac{3817}{288}+\frac{2}{3}(2+\log2)\zeta(2)-\frac{1}{6}\zeta(3)
  -\frac{N_f}{3}\biggl(\zeta(2)+\frac{71}{48}\biggr)
  +\frac{4}{3}\sum_{m_l \leq m_f} \Delta\biggl(\frac{m_l}{m_f}\biggr).
\label{eq:K}
\end{equation}
The first four terms in $K_f$ represent the QCD contribution with $N_f$
massless quarks, while the sum is the correction due to the
$N_f$ nonvanishing quark masses.

  Combining eqs.\ (\ref{eq:mrun}), (\ref{eq:f}) and eqs.
(\ref{eq:Astransform})-(\ref{eq:mtopole}), one obtains the
relation between the $\overline{\mbox{MS}}$ quark mass $m_f(M_H)$ renormalized
at $M_H$ and evaluated for the $N$-flavor theory and the pole quark
mass $m_f$ (Surguladze, 1994b)
\begin{eqnarray}
\lefteqn{m_f^{(N)}(M_H)=m_f\biggl\{1
   -\frac{\alpha_s^{(N)}(M_H)}{\pi}
      \biggl(\frac{4}{3}+\gamma_0\log\frac{M_H^2}{m_f^2}\biggr)}
                                                           \nonumber\\
 && \quad \hspace{-7mm}
      -\biggl(\frac{\alpha_s^{(N)}(M_H)}{\pi}\biggr)^2\biggl[K_f
        +\sum_{m_f<m_{f'}<M_H}\delta(m_f,m_{f'})
        -\frac{16}{9}
        +\biggl(\gamma_1^{(N)}-\frac{4}{3}\gamma_0
          +\frac{4}{3}\beta_0^{(N)}\biggr)\log\frac{M_H^2}{m_f^2}
                                                           \nonumber\\
 && \quad \hspace{19mm}
      +\frac{\gamma_0}{2}(\beta_0^{(N)}-\gamma_0)\log^2\frac{M_H^2}{m_f^2}
                                                     \biggr]\biggr\}.
\label{eq:mMHtopole}
\end{eqnarray}
Note that $N$ is specified according to the size of $M_H$ and has
no correlation with the quark mass $m_f$. Thus, for instance, one can
apply eq.\ (\ref{eq:mMHtopole}) to the charm mass $m_c^{(5)}(M_H)$
evaluated for five-flavor
theory.

Substituting eqs.
(\ref{eq:mMHtopole}), (\ref{eq:K}) and appropriate $\beta$-function and
mass anomalous dimension coefficients (see section 2)
into eq.\ (\ref{eq:GHtot}),
one obtains the decay rate in terms of the pole quark masses
\begin{eqnarray}
\lefteqn{\Gamma(H\rightarrow \mbox{\small hadrons})
     =\frac{3\sqrt{2}G_FM_H}{8\pi}\sum_{f=u,d,s,...}m_f^2\biggl\{1
  +\frac{\alpha_s^{(N)}(M_H)}{\pi}
   \biggl(3-2\log\frac{M_H^2}{m_f^2}\biggr)}
                                                             \nonumber\\
 && \quad
  +\biggl(\frac{\alpha_s^{(N)}(M_H)}{\pi}\biggr)^2
   \biggl[\frac{697}{18}
    -\biggl(\frac{73}{6}+\frac{4}{3}\log 2\biggr)\zeta(2)
          -\frac{115}{6}\zeta(3)
          -N\biggl(\frac{31}{18}-\zeta(2)-\frac{2}{3}\zeta(3)\biggr)
                                                             \nonumber\\
 && \quad
    -\biggl(\frac{87}{4}-\frac{13}{18}N\biggr)\log\frac{M_H^2}{m_f^2}
    -\biggl(\frac{3}{4}-\frac{1}{6}N\biggr)\log^2\frac{M_H^2}{m_f^2}
    -\frac{8}{3}\sum_{m_l<M_H}\Delta\biggl(\frac{m_l}{m_f}\biggr)
                                                    \biggr]\biggr\}.
\label{eq:Polemassresf}
\end{eqnarray}

Recall, that at the beginning we have neglected terms which are
suppressed by powers $m_f^2/M_H^2$. Such corrections to the
decay rate, in general, may not be entirely negligible
and have to be taken into account in
precise numerical analyses. Presently those corrections due to
 the nonvanishing quark masses
have also been calculated. For the explicit results, we refer
to the original works (Surguladze, 1994a,b; Kniehl, 1995a;
Chetyrkin and Kwiatkowski, 1995). In the next
section we give the results for the quark mass corrections to the
correlation functions $\Pi$.

The full analytical result
for the decay rate of $H \rightarrow q_f\overline{q}_f$
in terms of pole quark masses, including the
leading order (two-loop) QCD corrections has been obtained independently
by several groups: Braaten and Leveille (1980), Inami and Kubota (1981),
and Dreess and Hikasa (1990). In the work by Sakai (1980)
the two-loop result  has been obtained in the zero quark mass limit.
\begin{equation}
\Gamma_{H\rightarrow q_f\overline{q}_f}
      =\frac{3\sqrt{2}G_FM_H}{8\pi}m_f^2
           \biggl(1-\frac{4m_f^2}{M_H^2}\biggr)^{\frac{3}{2}}
             \biggl[1+\frac{\alpha_s(M_H)}{\pi}
                       \delta^{(1)}(\frac{m_f^2}{M_H^2})
               +O(\alpha_s^2)\biggr],
\label{eq:2loop}
\end{equation}
where
\begin{displaymath}
\delta^{(1)}=\frac{4}{3}\biggl[\frac{a(\eta)}{\eta}
     +\frac{3+34\eta^2-13\eta^4}{16\eta^3}\log\omega
         +\frac{21\eta^2-3}{8\eta^2}\biggr],
\end{displaymath}
\begin{displaymath}
a(\eta)=(1+\eta^2)\biggl[4Li_2(\omega^{-1})+2Li_2(-\omega^{-1})
          -\log\omega\log\frac{8\eta^2}{(1+\eta)^3}\biggr]
           -\eta\log\frac{64\eta^4}{(1-\eta^2)^3},
\end{displaymath}
\begin{displaymath}
\omega=\frac{1+\eta}{1-\eta},
\hspace{5mm}
\eta=\biggl(1-\frac{4m_f^2}{M_H^2}\biggr)^{\frac{1}{2}}
\end{displaymath}
and the Spence function is defined as usual
\begin{displaymath}
Li_2(x) = -\int_{0}^{x}dx\frac{\log(1-x)}{x}
              =\sum_{n=1}^{\infty}\frac{x^n}{n^2}.
\end{displaymath}
    The expansion of the r.h.s of eq.\ (\ref{eq:2loop}) in a power series
in terms of small $m_f^2/M_H^2$ has the following form
\begin{eqnarray}
\lefteqn{\hspace{-9mm}\Gamma_{H\rightarrow q_f\overline{q}_f}
                  =\frac{3\sqrt{2}G_FM_H}{8\pi}m_f^2
        \biggl\{\biggl(1-6\frac{m_f^2}{M_H^2}+...\biggr)} \nonumber\\
 && \quad
     \hspace{-5mm} +\frac{\alpha_s(M_H)}{\pi}
                   \biggl[3-2\log\frac{M_H^2}{m_f^2}
              -\frac{m_f^2}{M_H^2}\biggl(8-24\log\frac{M_H^2}{m_f^2}\biggr)
              +...\biggr]+O(\alpha_s^2)\biggr\},
\label{eq:2loopexpan}
\end{eqnarray}
where the periods cover higher order terms $\sim (m_f/M_H)^{2k}$, $k=2,3...$
One can see that the leading terms agree with the result
(\ref{eq:Polemassresf}).

Numerically the $\overline{\mbox{MS}}$ high order QCD corrections for
the considered process are large and reduce the decay rates by about 40\%.

\vspace{6mm}

\renewcommand{\thesection}{\Roman{section}}
\section{\bf Quark mass corrections to the correlation functions}
\renewcommand{\thesection}{\arabic{section}}
\setcounter{equation}{0}

In the previous section we neglected
all quark masses in the corresponding Feynman diagrams in comparison with
the momentum scale of the problem. In other words, we have calculated
the leading term in the expansion in terms of small $m_f^2/s$
(for the Higgs boson decay, $s=M_H^2$) in the limit of infinitely heavy
top quark, $m_t\rightarrow \infty$. However, in the real world
quarks are massive and the leading term in the above expansion may not
always give a satisfactory approximation. On the other hand, starting at
$O(\alpha_s^2)$, virtual heavy quark can also appear
in certain topological type of Feynman diagrams
(Fig.\ 2 and Fig.\ 3) regardless of the momentum scale of the problem.

\vspace{32mm}

\begin{center}
FIG.\ 2. \ $O(\alpha_s^2)$ Feynman diagrams responsible for the
virtual heavy quark contribution
\end{center}

\vspace{32mm}

\begin{center}
FIG.\ 3. \ $O(\alpha_s^2)$ Feynman diagrams responsible for the
contribution due to the top-bottom mass splitting
\end{center}

According to the decoupling
theorem (Appelquist and Carazzone, 1975), virtual quarks much heavier than
the momentum scale of the problem decouple. However, for instance, in the
process of Z boson decay the effect of the top quark may not be entirely
negligible since $m_t$ is not much greater than $M_Z$. A similar role
could be played by the charm quark in the hadronic decay of the tau-lepton.
The evaluation of the virtual top quark contribution (Fig.\ 2) to the decay
rate $Z\rightarrow \mbox{hadrons}$ and related quantities has been done in
Kniehl (1990), Soper and Surguladze (1994), and Hoang, Jezabek, K\"{u}hn
and Teubner (1994) without using large or small mass approximations.
The correction turned out to be moderate and in good agreement with
the results obtained with the help of the large mass expansion technique
(Chetyrkin, 1993a). The contribution of the diagrams in Fig.\ 2, in the
presence
of a virtual heavy quark, to the two-point correlation function of the
electromagnetic quark currents has been evaluated previously by
Wetzel and Bernreuther (1981).
Kniehl and K\"{u}hn (1989, 1990) have calculated the $O(\alpha_s^2)$ correction
to the decay rate $Z\rightarrow \mbox{hadrons}$ due to the large mass splitting
in the top-bottom doublet (Fig.\ 3). This correction turned out to be large
and important.

In this section we consider only the
leading correction in the expansion in terms of small quark mass.
For the calculations of virtual heavy quark contributions
we refer the reader to the above mentioned original works
(see also Kniehl, 1994b, 1995b).
The discussion in this section is based on the works
by Surguladze (1994a,b,c).

Let us expand the full two-point correlation function, defined by eq.\
(\ref{eq:pifunction})
in the vector channel and by eq.\ (\ref{eq:pifunctions}) in the
scalar and pseudoscalar channels,
in powers of $m_f^2/Q^2$ in the ``deep'' Euclidean region
\begin{equation}
\biggl(\frac{1}{m_f^2Q^2}\biggr)^d\Pi(Q^2,m_f,m_{\mbox{\tiny V}})
   =\Pi_1(Q^2)+\frac{m_f^2}{Q^2}\Pi_{m_f^2}(Q^2)
      +\sum_{{\mbox{\tiny V}}
     =u,d,s,c,b}\frac{m_{\mbox{\tiny V}}^2}{Q^2}\Pi_{m_{\mbox{\tiny V}}^2}(Q^2)
        +...,
\label{eq:Piexpan}
\end{equation}
where $d=0$ in the vector channel and $d=1$ in the scalar and pseudoscalar
channels.
The last term in the above expansion is due to the Feynman diagrams
containing a virtual fermionic loop. Note however that in the vector
channel the contribution
from the diagrams in Fig.\ 3 vanishes according to Furry's theorem
(Furry, 1937).

In order to evaluate the coefficient functions in the r.h.s
of eq.\ (\ref{eq:Piexpan}), it is
sufficient to write the diagrammatic representation for
$\Pi(Q^2,m_f^{\mbox{\tiny B}},m_{\mbox{\tiny V}}^{\mbox{\tiny B}})$
 up to the desired level of perturbation
theory and apply the appropriate projector. To $O(\alpha_s^2)$
one has
\begin{eqnarray}
\lefteqn{\hspace{-9mm}\Pi_{m_{f}^{2n}
            m_{\mbox{\tiny V}}^{2k}}(Q^2,\alpha_s) =} \nonumber\\
 &&
     \frac{1}{(2n)!(2k)!}
       \biggl(\frac{d}{dm_f^{\mbox{\tiny B}}}\biggr)^{2n}
        \biggl(\frac{d}{dm_{\mbox{\tiny V}}^{\mbox{\tiny B}}}\biggr)^{2k}
        \biggl\{\frac{\Pi(Q^2,m_f^{\mbox{\tiny B}}
     ,m_{\mbox{\tiny V}}^{\mbox{\tiny B}},\alpha_s^{\mbox{\tiny B}})}
                     {(m_f^{\mbox{\tiny B}})^{2d} Q^{2(d-n-k)}}\biggr\}
      _{\stackrel{m_f^{\mbox{\tiny B}}=m_{\mbox{\tiny V}}^{\mbox{\tiny B}}=0}
 {\alpha_s^{\mbox{\tiny B}}\rightarrow Z_{\alpha}\alpha_s}}
                (Z_m^2)^{(1+d)},
\label{eq:Proj}
\end{eqnarray}
where $n,k=0,1$, $n+k\leq 1$, and superscript ``B''
denotes the bare quantities. The mass renormalization constant
$Z_m=m_f^{\mbox{\tiny B}}/m_f$ can be obtained from eq.\ (\ref{eq:mrenorm}).
The Feynman diagrams contributing to the
$\Pi_{m_{f}^{2n}m_{\mbox{\tiny V}}^{2k}}$ are the same as
for the calculation of $\Pi_1$ (see eq.\ (\ref{eq:pidiagram}))
but with massive fermion propagators.
The calculations of all  one-, two- and three-loop diagrams have been done
using the program {\small HEPL}oops (Surguladze, 1992).

The obtained expressions for $\Pi_i$ at each order of $\alpha_s$
are polynomials with respect to $1/\varepsilon$ and
$\log\mu_{\overline{\mbox{\tiny MS}}}^2/Q^2$. The poles can be removed by an
additive renormalization. We note that there are no terms like
$(1/\varepsilon^n)(\log\mu_{\overline{\mbox{\tiny MS}}}^2/Q^2)^k$.
They appear only at higher orders $\sim m_f^2 m_f^4/Q^4$ and represent
infrared mass logarithms. The corresponding prescription similar
to the Bogolyubov ultraviolet $R$-operation has been worked out in the
work by Chetyrkin, Gorishny and Tkachov (1982), Tkachov (1983b,c), and
Gorishny, Larin and Tkachov (1983).
(see also Tkachov, 1991, 1993 and references
therein). The infrared mass singularities have been studied earlier by
Marciano (1975). In the present paper we consider only the terms
$\sim m_f^2/Q^2$ which are sufficient for most of the
phenomenologically interesting applications.

In the vector channel we obtain the following $\overline{\mbox{MS}}$
 analytical result (Gorishny, Kataev and Larin, 1986; Surguladze, 1994c)
\begin{eqnarray}
\lefteqn{\hspace{-9mm}\Pi_{m_{f}^2}\left(
   \frac{\mu_{\overline{\mbox{\tiny MS}}}^{2}}{Q^2},\alpha_s\right)=
       \frac{N_F}{(4\pi)^2}\biggl\{-8
       -\biggl(\frac{\alpha_s}{\pi}\biggr)C_F \biggl(16
      +12\log\frac{\mu_{\overline{\mbox{\tiny MS}}}^{2}}{Q^2}\biggr)}
                                                                   \nonumber\\
 && \quad
      -\biggl(\frac{\alpha_s}{\pi}\biggr)^2
    \biggl[  C_F^2\biggl(\frac{1667}{24}
                         -\frac{5}{3}\zeta(3)-\frac{70}{3}\zeta(5)
            +\frac{51}{2}\log\frac{\mu_{\overline{\mbox{\tiny MS}}}^{2}}{Q^2}
            +9\log^2\frac{\mu_{\overline{\mbox{\tiny MS}}}^{2}}{Q^2} \biggr)
                                                                   \nonumber\\
 && \quad \hspace{1cm}
            +C_FC_A\biggl(\frac{1447}{24}
                         +\frac{16}{3}\zeta(3)-\frac{85}{3}\zeta(5)
            +\frac{185}{6}\log\frac{\mu_{\overline{\mbox{\tiny MS}}}^{2}}{Q^2}
  +\frac{11}{2}\log^2\frac{\mu_{\overline{\mbox{\tiny MS}}}^{2}}{Q^2} \biggr)
                                                                   \nonumber\\
 && \quad \hspace{1cm}
            -C_FTN\biggl(\frac{95}{6}
            +\frac{26}{3}\log\frac{\mu_{\overline{\mbox{\tiny MS}}}^{2}}{Q^2}
  +2\log^2\frac{\mu_{\overline{\mbox{\tiny MS}}}^{2}}{Q^2} \biggr)
                                                  \biggr]\biggr\},
\label{eq:PimanalytV1}
\end{eqnarray}
\begin{equation}
\hspace{-66mm}\Pi_{m_{\mbox{\tiny V}}^2}
      \left(\frac{\mu_{\overline{\mbox{\tiny MS}}}^{2}}{Q^2},\alpha_s\right)
   =\frac{N_F}{(4\pi)^2}\biggl(\frac{\alpha_s}{\pi}\biggr)^2C_FT
   \biggl[\frac{64}{3}-16\zeta(3)\biggr].
\label{eq:PimanalytV2}
\end{equation}

The contribution to the physical process, in particular to the
decay rate of $Z\rightarrow \mbox{hadrons}$ can be obtained
 simply by taking the
imaginary part in the r.h.s. of eqs.\ (\ref{eq:PimanalytV1})
and (\ref{eq:PimanalytV2}) at $Q^2=-s+i0$. We note, that
the  $\Pi_{m_{f}^2}$ and $\Pi_{m_{\mbox{\tiny V}}^2}$ turned out to be finite.
No overall subtraction is necessary. Moreover, one can see that the
imaginary part or the contribution to the decay rate
vanishes at the parton level. This can be checked by the calculation
of the  parton contribution in the vector channel with explicit dependence
on quark mass. Indeed, calculating the trivial fermionic loop we obtain
\begin{equation}
\Pi_{\mbox{\scriptsize parton}}
  (\frac{\mu_{\overline{\mbox{\tiny MS}}}^{2}}{-q^2},\frac{m_f^2}{-q^2})
 =\frac{N_F}{(4\pi^2)}
    \biggl[\frac{4}{3}\frac{1}{\varepsilon}
   -8\int_{0}^{1}x(1-x)\log\frac{m_f^2-x(1-x)q^2}
      {\mu_{\overline{\mbox{\tiny MS}}}^2}dx
    \biggr].
\label{eq:1lPi}
\end{equation}
Taking the discontinuity under the integral and then evaluating the trivial
integral with the $\Theta$ function, we obtain
\begin{equation}
\frac{1}{2\pi i}\mbox{disc}\Pi_{\mbox{\scriptsize parton}}
       (\frac{\mu_{\overline{\mbox{\tiny MS}}}^{2}}{-q^2},\frac{m_f^2}{-q^2})
 =\frac{N_F}{(4\pi^2)}\biggl(1+\frac{2m_f^2}{q^2}\biggr)
       \sqrt{1-\frac{4m_f^2}{q^2}}
       =\frac{N_F}{(4\pi^2)}O\biggl(\frac{m_f^4}{q^4}\biggr).
\label{eq:1lPidisc}
\end{equation}
The $\sim m_f^2/Q^2$ contribution to the Adler $D$-function can be obtained
 from eqs.\ (\ref{eq:PimanalytV1})
and (\ref{eq:PimanalytV2}) by differentiating with respect to $Q^2$.

There is some confusion concerning the above results in the literature.
Initially, the corrections $\sim m_f^2/Q^2$
in the vector channel have been calculated by Gorishny, Kataev and Larin
(1986).
Later, in the similar calculations (Surguladze, 1989a),
a slightly different result was obtained, which was confirmed in further
publications (see, e.g.,  Kataev, 1990, 1991).
However, in the recent works (Chetyrkin and Kwiatkowski, 1993;
Surguladze, 1994c),
the initial result of Gorishny, Kataev and Larin (1986) has been confirmed.
Unfortunately, in the analysis of the mass corrections to the
$Z$ decay rates (Chetyrkin and K\"{u}hn, 1990) the incorrect result
was used. Fortunately, the main conclusions of Chetyrkin and K\"{u}hn (1990)
are not affected.
Summarizing, we note that the results (\ref{eq:PimanalytV1})
and (\ref{eq:PimanalytV2}) (Gorishny, Kataev and Larin, 1986;
Chetyrkin and Kwiatkowski, 1993;
Surguladze, 1994c) seem now to be reliable.

  In the scalar channel the result for the standard SU$_{\mbox{\scriptsize
c}}$(3) gauge
group reads (Surguladze, 1994b)
\begin{eqnarray}
\lefteqn{\hspace{-1mm}\Pi_{m_f^2}\biggl(
           \frac{\mu_{\overline{\mbox{\tiny MS}}}^2}{Q^2},\alpha_s\biggr)=
       -\frac{1}{4\pi^2}\biggl\{12+9
          \log\frac{\mu_{\overline{\mbox{\tiny MS}}}^2}{Q^2}} \nonumber\\
 && \quad
       +\biggl(\frac{\alpha_s}{\pi}\biggr)\biggl(94-36\zeta(3)
      +60\log\frac{\mu_{\overline{\mbox{\tiny MS}}}^{2}}{Q^2}
      +18\log^2\frac{\mu_{\overline{\mbox{\tiny MS}}}^{2}}{Q^2}\biggr)
                                                                   \nonumber\\
 && \quad
      +\biggl(\frac{\alpha_s}{\pi}\biggr)^2
    \biggl[\frac{17245}{16}
           -\frac{1690}{3}\zeta(3)-3\zeta(4)+\frac{385}{3}\zeta(5)
                                                                   \nonumber\\
 && \quad \hspace{9mm}
     +\biggl(\frac{7149}{8}-249\zeta(3)\biggr)
               \log\frac{\mu_{\overline{\mbox{\tiny MS}}}^{2}}{Q^2}
      +\frac{1113}{4}\log^2\frac{\mu_{\overline{\mbox{\tiny MS}}}^{2}}{Q^2}
      +\frac{81}{2}\log^3\frac{\mu_{\overline{\mbox{\tiny MS}}}^{2}}{Q^2}
                                                                   \nonumber\\
 && \quad \hspace{9mm}
      -N\biggl(\frac{817}{24}-6\zeta(3)
      +\biggl(\frac{313}{12}-6\zeta(3)\biggr)
               \log\frac{\mu_{\overline{\mbox{\tiny MS}}}^{2}}{Q^2}
      +\frac{15}{2}\log^2\frac{\mu_{\overline{\mbox{\tiny MS}}}^{2}}{Q^2}
      +\log^3\frac{\mu_{\overline{\mbox{\tiny MS}}}^{2}}{Q^2}\biggr)\biggr]
                                                                   \nonumber\\
 && \quad \hspace{77mm}
           +\mbox{\small ``simple poles''} \biggr\},
\label{eq:PimanalytS1}
\end{eqnarray}
\begin{equation}
\hspace{-42mm}\Pi_{m_{\mbox{\tiny V}}^2}
      \left(\frac{\mu_{\overline{\mbox{\tiny MS}}}^{2}}{Q^2},\alpha_s\right)
   =\frac{1}{4\pi^2}\biggl(\frac{\alpha_s}{\pi}\biggr)^2
   \biggl[\frac{8}{3}+6\log\frac{\mu_{\overline{\mbox{\tiny MS}}}^{2}}{Q^2}
   +\mbox{\small ``simple pole''}\biggr],
\label{eq:PimanalytS2}
\end{equation}
where under the ``simple pole'' we mean number$/\varepsilon^k$ with
no dependence on $\log\mu^2/Q^2$. The ``simple poles'' have no
imaginary part and consequently will not contribute to the observable
quantities at the given order of $\alpha_s$.
Note that the $\Pi_{m_{\mbox{\tiny V}}^2}$ in eq.
(\ref{eq:PimanalytS2}) does not include the contribution from the
triangle anomaly type graphs pictured in Fig.\ 3. Those graphs make
the following additional contribution to $\Pi$ in eq.\ (\ref{eq:Piexpan})
(Surguladze, 1994b)
\begin{equation}
+\sum_{f'=u,d,s,c,b}\frac{m_{f'}^2}{Q^2} \times
   \frac{1}{4\pi^2}\biggl(\frac{\alpha_s}{\pi}\biggr)^2
   \biggl[\frac{118}{3}-20\zeta(3)-10\zeta(5)
                     +12\log\frac{\mu_{\overline{\mbox{\tiny MS}}}^{2}}{Q^2}
   +\mbox{\small ``simple pole''}\biggr].
\label{eq:PimanalytS3}
\end{equation}
The above results are relevant for the decay rate
of the standard model Higgs boson into a quark antiquark pair,
calculated in the previous section in the massless quark limit.
Corrections $\sim m_f^2/M_H^2$ can be obtained
from eqs.\ (\ref{eq:PimanalytS1}), (\ref{eq:PimanalytS2})
 and (\ref{eq:PimanalytS3}) (Surguladze, 1994b).

In the pseudoscalar channel we define the quark currents as
$j_f=m_f\overline{q}_fi\gamma_5 q_f$. We also define the
 $\gamma_5$ matrix
in $D$-dimensional space-time as an object with the following properties
\begin{equation}
\{\gamma_5,\gamma_\mu\}=0, \hspace{5mm} \gamma_5\gamma_5=1.
\label{eq:gamma5def}
\end{equation}
The above definition causes no problems in dimensional regularization
when there are two $\gamma_5$ matrices in a closed fermionic loop.
We obtain (Surguladze, 1994a)
\begin{eqnarray}
\lefteqn{\hspace{-4mm}\Pi_{m_{f}^2}\left(
   \frac{\mu_{\overline{\mbox{\tiny MS}}}^{2}}{Q^2},\alpha_s\right)=
       -\frac{1}{4\pi^2}\biggl\{3
                    \log\frac{\mu_{\overline{\mbox{\tiny MS}}}^{2}}{Q^2}
       +\biggl(\frac{\alpha_s}{\pi}\biggr)\biggl(6-12\zeta(3)
      +4\log\frac{\mu_{\overline{\mbox{\tiny MS}}}^{2}}{Q^2}
      +6\log^2\frac{\mu_{\overline{\mbox{\tiny MS}}}^{2}}{Q^2}\biggr)}
                                                                   \nonumber\\
 && \quad
      +\biggl(\frac{\alpha_s}{\pi}\biggr)^2
    \biggl[-\frac{6713}{144}
           -116\zeta(3)-\zeta(4)+\frac{235}{3}\zeta(5)
                                                                   \nonumber\\
 && \quad \hspace{1cm}
     +\biggl(\frac{1429}{24}-83\zeta(3)\biggr)
               \log\frac{\mu_{\overline{\mbox{\tiny MS}}}^{2}}{Q^2}
      +\frac{155}{4}\log^2\frac{\mu_{\overline{\mbox{\tiny MS}}}^{2}}{Q^2}
      +\frac{27}{2}\log^3\frac{\mu_{\overline{\mbox{\tiny MS}}}^{2}}{Q^2}
                                                                   \nonumber\\
 && \quad \hspace{5mm}
      -N\biggl(-\frac{31}{72}-\frac{2}{3}\zeta(3)
      +\biggl(\frac{9}{4}-2\zeta(3)\biggr)
               \log\frac{\mu_{\overline{\mbox{\tiny MS}}}^{2}}{Q^2}
      +\frac{7}{6}\log^2\frac{\mu_{\overline{\mbox{\tiny MS}}}^{2}}{Q^2}
+\frac{1}{3}\log^3\frac{\mu_{\overline{\mbox{\tiny
MS}}}^{2}}{Q^2}\biggr)\biggr]
                                                                   \nonumber\\
 && \quad \hspace{77mm}
           +\mbox{\small ``simple poles''} \biggr\},
  \label{eq:PimanalytPS1}
\end{eqnarray}
\begin{equation}
\hspace{-43mm}\Pi_{m_{\mbox{\tiny V}}^2}
      \left(\frac{\mu_{\overline{\mbox{\tiny MS}}}^{2}}{Q^2},\alpha_s\right)
   =\frac{1}{4\pi^2}\biggl(\frac{\alpha_s}{\pi}\biggr)^2
   \biggl[\frac{8}{3}+6\log\frac{\mu_{\overline{\mbox{\tiny MS}}}^{2}}{Q^2}
   +\mbox{\small ``simple pole''}\biggr].
\label{eq:PimanalytPS2}
\end{equation}
The result for the pseudoscalar channel is relevant, for instance,
for the decay rates of the minimal supersymmetric version of the
Higgs particle into a quark antiquark pair (see Surguladze, 1994a).

Finally, we present the results of calculation of the $\sim m_f^2/Q^2$
 corrections to the correlation function in the axial channel
(Soper and Surguladze, 1994; Surguladze, 1994c).
We use the following definition of the correlation function
\begin{equation}
i\int d^4x e^{iqx}<Tj_{\mu}^f(x)j_{\nu}^f(0)>_0=
  g_{\mu\nu}Q^2\Pi(Q,m_f)-Q_{\mu}Q_{\nu}\Pi'(Q,m_f),
\label{eq:AxialPi}
\end{equation}
where $j_{\mu}^f=\overline{q}_f\gamma_{\mu}\gamma_5q_f$.
Note that in the axial channel the correlation function is
not transverse in contrast to the vector channel. However, for the
decay rate of the $Z$-boson only the $\sim g_{\mu\nu}$  part in
eq.\ (\ref{eq:AxialPi}) is relevant.

The expansions of $\Pi$ and $\Pi'$  in terms of small $m_f^2/Q^2$ has the same
form as in the vector channel ( eq.\ (\ref{eq:Piexpan}) ).
The coefficient functions in this expansion can be calculated according
to eq.\ (\ref{eq:Proj}) in the vector channel.
In the calculations of one-, two- and three-loop Feynman diagrams
the program {\small HEPL}oops (Surguladze, 1992) was used. The final results
for the SU$_{\mbox{\scriptsize c}}$(3) gauge group
 read (Soper and Surguladze, 1994;
Surguladze, 1994c)

\newpage
\begin{eqnarray}
\lefteqn{\hspace{-2mm}\Pi_{m_{f}^2}\left(
      \frac{\mu_{\overline{\mbox{\tiny MS}}}^{2}}{Q^2},\alpha_s\right)=
       \frac{1}{4\pi^2}\biggl\{6+6
             \log\frac{\mu_{\overline{\mbox{\tiny MS}}}^{2}}{Q^2} }
                                                                   \nonumber\\
 && \quad
       +\biggl(\frac{\alpha_s}{\pi}\biggr)\biggl(\frac{107}{2}-24\zeta(3)
      +22\log\frac{\mu_{\overline{\mbox{\tiny MS}}}^{2}}{Q^2}
      +6\log^2\frac{\mu_{\overline{\mbox{\tiny MS}}}^{2}}{Q^2}\biggr)
                                                                   \nonumber\\
 && \quad
      +\biggl(\frac{\alpha_s}{\pi}\biggr)^2
    \biggl[\frac{3241}{6}
           -387\zeta(3)-\frac{3}{2}\zeta(4)+165\zeta(5)
                                                                   \nonumber\\
 && \quad \hspace{1cm}
     +\biggl(\frac{8221}{24}-117\zeta(3)\biggr)
               \log\frac{\mu_{\overline{\mbox{\tiny MS}}}^{2}}{Q^2}
      +\frac{155}{2}\log^2\frac{\mu_{\overline{\mbox{\tiny MS}}}^{2}}{Q^2}
      +\frac{19}{2}\log^3\frac{\mu_{\overline{\mbox{\tiny MS}}}^{2}}{Q^2}
                                                                   \nonumber\\
 && \quad \hspace{5mm}
      -N\biggl(\frac{857}{36}-\frac{32}{3}\zeta(3)
      +\biggl(\frac{151}{12}-4\zeta(3)\biggr)
               \log\frac{\mu_{\overline{\mbox{\tiny MS}}}^{2}}{Q^2}
      +\frac{8}{3}\log^2\frac{\mu_{\overline{\mbox{\tiny MS}}}^{2}}{Q^2}
      +\frac{1}{3}\log^3\frac{\mu_{\overline{\mbox{\tiny
MS}}}^{2}}{Q^2}\biggr)\biggr]
                                                                   \nonumber\\
 && \quad \hspace{77mm}
           +\mbox{\small ``simple poles''} \biggr\}
  \label{eq:PimanalytAT1}
\end{eqnarray}
\begin{equation}
\hspace{-79mm}\Pi_{m_{\mbox{\tiny V}}^2}
      \left(\frac{\mu_{\overline{\mbox{\tiny MS}}}^{2}}{Q^2},\alpha_s\right)
   =\frac{1}{4\pi^2}\biggl(\frac{\alpha_s}{\pi}\biggr)^2
   \biggl[\frac{32}{3}-8\zeta(3)\biggr]
\label{eq:PimanalytAT2}
\end{equation}
\begin{eqnarray}
\lefteqn{\hspace{-22mm}\Pi'_{m_{f}^2}\left(
                 \frac{\mu_{\overline{\mbox{\tiny
MS}}}^{2}}{Q^2},\alpha_s\right)=
       \frac{1}{4\pi^2}\biggl\{-6
       +\biggl(\frac{\alpha_s}{\pi}\biggr)\biggl(-12
      -12\log\frac{\mu_{\overline{\mbox{\tiny MS}}}^{2}}{Q^2}\biggr)}
                                                                   \nonumber\\
 && \quad
      +\biggl(\frac{\alpha_s}{\pi}\biggr)^2
    \biggl[-\frac{4681}{24}
           -34\zeta(3)+115\zeta(5)
                                                                   \nonumber\\
 && \quad \hspace{1cm}
     -\frac{215}{2}
               \log\frac{\mu_{\overline{\mbox{\tiny MS}}}^{2}}{Q^2}
      -\frac{57}{2}\log^2\frac{\mu_{\overline{\mbox{\tiny MS}}}^{2}}{Q^2}
                                                                   \nonumber\\
 && \quad \hspace{9mm}
      -N\biggl(-\frac{55}{12}-\frac{8}{3}\zeta(3)
      -\frac{11}{3}
               \log\frac{\mu_{\overline{\mbox{\tiny MS}}}^{2}}{Q^2}
      -\log^2\frac{\mu_{\overline{\mbox{\tiny MS}}}^{2}}{Q^2}\biggr)\biggr]
                                                                   \nonumber\\
 && \quad \hspace{67mm}
           +\mbox{\small ``simple poles''} \biggr\}
  \label{eq:PimanalytAL1}
\end{eqnarray}
\begin{equation}
\Pi'_{m_{\mbox{\tiny V}}^2}=\Pi_{m_{\mbox{\tiny V}}^2}
\label{eq:PimanalytAL2}
\end{equation}

The results given in this section can be tested using the
renormalization group. Namely, the relations similar to
eqs.\ (\ref{eq:relations1}), (\ref{eq:relations2}) and (\ref{eq:relations3})
can be obtained here (Surguladze, 1994a,b,c).
In fact, in the vector channel, one can obtain the $O(\alpha_s^3)$
logarithmic terms without actual calculation of the corresponding
four-loop diagrams. On the other hand, the leading logarithmic terms
in $\Pi$-function form the corresponding contribution to the decay rates
of, for instance, the Z-boson (Chetyrkin and K\"{u}hn, 1990;
Chetyrkin, K\"{u}hn and Kwiatkowski, 1992; Surguladze, 1994c).
In the axial channel the situation is more complicated. Here,
because the renormalization group equation similar to
eq.\ (\ref{eq:RGEDscalar}) is no longer a homogeneous one,
the renormalization group approach is restricted to $O(\alpha_s^2)$.

\renewcommand{\thesection}{\Roman{section}}
\section{\bf Two-loop coefficient functions of $\mbox{dim}=4$
 power corrections}
\renewcommand{\thesection}{\arabic{section}}
\setcounter{equation}{0}

In this section we outline the
calculations of the two-loop coefficient functions of $\mbox{dim}=4$
power corrections. We consider the
contributions which appear in the short distance expansion of the
correlation function of two flavor-diagonal vector, scalar and
pseudoscalar currents constructed from light quark fields.
 The methods and corresponding references are given in the earlier
 sections.
The corrections for the vector channel have been evaluated
in  Loladze, Surguladze and Tkachov (1984, 1985) and
 Surguladze and Tkachov (1988).
In the scalar and pseudoscalar channels, the calculation has
been done in Surguladze and Tkachov (1990). The
 calculation for vector and axial vector channels has been
done in Chetyrkin, Gorishny and Spiridonov (1985), where the
 previous results for the vector channel have been confirmed
 and the calculation was extended for flavor non-diagonal currents
 as well. The three-loop
 correction to the coefficient function of gluon condensate in the
 scalar channel has also been computed in
Surguladze and Tkachov (1989b).
 For the calculation
 of dimension 8 terms in the operator product expansion see also
 Broadhurst and Generalis (1985).
 Here we follow the work by Surguladze and Tkachov (1990).

Consider first the T-product of flavor diagonal vector
currents of light quarks
\begin{equation}
{\cal T}_{\mu\nu}^{f'}(Q)=i\int d^{4}x e^{iqx} T
J_{\mu}^{f'}(x)J_{\nu}^{f'}(0),
\label{eq:Pid4V}
\end{equation}
where $J^{f'}_{\mu}=\overline{q}_{f'}\gamma_{\mu}q_{f'}$.
Taking into account the current conservation and operator product expansion
 technique (Wilson, 1969) for large momentum transfer
($Q^2 \rightarrow \infty$) we write
\begin{equation}
{\cal T}_{\mu\nu}^{f'}(Q)=(g_{\mu\nu}Q^2-Q_{\mu}Q_{\nu})
 \biggl\{C_{0}+\frac{1}{Q^4}\biggl[C_{G^2}(Q^2)(G_{\mu\nu}^{a})^2+
   \sum_{f'}C_{\overline{q}q}^{f'}(Q^2)m_{f'}\overline{q}_{f'}q_{f'}\biggr]
   +\cdots \biggr\},
\label{eq:Pid4VOPE}
\end{equation}
where $C_{0}$ is the coefficient function of the unity operator including
the terms $\sim m_f^2/Q^2$ discussed in the previous section.
The period covers the operators of higher twists. For the scalar and
 pseudoscalar channels the transverse factor in the above equation is absent.
To simplify the calculation, we contract over the Lorentz indices
$\mu$ and $\nu$. Then the expressions for $C_{i}$ defined in eq.
 (\ref{eq:Pid4VOPE}) coincide with the ones in eq.\ (\ref{eq:CFexplicit})
 if ${\cal T}(Q)$ is replaced by
${\cal T}_{\mu\mu}^{f'}(Q^2)/(D-1)Q^2$, where $D=4-2\varepsilon$.
Let us rewrite eqs.\ (\ref{eq:CFexplicit}) in a somewhat symbolic
 diagrammatic representation to $O(\alpha_s^2)$.

\vspace{3mm}

\begin{displaymath}
C_{G^2}=\pi_{1}\frac{\alpha_s}{\pi}\biggl\{2\hspace{16mm}+4\hspace{16mm}
        +\frac{\alpha_s}{\pi}
       \biggl[\hspace{17mm}+...+\hspace{17mm}+...
                        +\hspace{17mm}+...\biggr]\biggr\}
\end{displaymath}

\vspace{3mm}

\begin{displaymath}
\hspace{-8mm}C_{\overline{q}q}^{f}
    =\pi_{2}^{f}\biggl\{2\hspace{23mm}+\frac{\alpha_s}{\pi}
       \biggl[2\hspace{23mm}+4\hspace{23mm}+2\hspace{23mm}+...\biggr]
\end{displaymath}

\vspace{2mm}

\begin{equation}
  \hspace{93mm}
      +\biggl(\frac{\alpha_s}{\pi}\biggr)^2\biggl[\hspace{23mm}+...\biggr]
      \biggr\}.
\label{eq:OPEdiagrammer}
\end{equation}

\vspace{3mm}

The total number of two-loop graphs contributing to $C_{G^2}$ is 30 and to
$C_{\overline{q}q}^{f}$ is 38. There is a simple rule for generating
the appropriate graphs at $O(\alpha_{s}^n)$. One should take
 the graphs contributing to $O(\alpha_{s}^{(n+1)})$ in the
unity operator and disconnect one fermion line in all possible ways
 for the coefficient function $C_{\overline{q}q}^{f}$. For the
coefficient function  $C_{G^2}$ it is necessary to write all the
diagrams with one disconnected gluon line (relevant for the
projector ${\cal P}_{1}$), all the diagrams with one disconnected
ghost line (relevant for the projector ${\cal P}_{4}$) and all
the diagrams with disconnected gluon-ghost-ghost vertex
 (relevant for the projector ${\cal P}_{5}$). To see this,
recall eqs.\ (\ref{eq:ElProj}). Acting with the projectors
 ${\cal P}_{j}$ on the appropriate
 diagrams, the calculations are reduced to the evaluation of one-
and two-loop propagator type massless Feynman integrals.
In the original calculation (Loladze, Surguladze and Tkachov, 1984, 1985;
Surguladze and Tkachov, 1988, 1990)
 all Feynman integrals have been evaluated analytically using
the {\small REDUCE} (Hearn, 1973) program LOOPS (Surguladze and Tkachov,
1989a).

The $\overline{\mbox{MS}}$ results for the projectors ${\cal P}_{j}$
in the vector channel read
\begin{equation}
{\cal P}_{1}[{\cal T}_{\mu\mu}^{f'}]=\frac{1}{Q^4}C_{F}\frac{N_{F}}{N_{A}}
   \frac{\alpha_{s}^{\mbox{\tiny B}}}{\pi}\biggl\{48-32\varepsilon
       +\frac{\alpha_{s}^{\mbox{\tiny B}}}{\pi}\biggl[C_{F}(-12)
      +C_{A}(\frac{18}{\varepsilon}-42+72\zeta(3)\biggr)\biggr]
                   +O(\alpha_{s}^2)\biggr\}
\label{eq:P1}
\end{equation}
\begin{equation}
{\cal P}_{4}[{\cal T}_{\mu\mu}^{f'}]=\frac{1}{Q^4}C_{F}\frac{N_{F}}{N_{A}}C_{A}
   \biggl(\frac{\alpha_{s}^{\mbox{\tiny B}}}{\pi}\biggr)^2
       \biggl(\frac{3}{\varepsilon}-9+12\zeta(3)\biggr)
                  +O(\alpha_{s}^3),
\label{eq:P4}
\end{equation}
\begin{equation}
{\cal P}_{5}[{\cal T}_{\mu\mu}^{f'}]=0+O(\alpha_{s}^3),
\label{eq:P5}
\end{equation}
\begin{equation}
\biggl({\cal P}_{2}^{f\neq f'}+\frac{1}{D}{\cal P}_{3}^{f\neq f'}
\biggr)[{\cal T}_{\mu\mu}^{f'}]=
     \frac{1}{Q^4}C_{F}T
   \biggl(\frac{\alpha_{s}^{\mbox{\tiny B}}}{\pi}\biggr)^2
       \biggl(\frac{24}{\varepsilon}-60+96\zeta(3)\biggr)
                  +O(\alpha_{s}^3),
\label{eq:P231}
\end{equation}
\begin{eqnarray}
\lefteqn{\hspace{-12mm}
\biggl({\cal P}_{2}^{f=f'}+\frac{1}{D}{\cal P}_{3}^{f=f'}\biggr)
[{\cal T}_{\mu\mu}^{f'}]=} \nonumber\\
 && \quad
   \frac{1}{Q^4}\biggl\{6
         +\frac{\alpha_{s}^{\mbox{\tiny B}}}{\pi}C_{F}\biggl(\frac{3}{2}
                                         +\frac{11}{4}\varepsilon\biggr)
         +\biggl(\frac{\alpha_{s}^{\mbox{\tiny B}}}{\pi}\biggr)^2C_{F}
        \biggl[C_{F}\frac{387}{16}+C_{A}\biggl(\frac{11}{8\varepsilon}
                                         +\frac{7}{16}\biggr)
                                                               \nonumber\\
 && \quad \hspace{7mm}
       +T\biggl(\frac{3}{4\varepsilon}-\frac{15}{4}+6\zeta(3)\biggr)
          -TN\biggl(\frac{1}{2\varepsilon}+\frac{7}{4}\biggr)\biggr]
                  +O(\alpha_{s}^3) \biggr\}.
\label{eq:P232}
\end{eqnarray}
The vanishing of ${\cal P}_{5}[{\cal T}_{\mu\mu}^{f'}]$
at the two-loop level is the consequence of gauge invariance,
as was shown by Spiridonov (1987).
Combining eqs.\ (\ref{eq:piexplicit}) and (\ref{eq:CFexplicit})
with the  above results and renormalizing the bare coupling via eq.
(\ref{eq:AsR}) we obtain $\overline{\mbox{MS}}$ $O(\alpha_{s}^2)$
analytical  expressions for the coefficient functions in the vector channel
(Surguladze and Tkachov, 1990)
\begin{equation}
C_{G^2}(Q^2)=\frac{1}{Q^4}C_{F}\frac{N_{F}}{N_{A}}\frac{1}{6}
    \frac{\alpha_{s}}{\pi}\biggl[1
        +\frac{\alpha_{s}}{\pi}\biggl(\frac{C_{A}}{2}-\frac{C_{F}}{4}\biggr)
                +O(\alpha_{s}^2) \biggr],
\label{eq:Cgganalytic}
\end{equation}
\begin{equation}
C_{\overline{q}q}^{f=f'}(Q^2)=\frac{1}{Q^4}
    \biggl\{2+\frac{\alpha_{s}}{\pi}\frac{C_{F}}{2}\biggl[1
        +\frac{\alpha_{s}}{\pi}
    \biggl(\frac{129}{8}C_{F}-\frac{25}{18}C_{A}-\frac{5}{9}TN
        +T(-3+4\zeta(3))\biggr)+O(\alpha_{s}^2)\biggr]\biggr\}
\label{eq:Cqqanalytic1}
\end{equation}
\begin{equation}
C_{\overline{q}q}^{f \neq f'}(Q^2)=\frac{1}{Q^4}
    \biggl(\frac{\alpha_{s}}{\pi}\biggr)^2C_{F}T
    \biggl(-\frac{3}{2}+2\zeta(3)\biggr)+O(\alpha_{s}^3).
\label{eq:Cqqanalytic2}
\end{equation}
The above results are gauge invariant. This statement was
checked by straightforward calculation in an arbitrary
 covariant gauge up to the term
$\sim \varepsilon$ (Surguladze and Tkachov, 1990). The
 dependence on the gauge parameter canceled. Thus, it
 is simplest to do the calculation in the Feynman gauge.
 For simplicity, we have omitted the terms
 $\sim \log(\mu_{\overline{\mbox{\tiny MS}}}^2/Q^2)$, taking
$\mu_{\overline{\mbox{\tiny MS}}}^2=Q^2$. The dependence on
 $\mu$ can be restored via the renormalization
group (see below).
Note that the coefficient function
 $C_{\overline{q}q}^{f \neq f'}$ is due
to the diagrams pictured in Fig.\ 4 with
disconnected fermion lines of the
virtual loop (see also Fig.\ 2).

\vspace{4cm}

\begin{center}
FIG.\ 4. \hspace{2mm} Two-loop diagrams forming
      $C_{\overline{q}q}^{f \neq f'}$
\end{center}

   Specifically for QCD with the SU$_{\mbox{\scriptsize c}}$(3)
symmetry group we obtain
\begin{equation}
C_{G^2}(Q^2)=\frac{1}{Q^4}\frac{1}{12}
    \frac{\alpha_{s}}{\pi}\biggl(1+\frac{\alpha_{s}}{\pi}\frac{7}{6}
                                                +O(\alpha_{s}^2)\biggr),
\label{eq:CgganalyticQCD}
\end{equation}
\begin{equation}
C_{\overline{q}q}^{f=f'}(Q^2)=\frac{1}{Q^4}
    \biggl\{2+\frac{2}{3}\frac{\alpha_{s}}{\pi}\biggl[1
        +\frac{\alpha_{s}}{\pi}
    \biggl(\frac{95}{6}+2\zeta(3)-\frac{5}{18}N\biggr)
                 +O(\alpha_{s}^2)\biggr]\biggr\},
\label{eq:Cqqanalytic1QCD}
\end{equation}
\begin{equation}
C_{\overline{q}q}^{f \neq f'}(Q^2)=\frac{1}{Q^4}
    \biggl(\frac{\alpha_{s}}{\pi}\biggr)^2
    \biggl(-1+\frac{4}{3}\zeta(3)\biggr)+O(\alpha_{s}^3).
\label{eq:Cqqanalytic2QCD}
\end{equation}
Note the very large $O(\alpha_{s}^2)$ coefficient in
eq.\ (\ref{eq:Cqqanalytic1QCD}). However, this coefficient
is renormalization scheme dependent and requires
 special analysis (see below).

   In the scalar and pseudoscalar channels the general
 expression for the coefficient functions eq.\ (\ref{eq:CF})
 takes the form
\begin{equation}
C_{i}(Q)=Z_{m}^2\sum_{j}\pi_{j}
\biggl[\frac{{\cal T}(Q)}{(m_{f}^{\mbox{\tiny B}})^2}\biggr](Z_{O}^{-1})_{ji},
\label{eq:CFscalar}
\end{equation}
where $Z_{m}=m_{f}^{\mbox{\tiny B}}/m_{f}$ is the quark
mass renormalization constant (see eq.\ (\ref{eq:mrenorm}) ).
The $\gamma^{5}$ matrix is defined within the dimensional regularization
according to eq.\ (\ref{eq:gamma5def}).
It is easy to see that in the calculations of $C_{G^2}$, two matrices
$i\gamma^{5}$ can be anticommuted over the fermion propagators and
 ``annihilate'' each other so that the results in both channels
 coincide.
 The calculational procedure is exactly the same as it was for
 the vector channel, except for the need of mass renormalization.
 The results for the coefficient functions $C_{G^2}$ and
$C_{\overline{q}q}^{f}$ in the $\overline{\mbox{MS}}$ -scheme are as follows
(Surguladze and Tkachov, 1986, 1990).

\noindent
In the (pseudo)scalar channel
\begin{equation}
C_{G^2}(Q^2)=\frac{1}{Q^2}C_{F}\frac{N_{F}}{N_{A}}\frac{1}{4}
    \frac{\alpha_{s}}{\pi}\biggl[1
        +\frac{\alpha_{s}}{\pi}\biggl(\frac{3}{2}C_{A}
                       +\frac{3}{4}C_{F}\biggr)
                +O(\alpha_{s}^2) \biggr].
\label{eq:CgganalyticS}
\end{equation}
In the scalar channel
\begin{eqnarray}
\lefteqn{\hspace{-30mm}C_{\overline{q}q}^{f=f'}(Q^2)=
 \frac{1}{Q^2}\biggl\{3+\frac{\alpha_{s}}{\pi}\frac{39}{4}C_{F}\biggl(1
        +\frac{\alpha_{s}}{\pi}
    \biggl[C_{F}\biggl(\frac{447}{208}-\frac{21}{13}\zeta(3)\biggr)
          +C_{A}\biggl(\frac{389}{144}+\frac{3}{26}\zeta(3)\biggr)} \nonumber\\
 && \quad \hspace{35mm} -\frac{5}{39}T-\frac{25}{36}TN\biggr]
                               +O(\alpha_{s}^2)\biggr)\biggr\},
\label{eq:Cqqanalytic1S}
\end{eqnarray}
\begin{equation}
C_{\overline{q}q}^{f \neq f'}(Q^2)=\frac{1}{Q^2}
    \biggl(\frac{\alpha_{s}}{\pi}\biggr)^2C_{F}T
    \biggl(-\frac{5}{4}\biggr)+O(\alpha_{s}^3)
\label{eq:Cqqanalytic2S}
\end{equation}
and in the pseudoscalar channel
\begin{eqnarray}
\lefteqn{\hspace{-30mm}C_{\overline{q}q}^{f=f'}(Q^2)=
 -\frac{1}{Q^2}\biggl\{1+\frac{\alpha_{s}}{\pi}\frac{17}{4}C_{F}\biggl(1
        +\frac{\alpha_{s}}{\pi}
    \biggl[C_{F}\biggl(\frac{583}{272}-\frac{45}{17}\zeta(3)\biggr)
     +C_{A}\biggl(\frac{2443}{816}+\frac{27}{34}\zeta(3)\biggr)} \nonumber\\
 && \quad \hspace{35mm} +\frac{5}{17}T-\frac{167}{204}TN\biggr]
                               +O(\alpha_{s}^2)\biggr)\biggr\}.
\label{eq:Cqqanalytic1PS}
\end{eqnarray}
The result for $C_{\overline{q}q}^{f \neq f'}$ coincides with the
 analogous one for the scalar channel.

  Let us turn to the renormalization group analysis of the
 above results.
 in this particular case it is possible
 to use the following trick (Surguladze and Tkachov, 1990).
Note first that the vacuum average of the renormalized operators $G^2$
 and $m\overline{q}{q}$ and their coefficient functions
 depend on the
renormalization parameter $\mu$ and therefore are not convenient
 for further analysis. However,
as was shown by Collins, Duncan and Joglekar (1977)
(see also Nielsen, 1977; Tarrach, 1982; Narison and Tarrach, 1983),
 the vacuum average of the trace of the energy-momentum tensor
\begin{equation}
<\Theta_{\alpha\alpha}>_{0}=-\frac{\beta(\alpha_s)}{2\beta_{0}}
                       <(G_{\mu\nu}^{a})^2>_{0}
    +\biggl(1-\frac{2\gamma_{m}(\alpha_s)}{\beta_{0}}\biggr)
         \sum_{f}<m_{f}\overline{q}_{f}q_{f}>_{0}
\label{eq:energymomentum}
\end{equation}
is renormalization group invariant. On the other hand, in
 the MS type schemes
the quark condensate $<m_{f}\overline{q}_{f}q_{f}>_{0}$ is
 renormalization group invariant to all orders of perturbation
 theory  (see, e.g., Tarrach, 1982). One can introduce the
 renormalization group invariant quantity
\begin{equation}
\Omega = -\frac{\beta(\alpha_s)}{\beta_{0}}<(G_{\mu\nu}^{a})^2>_{0}
        -\frac{4\gamma_{m}(\alpha_s)}{\beta_{0}}
          \sum_{f}<m_{f}\overline{q}_{f}q_{f}>_{0}
\label{eq:omega}
\end{equation}
so that the new coefficient functions defined from equation
\begin{displaymath}
\hspace{-45mm}
C_{G^2}\biggl(\frac{\mu^2}{Q^2},\alpha_s\biggr)<(G_{\mu\nu}^{a})^2>_{0}
+C_{\overline{q}q}^{f}\biggl(\frac{\mu^2}{Q^2},\alpha_s\biggr)
\sum_{f}<m_{f}\overline{q}_{f}q_{f}>_{0}
\end{displaymath}
\begin{equation}
\hspace{4cm} =\overline{C}_{G^2}\biggl(\frac{\mu^2}{Q^2},\alpha_s\biggr)\Omega
+\overline{C}_{\overline{q}q}^{f}\biggl(\frac{\mu^2}{Q^2},\alpha_s\biggr)
\sum_{f}<m_{f}\overline{q}_{f}q_{f}>_{0}
\label{eq:CFequation}
\end{equation}
should be the renormalization group invariants. This is true
 since the l.h.s of eq.\ (\ref{eq:CFequation}) is directly
 connected to the observables (Shifman, Vainshtein and
 Zakharov, 1979) and consequently is invariant.
{}From eqs.\ (\ref{eq:omega}) and (\ref{eq:CFequation})
 we find the invariant
coefficient functions corresponding to the invariant
 combinations of the gluon and quark condensates
\begin{equation}
\overline{C}_{G^2}\biggl(\frac{\mu^2}{Q^2},\alpha_s\biggr)
  =-\frac{\beta_{0}}{\beta(\alpha_s)}
              C_{G^2}\biggl(\frac{\mu^2}{Q^2},\alpha_s\biggr),
\label{eq:CFggnew}
\end{equation}
\begin{equation}
\overline{C}_{\overline{q}q}^{f}\biggl(\frac{\mu^2}{Q^2},\alpha_s\biggr)
 =C_{\overline{q}q}^{f}\biggl(\frac{\mu^2}{Q^2},\alpha_s\biggr)
 -\frac{4\gamma_{m}(\alpha_s)}{\beta(\alpha_s)}
   C_{G^2}\biggl(\frac{\mu^2}{Q^2},\alpha_s\biggr).
\label{eq:CFqqnew}
\end{equation}

Note that, in fact, there are terms of the type
 $m_{f}^2m_{f'}^2$ or/and
$m_{f}^4$ in the r.h.s. of eq.\ (\ref{eq:CFequation}). However,
 obviously these terms do not affect our equations for
invariant coefficient functions. The two-loop coefficient
functions for $\sim m^4$ terms have been calculated by Chetyrkin,
 Gorishny and Spiridonov (1985). The contributions from
such terms are negligible for phenomenological
 applications and will not be discussed here.

Now one can use the renormalization group invariance
 of the coefficient functions and write
\begin{equation}
\overline{C}_{i}\biggl(\frac{\mu^2}{Q^2},\alpha_s\biggr)
  =\overline{C}_{i}(1,\alpha_s(Q^2)).
\label{eq:CFRGnew}
\end{equation}

   Reevaluating the coefficient functions for the $u,d,s$
light quarks
($N=3$) we obtain the following results in the
 $\overline{\mbox{MS}}$ scheme.

\noindent
In the vector channel
\begin{equation}
\overline{C}_{G^2}(\alpha_s(Q^2))
   =\frac{1}{Q^4}\frac{1}{12}\biggl(1-\frac{\alpha_s(Q^2)}{\pi}
       0.6111+O(\alpha_{s}^{2})\biggr),
\label{eq:CggnewV}
\end{equation}
\begin{equation}
\overline{C}_{\overline{q}q}^{f=f'}(\alpha_s(Q^2))
   =\frac{1}{Q^4}2\biggl[1+0.4074\frac{\alpha_s(Q^2)}{\pi}
     \biggl(1+\frac{\alpha_s(Q^2)}{\pi}14.8180+O(\alpha_{s}^{2})\biggr)
          \biggr].
\label{eq:CqqnewV}
\end{equation}
In the scalar channel
\begin{equation}
\overline{C}_{G^2}(\alpha_s(Q^2))
   =\frac{1}{Q^2}\frac{1}{8}\biggl(1+\frac{\alpha_s(Q^2)}{\pi}
       3.7222+O(\alpha_{s}^{2})\biggr),
\label{eq:CggnewS}
\end{equation}
\begin{equation}
\overline{C}_{\overline{q}q}^{f=f'}(\alpha_s(Q^2))
   =\frac{1}{Q^2}3\biggl[1+4.4074\frac{\alpha_s(Q^2)}{\pi}
     \biggl(1+\frac{\alpha_s(Q^2)}{\pi}7.6879+O(\alpha_{s}^{2})\biggr)
          \biggr].
\label{eq:CqqnewS}
\end{equation}
   In the pseudoscalar channel
\begin{equation}
\overline{C}_{\overline{q}q}^{f=f'}(\alpha_s(Q^2))
   =-\frac{1}{Q^2}\biggl[1+5.4444\frac{\alpha_s(Q^2)}{\pi}
     \biggl(1+\frac{\alpha_s(Q^2)}{\pi}9.4559+O(\alpha_{s}^{2})\biggr)
          \biggr].
\label{eq:CqqnewPS}
\end{equation}
For all channels
\begin{equation}
\overline{C}_{\overline{q}q}^{f \neq f'}(\alpha_s(Q^2))
=C_{\overline{q}q}^{f \neq f'}(\alpha_s(Q^2))
+O(\alpha_{s}^3).
\label{eq:Cqqffnew}
\end{equation}
 Note again very large $O(\alpha_{s}^2)$ corrections
 for the coefficient functions of quark condensates in
the $\overline{\mbox{MS}}$-scheme. The running
coupling is evaluated at the typical hadronic mass scale.
Presently the $O(\alpha_{s})$ corrections have also been calculated
 for the $\mbox{dim}=6$ operators
 (Lanin, Spiridonov, and Chetyrkin (1986)). We also mention
the calculations in the case of heavy quark currents
(see, e.g., Broadhurst et al, 1994 and references therein).

\vspace{2cm}

\renewcommand{\thesection}{\Roman{section}}
\section{\bf $R(s)$ in electron-positron annihilation to
                                         $O(\alpha^{3}_{s})$}
\renewcommand{\thesection}{\arabic{section}}
\setcounter{equation}{0}

In this section we present an outline of
the evaluation of the corrections up to $O(\alpha^{3}_{s})$
to the total cross section in the process
$e^{+}e^{-}\rightarrow \mbox{hadrons}$ (Fig.\ 5) in the limit of zero
light quark masses and infinitely large top mass.
 We also mention the QCD evaluation of
the hadronic decay rates of the Z boson and the
relevant quark mass effects.

\vspace{6cm}

\begin{center}
FIG.\ 5. \hspace{2mm} The process $e^{+}e^{-} \rightarrow \mbox{hadrons}$.
     The shaded bulb includes any interactions of quarks and gluons (or ghosts)
     allowed in QCD. The dots cover any relevant number of gluon and quark
     propagators.
\end{center}

These calculations were first attempted by Gorishny, Kataev and Larin (1988).
However, it was shown that those results were incorrect. Indeed,
about 4 years ago an independent calculation of the above quantity
was completed (Surguladze and Samuel, 1991a,b).
The result is much smaller and has the opposite sign compared with
the old 1988 result.
This finding was confirmed shortly after that
by Gorishny, Kataev and Larin (1991).

In the process shown in Fig.\ 5 an electron-positron pair annihilates
 producing either a photon or a $Z$-boson, which further produces
 quark-antiquark pairs
(in QED) plus gluons (if strong interactions are ``switched on'').
Finally, quarks through hadronization form hadronic final
 states with probability equal to one (confinement hypothesis)
 and the total cross-section is given by
\begin{equation}
\sigma_{\mbox{\scriptsize{tot}}}(e^{+}e^{-}\rightarrow
\mbox{hadrons})=\frac{4\pi\alpha^2}{3s}
    3\sum_{f}Q_{f}^2(1+\delta_{\mbox{\tiny QCD}}),
\label{eq:sigmatot}
\end{equation}
where
 $s$ is the total centre-of-mass energy squared, $Q_f$ is the
 electric charge of the participating at the given energy
 quark flavor $f$, factor 3 stands for the number of color degrees of freedom
 and
$\delta_{\mbox{\tiny QCD}}$ stands for the strong interaction contributions.
 The hadronic production in electron-positron annihilation
 is usually characterized
 in terms of the $R$-ratio
-the total hadronic cross section normalized by the
 muon pair production  cross section
\begin{equation}
R(s) = \frac{\sigma_{\mbox{\scriptsize{tot}}}(e^{+}e^{-}\rightarrow
\mbox{hadrons})}
   {\sigma(e^{+}e^{-}\rightarrow \mu^{+}\mu^{-})}
   = 3\sum_{f}Q_{f}^2(1+\delta_{\mbox{\tiny{QCD}}}).
\label{eq:Rratio}
\end{equation}
The above expressions are relevant at energies much less than the Z mass
($\sqrt{s} \ll M_Z$) corresponding to, for instance,
PEP/PETRA energy range. At LEP the effects of the Z boson become important.
 The corresponding $R$-ratio is defined as a ratio of the hadronic and
 electronic widths of the Z boson
\begin{equation}
R_{Z} = \frac{\Gamma(Z \rightarrow \mbox{hadrons})}
   {\Gamma(Z \rightarrow e^{+}e^{-})}.
\label{eq:Rzratio}
\end{equation}
Note that the total hadronic width of the Z boson in the above equation
is the sum of the vector and axial current induced decay rates.
Strictly speaking, those rates get different strong interaction
contributions. In the present section we calculate the QCD corrections
in the vector channel - $\delta_{\mbox{\tiny QCD}}$ in the limit of massless
light quarks
and the infinitely large top mass. This quantity is, in fact, relevant
for the axial part as well. To get the complete axial decay rate,
additional contributions are necessary. For details see the original
works:
Kniehl and K\"{u}hn (1990),
Kniehl (1990),
Chetyrkin and K\"{u}hn (1990),
Chetyrkin, K\"{u}hn and Kwiatkowski (1992),
Chetyrkin (1993a),
Soper and Surguladze (1994),
Surguladze (1994c), the review articles by Kniehl (1994b, 1995b),
Soper and Surguladze (1995) and
also section 4 of the present paper.

\renewcommand{\thesection}{\arabic{section}}
\subsection{\tenbf $R(s)$ via renormalization constants}

The vacuum polarization function $\Pi(Q^2)$ defined in
eq.\ (\ref{eq:pifunction}) has a cut along the negative $Q^2$ axis
in the massless case. The ratio $R(s)$ can be found taking the imaginary
part of $\Pi(s+i0)$, according to eq.\ (\ref{eq:Rsdefin}). Alternatively,
$R(s)$ can also be found from eq.\ (\ref{eq:Ddefin}), which in combination
with eq.\ (\ref{eq:RGDsolut}) gives to $O(\alpha_s^3)$
\begin{equation}
R(s)=R_0+\frac{\alpha_s(s)}{\pi}R_1
+\biggl(\frac{\alpha_s(s)}{\pi}\biggr)^2 R_2
+\biggl(\frac{\alpha_s(s)}{\pi}\biggr)^3
     \biggl(R_3-\frac{\pi^2\beta_0^2}{3}R_{1}\biggr).
\label{eq:Rexpans}
\end{equation}
The origin of the large and negative scheme-scale independent term
$R_{1}\pi^2\beta_0^2/3$ can be understood if one takes into account
the presence of $\sim \log^3\mu^2/s$ terms at $O(\alpha_s^3)$
in the $\Pi$-function and
\begin{displaymath}
\frac{1}{\pi}\mbox{Im}\log^3(s+i0)=-3\log^2 s+\pi^2.
\end{displaymath}
The leading QCD term $R_1$ at $O(\alpha_s^3)$ is due to the
coupling renormalization. Note, that the $R_i$ in the above
equation are the perturbative coefficients of the $D(Q^2)$ function
defined in eq.\ (\ref{eq:Ddefin}).
For the discussion of the procedure of analytical continuation
and the origin of additional $\sim \pi^2$ terms, see also
Krasnikov and Pivovarov (1982), Pennington and Ross (1982),
Radyushkin (1982), and Pivovarov (1992a).

    Substituting eq.\ (\ref{eq:Piexpans}) with the renormalized
strong coupling into eq.\ (\ref{eq:Rsdefin}) and taking into account
the relations (\ref{eq:Pirelat0}), (\ref{eq:Pirelat1}), (\ref{eq:Pirelat2}),
(\ref{eq:Pirelat3}) we obtain
\begin{eqnarray}
\lefteqn{\hspace{-9mm}R(s) = -\frac{3}{4}\biggl\{
   Z_{1,-1}+\frac{\alpha_s(\mu)}{\pi}(2Z_{2,-1})
     +\left(\frac{\alpha_s(\mu)}{\pi}\right)^2
         \biggl(3Z_{3,-1}-\beta_0\Pi_{2,0}
                 +2\beta_0Z_{2,-1}\log\frac{\mu^2}{s}\biggr) }
                                                             \nonumber\\
 && \quad \hspace{-7mm}
      +\left(\frac{\alpha_s(\mu)}{\pi}\right)^3
          \biggl[4Z_{4,-1}-2\beta_0\Pi_{3,0}-\beta_1\Pi_{2,0}
                +2\beta_0^2\Pi_{2,1}-\frac{2\pi^2\beta_0^2}{3}Z_{2,-1}
                                                               \nonumber\\
 && \quad
   +(6\beta_0Z_{3,-1}+2\beta_1Z_{2,-1}-2\beta_0^2\Pi_{2,0})\log\frac{\mu^2}{s}
   +2\beta_0^2Z_{2,-1}\log^2\frac{\mu^2}{s}\biggr]+O(\alpha_s^4)\biggr\}.
\label{eq:Rmain}
\end{eqnarray}
Note that the appearance of perturbative coefficients of the renormalization
constant in the above equation is totally due to the relations
(\ref{eq:Pirelat3}). In fact, $Z_{\Pi}$ has only simple poles
and hence has no imaginary part.
The latter is the specific feature of the MS prescription.
The expression (\ref{eq:Rmain}) exhibits one of the main ideas of this
calculation. Namely, in order to calculate the
$l$-loop contribution to $R$, it suffices to calculate the $l$-loop counterterm
$Z_{\Pi}$ to the bare quantity $\Pi^{\mbox{\scriptsize B}}$,
 and the $l-1$ -loop approximation to $\Pi^{\mbox{\scriptsize B}}$.
In other words, the minimal information necessary to obtain
the four-loop $R(s)$ is contained in the divergent part of one-loop diagram,
two-loop diagrams calculated up to $\sim \varepsilon$ terms, three-loop
diagrams calculated up to the finite parts in the limit
$\varepsilon \rightarrow 0$ and only a leading $\sim 1/\varepsilon$ terms
in the overall counterterms of the four-loop diagrams.
In fact, as we demonstrate
in the next subsection, using the infrared rearrangement procedure
(Vladimirov, 1980; Chetyrkin and Tkachov, 1982)
one can complete the entire calculation dealing effectively only with
three-loop diagrams. We mention once again that,
through the procedure of infrared rearangement, within the MS prescription,
the problem of calculation of the counterterms to
arbitrary $l$ -loop diagrams with an arbitrary number of
masses and external momenta
can be reduced to the
calculation of $l-1$ -loop propagator type massless integrals up
to finite terms. In our case $l=4$. On the other hand,
the recursive type algorithms for multiloop Feynman integrals
(Chetyrkin and Tkachov, 1981; Tkachov, 1981, 1983) and their computer
implementation (Surguladze and Tkachov, 1989; Surguladze, 1989b,c, 1992;
Gorishny, Larin, Surguladze and Tkachov, 1989) allow one to calculate
propagator type Feynman diagrams to three-loop level.

\renewcommand{\thesection}{\arabic{section}}
\subsection{\tenbf Full calculational procedure with a typical four-loop
 diagram}

    In this subsection we demonstrate the full calculational procedure for a
typical
four-loop diagram pictured in Fig.\ 6, which contributes to the photon
renormalization constant $Z_{\Pi}$ and hence to the $R$-ratio.
To simplify the description, in some cases we will
avoid complicated equations, substituting for them their graphical
representation.
\newpage

\mbox{}

\vspace{33mm}

\begin{center}
FIG.\ 6. \hspace{2mm} A typical four-loop nonplanar type
                         diagram contributing to $R(s)$
\end{center}

We need to evaluate the counterterm to the diagram pictured in Fig.\ 6.
In other words, we should evaluate  $-{\cal K}R'$ for this diagram.
A simple power counting shows that the given diagram diverges as
\begin{displaymath}
G\sim\lim_{Q \rightarrow \infty}Q^{4D-14}
\end{displaymath}
and the superficial degree of divergence is 2. Using the fact that
the counterterm has only a polynomial dependence on the external momenta $Q$
within the MS prescription, one can remove such a dependence
by differentiating the diagram twice with respect to $Q$ and then set the
external momentum to zero. At the next step, since there is no
dependence on the external momentum, one can introduce a new fictitious
external momentum flowing through one of the diagram lines. This line should
be chosen in a way that simplifies the topology of the
diagram and avoids infrared divergences. The above procedure for the
diagram in Fig.\ 6 is displayed in the following graphical equation
\vspace{1cm}
\begin{displaymath}
Z\supset{\cal K}R'\biggl\{
   \biggl(\frac{\partial}{\partial Q_{\mu}}\biggr)^2
 \hspace{43mm} \biggr\}_{Q=0}
\end{displaymath}
\vspace{2cm}
\begin{equation}
\sim{\cal K}R'\biggl\{4(2-D)\hspace{37mm}+2\hspace{37mm}\biggr\}
\label{eq:IR}
\end{equation}
\vspace{2cm}
\begin{displaymath}
=\hspace{5mm}{\cal K}\biggl\{4(2-D)\hspace{36mm}+2\hspace{36mm}\biggr\},
\end{displaymath}

\vspace{2cm}

\noindent
where the dot and dashes on the lines result from differentiating the
corresponding fermion propagators
\begin{displaymath}
 \biggl(\frac{\partial}{\partial Q_{\mu}}\biggr)^2
    \biggl[-\hspace{-2mm}-\hspace{-2mm}
   \stackrel{P+Q}{\longleftarrow}\hspace{-2mm}-\hspace{-2mm}-
           \biggr]_{Q=0} \equiv 2(2-D)
          \biggl[\longleftarrow \hspace{-2mm}
                 \stackrel{P}{\bullet}\hspace{-2mm}\longleftarrow
                          \biggr]=2(2-D)\frac{\hat{P}}{P^4},
\end{displaymath}
\begin{displaymath}
   \frac{\partial}{\partial Q_{\mu}}
        \biggl[-\hspace{-2mm}-\hspace{-2mm}
   \stackrel{P+Q}{\longleftarrow}\hspace{-2mm}-\hspace{-2mm}-
           \biggr]_{Q=0}\equiv
 \biggl[\longleftarrow \hspace{-2mm}
          \hspace{-2mm}\stackrel{P}{\setminus}\longleftarrow\biggr]
                     =-\frac{\hat{P}}{P^2}\gamma^{\mu}\frac{\hat{P}}{P^2}
\end{displaymath}
Boxes contain the corresponding three-loop propagator type subgraphs
with subtracted divergences - complete $R$-operation (Fig.\ 7).
The dotted lines mean that this line is temporarily ``torn''.
 After the evaluation of boxes,
 the parts of the torn line should be pasted and a trivial fourth loop
 integration should be done, taking into account the corresponding
 exponents of the propagators due to the three-, two- and one-loop
 ``box'' insertions.
The above procedure gives a great simplification of the problem.
Indeed, the evaluation of the four-loop counterterm is reduced to the
evaluation of three-, two- and one-loop graphs.

\vspace{15mm}
\begin{displaymath}
-\fbox{{\bf A}}-\equiv R \biggl\{ \hspace{19mm} \biggr\} =
\hspace{19mm}-\biggl(\hspace{17mm}\biggr)\hspace{19mm}
            -\biggl(\hspace{17mm}\biggr) \hspace{17mm}
\end{displaymath}
\vspace{2cm}
\begin{displaymath}
\hspace{-43mm}-\fbox{{\bf B}}-\equiv R \biggl\{ \hspace{19mm} \biggr\} =
  \hspace{19mm}-\biggl(\hspace{17mm}\biggr)\hspace{19mm}
\end{displaymath}

\vspace{1cm}

\begin{center}
FIG.\ 7. \hspace{2mm}  Complete $R$-operation for the three-loop
                                          subgraphs
\end{center}

The complete $R$-operation of the three-loop diagram insertions corresponding
to the ones at the r.h.s. of eq.\ (\ref{eq:IR}) is given in Fig.\ 7.
Graphs in the brackets correspond to two- and three-loop
counterterms. There is no one-loop divergent subgraph in
this particular diagram. Thus,
\begin{displaymath}
\biggl( G_{i} \biggr) \equiv {\cal K}R' \biggl\{ G_{i} \biggr\},
\end{displaymath}
where $G_{i}$ is any divergent subgraph of the given diagram.
It is easy to recognize that the two-loop subgraph in Fig.\ 7 does not have
subdivergences (only an overall one) and the corresponding counterterm
is simply the pole part of this subgraph

\vspace{7mm}

\begin{displaymath}
{\cal K}R' \biggl\{ \hspace{3cm} \biggr\} =
                          {\cal K}\biggl\{ \hspace{3cm} \biggr\}.
\end{displaymath}

\vspace{7mm}

Analogously,  because of the topology, the three-loop counterterm
does not have a subdivergence and the
corresponding counterterm is the pole part of this diagram

\vspace{7mm}

\begin{displaymath}
{\cal K}R' \biggl\{ \hspace{3cm} \biggr\} =
                          {\cal K}\biggl\{ \hspace{3cm} \biggr\}.
\end{displaymath}

\vspace{7mm}

If, in general, a diagram contains divergent subgraphs,
 then the recursive formula (\ref{eq:KR'}) should be used.

   As a result of the above manipulations, we managed to reduce the
problem of calculation of the counterterm to the four-loop diagram
pictured in Fig.\ 6 to the calculation of several three-,
two- and one-loop diagrams shown in Fig.\ 7. Note, however that
the ``dots'' and ``dashes'' on the diagram lines make their
evaluation significantly more difficult. The computer programs for
analytical programming systems capable of handling such calculations are
the {\small SCHOONSCHIP} program {\small MINCER}
(Gorishny, Larin, Surguladze and Tkachov, 1989; Surguladze, 1989b,c)
and the {\small FORM} program {\small HEPL}oops (Surguladze, 1992).
The latter is especially
well-suited for large scale calculations and is much more efficient
than the {\small MINCER} program.

    It is important to stress that, in fact, it is sufficient to evaluate
only the ${\cal K}R'$ for the relevant three-loop subgraphs.
 In  other words, it is not necessary to calculate separately
 three-loop counterterms
similar to the graph in the last brackets for the box A in Fig.\ 7.
Indeed, a more detailed analysis gives
\begin{equation}
R[G]=R'[G]-(1-D/2){\cal K}\left(\frac{1}{1-D/2}R'[G]\right),
\label{eq:CTrelation}
\end{equation}
where G is the corresponding three-loop subgraph. The above relation allows
simple computer implementation and facilitates calculations considerably.

  The complete $R$-operation for each four-loop diagram  generally
has the form
\begin{displaymath}
\left(\frac{\mu^2}{Q^2}\right)^{4\varepsilon}f_4(\varepsilon)
-\sum_{l=1}^{3}\left(\frac{\mu^2}{Q^2}\right)^{(4-l)\varepsilon}
c_l(1/\varepsilon)f_{4-l}(\varepsilon),
\end{displaymath}
where $f_i(\varepsilon)$ is the result of the calculation of the corresponding
Feynman graphs including the last trivial loop integration and $c_l$ are
the $l$-loop counterterms, polynomials in $1/\varepsilon$.
As we already mentioned, in the MS type renormalization scheme, the
counterterm for a particular diagram is a polynomial in dimensional
parameters (see, e.g., the textbook by Collins, 1984 and references therein).
Thus, the terms of the
type $(1/\varepsilon)^n\ln^m(\mu^2/Q^2)$, which appear due to the expansion
of the factors $(\mu^2/Q^2)^{l\varepsilon}$ into the Laurent series in
$\varepsilon$, must be canceled in the final answer for the particular
diagram.  This can be used to test the calculations at the graph-by-graph
level. Recall, that we calculate the counterterm $Z_{4,i}$
to the four-loop diagram.

    Finally, for the contribution to the $Z_{\Pi}$ of the diagram pictured in
Fig.\ 6 we obtain the following result
\begin{displaymath}
\left(\frac{\alpha_s}{4\pi}\right)^3\hspace{2mm}N_FC_F(C_F-C_A)(C_F-C_A/2)
\hspace{2mm} \frac{1}{3-2\varepsilon}
\hspace{2mm}
\biggl[4\frac{1}{\varepsilon^3}-26\frac{1}{\varepsilon^2}
+\frac{65}{4}\frac{1}{\varepsilon}-40\zeta(3)\frac{1}{\varepsilon}\biggr].
\end{displaymath}
The {\small CPU} time for the above diagram on a 0.8 MFlop {\small IBM}
compatible
mainframe was over 6 hours. The extended version of the program
{\small MINCER} for the system {\small SCHOONSCHIP} was used.
 Note that the above result,
 as well as the total result for the photon renormalization constant
 does not depend on any modification of the
 minimal subtraction prescription.

\renewcommand{\thesection}{\arabic{section}}
\subsection{\tenbf Four-loop results}

    In this subsection, we present results and some of the details
of the $O(\alpha_s^3)$ QCD evaluation of the ratio $R(s)$ in
electron-positron annihilation (Surguladze and Samuel, 1991a,b).

    The total number of topologically distinct Feynman diagrams contributing
to $Z_{1,i}$ is 1, to $Z_{2,i}$ is 2, to $Z_{3,i}$ is 17 and to
$Z_{4,i}$ is 98. However, after application of the infrared rearrangement
procedure which,
as discussed above, involves differentiation twice with respect to the external
momentum of the diagram, the number of four-loop graphs which need to be
calculated increases to approximately 250. Furthermore, there are one-,
two- and three-loop diagrams, approximately 600, which need to be
calculated to subtract subdivergences (evaluate $R'$) for all
four-loop diagrams.

   All analytical calculations of the four-loop diagrams have been done by
using the program, which is an extended version (Surguladze, 1989c) of
the program {\small MINCER} (Gorishny, Larin, Surguladze and Tkachov, 1989;
Surguladze, 1989b). This  version
includes new subprograms for 4th loop integration and
for ultraviolet renormalization.
Evaluation of one- and two-loop counterterms has
been done by using the  program LOOPS
(Surguladze and Tkachov, 1989a). The above programs
are written on the algebraic programming systems {\small SCHOONSCHIP}
(Veltman, 1967; Strubbe, 1974)
and {\small REDUCE} (Hearn, 1973) respectively. The  full calculation
took over 700 hours of {\small CPU} time on three {\small IBM} compatible
0.8 MFlop EC-1037 mainframes with
the {\small SCHOONSCHIP} system.
We have also recalculated some of the difficult
four-loop diagrams with {\small HEPL}oops - a new program for
analytical multiloop calculations (Surguladze, 1992).
The status of these and some other programs has been reviewed recently
in Surguladze (1994d).

   In the diagram calculations
the Feynman gauge is used. The momentum integrations are performed
within the $\overline{\mbox{MS}}$ modification (Bardeen, Buras, Duke and Muta,
1978)
of the minimal subtraction prescription ('t Hooft, 1973),
which amounts to formally
setting $\gamma=\zeta(2)=\log4\pi=0$. A discussion of the scheme dependence
of the results is given at the end of this section and in section 9.
The full graph-by-graph results will be published elsewhere.

  The analytical result for the four-loop photon renormalization constant
reads
\begin{eqnarray}
\lefteqn{Z_{\mbox{\scriptsize ph}}
              \equiv 1+\frac{\alpha}{4\pi}Z_{\Pi}=} \nonumber\\
  &&  1 +N_F\frac{\alpha}{4\pi}
        \sum_{f}Q_{f}^2\biggl\{ -\frac{4}{3}\frac{1}{\varepsilon}
   +\frac{\alpha_s}{4\pi}\biggl[\frac{1}{\varepsilon}(-2C_F)\biggr] \nonumber\\
  && \quad
   +\biggl(\frac{\alpha_s}{4\pi}\biggr)^2\biggl[\frac{1}{\varepsilon^2}
   \biggl(\frac{22}{9}C_FC_A-\frac{8}{9}NTC_F\biggr)
   +\frac{1}{\varepsilon}\biggl(\frac{2}{3}C_F^2-\frac{133}{27}C_FC_A
                      +\frac{44}{27}NTC_F\biggr)\biggr] \nonumber\\
  && \quad  +\biggl(\frac{\alpha_s}{4\pi}\biggr)^3
                \biggl[\frac{1}{\varepsilon^3}
    \biggl(-\frac{121}{27}C_FC_A^2+\frac{88}{27}NTC_FC_A
                         -\frac{16}{27}N^2T^2C_F\biggr) \nonumber\\
  && \quad
      +\frac{1}{\varepsilon^2}\biggl(-\frac{11}{9}C_F^2C_A+\frac{2381}{162}
         C_FC_A^2-\frac{14}{9}NTC_F^2-\frac{778}{81}NTC_FC_A
         +\frac{88}{81}N^2T^2C_F\biggr) \nonumber\\
  && \quad  +\frac{1}{\varepsilon}\biggl(\frac{23}{2}C_F^3
         +\biggl(-\frac{430}{27}+\frac{88}{9}\zeta(3)\biggr)C_F^2C_A
         +\biggl(-\frac{5815}{972}
                -\frac{88}{9}\zeta(3)\biggr)C_FC_A^2 \nonumber\\
  && \quad
           +\biggl(\frac{338}{27}-\frac{176}{9}\zeta(3)\biggr)NTC_F^2
         +\biggl(\frac{769}{243}
          +\frac{176}{9}\zeta(3)\biggr)NTC_FC_A
         +\frac{308}{243}N^2T^2C_F\biggr)\biggr] \nonumber\\
  && \quad   + O(\alpha_s^4) \biggr\}
         +\frac{\alpha}{4\pi}
          \biggl(\frac{\alpha_s}{4\pi}\biggr)^3\biggl(\sum_{f}Q_f\biggr)^2
          \biggl(\frac{d^{abc}}{4}\biggr)^2\biggl(-\frac{176}{9}
            +\frac{128}{3}\zeta(3)\biggr)\frac{1}{\varepsilon}.
\label{eq:Zanalres}
\end{eqnarray}
It should be stressed that the Riemann  $\zeta$-functions $\zeta(4)$ and
$\zeta(5)$, which appear at the individual graph level cancel in the
above expression. Moreover, as we have observed, the $\zeta(4)$ has
disappeared within each gauge invariant set of diagrams. Note
that $\zeta(3)$ disappears
for QED ($C_F=1, C_A=0, T=1$) except the last term, which comes
from the ``light-by-light'' type diagrams (Fig.\ 8).
The diagrams pictured in Fig.\ 8 are some of the most complicated ones
and the computation of each of them requires over 80h of {\small CPU} time.
Note, however, that the second and fourth diagrams in Fig.\ 8 differ
correspondingly from the first and third ones only by the SU(N) group weights.
So, in fact, only two of them have been calculated.
The result (\ref{eq:Zanalres}) does not depend on the particular
modification of the minimal subtraction prescription.

\vspace{35mm}

\begin{center}
FIG.\ 8. \hspace{2mm}  ``Light-by-light'' type diagrams
\end{center}

\vspace{5mm}

   In order to evaluate $R(s)$ to $O(\alpha_s^3)$, besides
the four-loop $Z_{\Pi}$ we calculate the unrenormalized
hadronic vacuum polarization function $\Pi^{\mbox{\scriptsize B}}(Q^2)$
to the three-loop level. We get the following analytical result in the
$\overline{\mbox{MS}}$ scheme.
\newpage
\begin{eqnarray}
\lefteqn{\Pi^{\mbox{\scriptsize B}}
  \biggl(\frac{\mu_{\overline{\mbox{\tiny MS}}}^2}{Q^2}
                             ,\alpha_{s}^{\mbox{\tiny B}}\biggr)=}
                                                                  \nonumber\\
 &&  N_F\sum_{f}Q_{f}^2\biggl\{
       \biggl(\frac{\mu_{\overline{\mbox{\tiny MS}}}^2}
                                 {Q^2}\biggr)^\varepsilon
                \biggl[
      \frac{4}{3}\frac{1}{\varepsilon}
                   +\frac{20}{9}+\frac{112}{27}\varepsilon
          +\frac{656}{81}\varepsilon^2
                -\frac{28}{9}\zeta(3)\varepsilon^2       \biggr]   \nonumber\\
 && \quad
       +\biggl(\frac{\alpha_{s}^{\mbox{\tiny B}}}{4\pi}\biggr)
           \biggl(\frac{\mu_{\overline{\mbox{\tiny
MS}}}^2}{Q^2}\biggr)^{2\varepsilon}
              C_{F}\biggl[ 2\frac{1}{\varepsilon}
                                     +\frac{55}{3}-16\zeta(3)
           +\varepsilon \biggl(\frac{1711}{18}-\frac{152}{3}\zeta(3)
                                    -24\zeta(4) \biggr)  \biggr]   \nonumber\\
 && \quad
      +\biggl(\frac{\alpha_{s}^{\mbox{\tiny B}}}
                                               {4\pi}\biggr)^2
 \biggl(\frac{\mu_{\overline{\mbox{\tiny MS}}}^2}{Q^2}\biggr)^{3\varepsilon}
                    \biggl[ C_{F}^{2}\biggl(
                     -\frac{2}{3}\frac{1}{\varepsilon}-\frac{286}{9}
                             -\frac{296}{3}\zeta(3)+160\zeta(5) \biggr)
                                                                   \nonumber\\
 && \quad
   +C_{F}C_{A} \biggl(\frac{44}{9}\frac{1}{\varepsilon^2}
                    +\frac{1948}{27}\frac{1}{\varepsilon}
             -\frac{176}{3}\zeta(3)\frac{1}{\varepsilon}+\frac{50339}{81}
                                 -\frac{3488}{9}\zeta(3)-88\zeta(4)
                   -\frac{80}{3}\zeta(5) \biggr)                   \nonumber\\
 && \quad
    +NTC_F \biggl( -\frac{16}{9}\frac{1}{\varepsilon^2}
                        -\frac{704}{27}\frac{1}{\varepsilon}
                +\frac{64}{3}\zeta(3)\frac{1}{\varepsilon}-\frac{17668}{81}
                     +\frac{1216}{9}\zeta(3)+32\zeta(4) \biggr) \biggr]
                                                                 \biggr\}.
\label{eq:Pi3lres}
\end{eqnarray}
The above result depends on the particular modifications of the
minimal subtraction prescription, unlike the result for the renormalization
constant (\ref{eq:Zanalres}).

    Substituting the expressions for the relevant $Z_{i,j}$ and $\Pi_{i,j}$,
extracted by comparing eqs.\ (\ref{eq:Zanalres}) and (\ref{eq:Pi3lres})
to eqs.\ (\ref{eq:Zexpans}) and (\ref{eq:Piexpans}),
into eq.\ (\ref{eq:Rmain}) and recalling the values for
$\beta_0$ and $\beta_1$ from eq.\ (\ref{eq:beta}) we get the following
$\overline{\mbox{MS}}$ analytical result for $R(s)$ at the four-loop level
\begin{eqnarray}
\lefteqn{R^{\overline{\mbox{\scriptsize MS}}}(s)=} \nonumber\\
  &&  N_F\sum_{f}Q_{f}^2\biggl\{ 1
        +\biggl(\frac{\alpha_s(s)}{4\pi}\biggr)(3C_F)     \nonumber\\
  && \quad     +\biggl(\frac{\alpha_s(s)}{4\pi}\biggr)^2
          \biggl[C_F^2\biggl(-\frac{3}{2}\biggr)
             +C_FC_A\biggl(\frac{123}{2}-44\zeta(3)\biggr)
               +NTC_F(-22+16\zeta(3))\biggr]                    \nonumber\\
  && \quad   +\biggl(\frac{\alpha_s(s)}{4\pi}\biggr)^3
          \biggl[C_F^3\biggl(-\frac{69}{2}\biggr)
                +C_F^2C_A(-127-572\zeta(3)+880\zeta(5))           \nonumber\\
  && \quad \hspace{34mm}
           +C_FC_A^2\biggl(\frac{90445}{54}-\frac{10948}{9}\zeta(3)
                        -\frac{440}{3}\zeta(5)\biggr)             \nonumber\\
  && \quad \hspace{34mm}
           +NTC_F^2(-29+304\zeta(3)-320\zeta(5))                \nonumber\\
  && \quad \hspace{34mm}
           +NTC_FC_A\biggl(-\frac{31040}{27}+\frac{7168}{9}\zeta(3)
                                 +\frac{160}{3}\zeta(5)\biggr)    \nonumber\\
  && \quad \hspace{15mm}
           +N^2T^2C_F\biggl(\frac{4832}{27}
                                 -\frac{1216}{9}\zeta(3)\biggr)
              -\pi^2C_F\biggl(\frac{11}{3}C_A
              -\frac{4}{3}NT\biggr)^2\biggr]+O(\alpha_s^4)\biggr\}
                                                                   \nonumber\\
  && \quad      +\biggl(\frac{\alpha_s(s)}{4\pi}\biggr)^3
            \biggl(\sum_{f}Q_f\biggr)^2\biggl(\frac{d_{abc}}{4}\biggr)^2
         \biggl[\frac{176}{3}-128\zeta(3)\biggr]
                             +O(\alpha_s^4).
\label{eq:Ranalytic0}
\end{eqnarray}
The logarithmic contributions are absorbed in the running coupling by
taking $\mu^2=s$. Those contributions will be presented explicitly in
section 9. Note that $\zeta(5)$ appears in the final result due to
 the contributions from $\Pi_{3,0}$. The last term
$\sim (\sum_{f}Q_f)^2$ comes
 from the so called ``light-by-light'' type diagrams (Fig.\ 8).
For standard QCD with the SU$_{c}$(3) gauge group we obtain
\begin{eqnarray}
\lefteqn{\hspace{-26mm}
    R^{\overline{\mbox{\scriptsize MS}}}(s)=3\sum_{f}Q_{f}^2\biggl\{ 1
        +\frac{\alpha_s(s)}{\pi}
        +\biggl(\frac{\alpha_s(s)}{\pi}\biggr)^2
          \biggl[\frac{365}{24}-11\zeta(3)-N\biggl(\frac{11}{12}
                -\frac{2}{3}\zeta(3)\biggr)\biggr]}
                                                                \nonumber\\
 && \quad
             +\biggl(\frac{\alpha_s(s)}{\pi}\biggr)^3
        \biggl[\frac{87029}{288}-\frac{121}{8}\zeta(2)
           -\frac{1103}{4}\zeta(3)+\frac{275}{6}\zeta(5)
                                                                \nonumber\\
 && \quad \hspace{19mm} +N\biggl(-\frac{7847}{216}+\frac{11}{6}\zeta(2)
             +\frac{262}{9}\zeta(3)-\frac{25}{9}\zeta(5)\biggr)
                                                                 \nonumber\\
 && \quad \hspace{19mm}
        +N^2\biggl(\frac{151}{162}-\frac{1}{18}\zeta(2)
           -\frac{19}{27}\zeta(3)\biggr)\biggr]
                              \biggr\}                          \nonumber\\
 && \quad \hspace{-18mm}
        +\biggl(\sum_{f}Q_f\biggr)^2
        \biggl(\frac{\alpha_s(s)}{\pi}\biggr)^3
             \biggl[\frac{55}{72}-\frac{5}{3}\zeta(3)\biggr]
                    +O(\alpha_s^4).
\label{eq:Ranalytic1}
\end{eqnarray}
Finally, taking into account the values for the relevant Riemann $\zeta$
-functions, $\zeta(2)=\pi^2/6$, $\zeta(3)=1.2020569...$ and
$\zeta(5)=1.0369278...$ we obtain the numerical form
\begin{eqnarray}
\lefteqn{R^{\overline{\mbox{\scriptsize MS}}}(s)
   =3\sum_{f}Q_{f}^2\biggl[1+\frac{\alpha_s(s)}{\pi}
        +\biggl(\frac{\alpha_s(s)}{\pi}\biggr)^2(1.9857-0.1153N)}
                                                              \nonumber\\
 && \quad \hspace{27mm}
        +\biggl(\frac{\alpha_s(s)}{\pi}\biggr)^3
        (-6.6368-1.2001N-0.0052N^2)\biggr]
                                                              \nonumber\\
 && \quad \hspace{8mm}
        -\biggl(\sum_{f}Q_f\biggr)^2
          \biggl(\frac{\alpha_s(s)}{\pi}\biggr)^3 1.2395
          +O(\alpha_s^4).
\label{eq:Rnumerical}
\end{eqnarray}
Note that only 19 four-loop diagrams contribute
to the term $\sim N$ and 2 four-loop diagrams contribute to the term
$\sim N^2$. The most complicated diagrams are pictured in Fig.\ 9.
The {\small CPU} time for each of them was over 100h and
the intermediate expression had as many as $\sim$ $10^5-10^6$ terms.

\vspace{4cm}

\begin{center}
FIG.\ 9. \hspace{2mm} Some of the most complicated diagrams
\end{center}

\vspace{5mm}

It is known, that the perturbative coefficients for $R(s)$ are scheme
dependent. The above result was obtained in the modified minimal subtraction,
the so-called $\overline{\mbox{MS}}$ scheme introduced by Bardeen,
 Buras, Duke and
 Muta (1978). While the scheme-scale dependence problem will be discussed
 in section 9, here we present the results for a couple of other
 versions of the
minimal subtraction scheme. First, we  consider the so called G scheme
 (Chetyrkin and Tkachov, 1979, 1981;
 Chetyrkin, Kataev and Tkachov, 1980),
 which is convenient for practical
 multiloop calculations.
 The G scheme is defined in such a way that
 the trivial one-loop integral in this scheme is
\begin{displaymath}
\mu^{2\varepsilon}\int\frac{d^{4-2\varepsilon}p}{(2\pi)^{4-2\varepsilon}}
     \frac{1}{p^2(p-k)^2}=\frac{1}{(4\pi)^2} \hspace{1mm}
\biggl(\frac{\mu^2}{k^2}\biggr)^{\varepsilon}
 \hspace{1mm} \frac{1}{\varepsilon}.
\end{displaymath}
The result for $R(s)$  in this  scheme is
\begin{eqnarray}
\lefteqn{
  R^{\mbox{\scriptsize G}}(s)=3\sum_{f}Q_{f}^2\biggl[1+\frac{\alpha_s(s)}{\pi}
        +\biggl(\frac{\alpha_s(s)}{\pi}\biggr)^2(-3.514+0.218N)}
                                                              \nonumber\\
 && \quad \hspace{26mm}
        +\biggl(\frac{\alpha_s(s)}{\pi}\biggr)^3
        (-10.980-0.692N+0.029N^2)\biggr]
                                                              \nonumber\\
 && \quad \hspace{7mm}
        -\biggl(\sum_{f}Q_f\biggr)^2
          \biggl(\frac{\alpha_s(s)}{\pi}\biggr)^3 1.240
          +O(\alpha_s^4).
\label{eq:RGnumerical}
\end{eqnarray}
The parametrization of the running coupling in the above equation has the
same form as in eq.\ (\ref{eq:Asparametr}). However, the parameter $\Lambda$
has to be changed to some other parameter $\Lambda_G$.

Finally, in the original MS scheme ('t~Hooft, 1973) we get
\begin{eqnarray}
\lefteqn{
 R^{\mbox{\scriptsize MS}}(s)=3\sum_{f}Q_{f}^2\biggl[1+\frac{\alpha_s(s)}{\pi}
        +\biggl(\frac{\alpha_s(s)}{\pi}\biggr)^2(7.359-0.441N)}
                                                              \nonumber\\
 && \quad \hspace{26mm}
        +\biggl(\frac{\alpha_s(s)}{\pi}\biggr)^3
        (56.026-8.778N+0.176N^2)\biggr]
                                                              \nonumber\\
 && \quad \hspace{7mm}
        -\biggl(\sum_{f}Q_f\biggr)^2
          \biggl(\frac{\alpha_s(s)}{\pi}\biggr)^3 1.240
          +O(\alpha_s^4).
\label{eq:RMSnumerical}
\end{eqnarray}

   As one can see, starting from $O(\alpha_s^2)$ the results heavily
depend on the choice of the particular modifications of the
minimal subtraction scheme. This dependence, called renormalization
group ambiguity of perturbative results is an important problem and
deserves special consideration.
We will return to this issue in section 9.

Concluding this section, we mention once again that the results of the
above described calculation of the four-loop correction
to the $R(s)$ have been published in Surguladze and Samuel (1991a,b)
and independently
\footnote{See however the discussion in the last three paragraphs
of section 3 in the review article by Surguladze (1994d).}
in Gorishny, Kataev and Larin (1991) and hence, most likely, the
above results are reliable. Interesting relations between the radiative
corrections for different observables, found by
Brodsky and Lu (1994, 1995) serve, in particular,
as another confirmation of our results.

\vspace{7mm}

\renewcommand{\thesection}{\Roman{section}}
\section{\bf \hspace{2mm}
   $\Gamma(\tau^{-}\rightarrow \nu_{\tau}+\mbox{hadrons})$
                                      to $O(\alpha^{3}_{s})$}
\renewcommand{\thesection}{\arabic{section}}
\setcounter{equation}{0}

    The other important inclusive process for phenomenology and testing
 the Standard Model is the hadronic decay of the $\tau$  lepton (Fig.\ 10).
For a recent review see, for instance, Pich (1994a). For earlier references
see  Altarelli (1992), Marciano (1992), and Pich (1991).

\vspace{57mm}
\begin{center}
FIG.\ 10. \hspace{2mm} Hadronic decay of the $\tau$-lepton
\end{center}
\vspace{2mm}

   In this section, using our result of  four-loop calculation of the
$\sigma_{\mbox{\scriptsize{tot}}}(e^{+}e^{-} \rightarrow \mbox{hadrons})$
(Surguladze and Samuel, 1991a,b),
we evaluate the hadronic decay rate of the $\tau$ lepton
to $O(\alpha_{s}^3)$ in perturbative QCD (Pich, 1990;
Gorishny, Kataev and Larin, 1991; Samuel and Surguladze, 1991;
see also Braaten, Narison and Pich, 1991; Pich, 1992a,b;
Diberder and Pich, 1992a,b; Pivovarov, 1992b).
We also comment on the status of the nonperturbative corrections
to this quantity.

 We follow the method first suggested by Tsai (1971), Shankar (1977), and
Lam and Yan (1977) for theoretical evaluation of heavy lepton decay rates.
This method has been further developed for the $\tau$ lepton including the
higher order perturbative corrections and involving the operator product
expansion technique (Wilson, 1969) to analyze the nonperturbative
contributions (Schilcher and Tran, 1984; Braaten, 1988; Narison and Pich,
1988).
As was shown in the above works, combining the operator product expansion
technique and analyticity properties of the correlation function of quark
currents, the ratio
\begin{equation}
R_{\tau}=\frac{\Gamma(\tau^{-}\rightarrow\nu_{\tau}+\mbox{hadrons})}
         {\Gamma(\tau^{-}\rightarrow\nu_{\tau}e^{-}\overline{\nu}_{e})}
\label{eq:Rtaudef}
\end{equation}
is calculable in perturbative QCD. Strictly speaking, besides the QCD
perturbative parts the nonperturbative and weak contributions should be
included to estimate $R_{\tau}$. There are instanton
contributions as well. However, it was shown recently by
 Nason and Porrati (1993) (see also Kartvelishvili
 and Margvelashvili, 1995) that
these contributions are completely negligible due to the chiral
suppression factor $m_{u}m_{d}m_{s}/M_{\tau}^2$. The
$R_{\tau}$ can be written as the following sum
\begin{equation}
R_{\tau}=R_{\tau}^{\mbox{\scriptsize{pert}}}
    +R_{\tau}^{\mbox{\scriptsize{nonpert}}}
    +R_{\tau}^{\mbox{\scriptsize{weak}}}.
\label{eq:Rtausum}
\end{equation}

\renewcommand{\thesection}{\arabic{section}}
\subsection{\tenbf Perturbative QCD contributions}

The quantity $R_{\tau}^{\mbox{\scriptsize{pert}}}$
can be expressed as the following
integral over the invariant
mass of the hadronic decay products of the $\tau$ lepton
(Lam and Yan, 1977; Braaten, 1988)
\begin{equation}
R_{\tau}^{\mbox{\scriptsize{pert}}}
=\frac{3}{4\pi}\int_{0}^{M_{\tau}^2}\frac{ds}{M_{\tau}^2}
\biggl(1-\frac{s}{M_{\tau}^2}\biggr)^2
\biggl[\biggl(1+2\frac{s}{M_{\tau}^2}\biggr)\mbox{Im}\Pi^{T}(s+i0)
            +\mbox{Im}\Pi^{L}(s+i0)\biggr],
\label{eq:IntM}
\end{equation}
where $M_{\tau}$ is the mass of the $\tau$ lepton. The functions
$\Pi^T$ and $\Pi^L$ are the transverse and longitudinal parts of the
correlation function of weak currents of quarks coupled to W boson.
In fact, $\Pi^{T,L}$ are the appropriate combinations of vector and
axial parts corresponding to the
vector and axial currents of u, d, s light quarks (for details
see, e.g., Pich, 1994a). The expression for
 $R_{\tau}^{\mbox{\scriptsize{pert}}}$ in the form of
(\ref{eq:IntM}) is not quite useful. The problem is that the correlation
functions involved can not be calculated at low energies because of the
large nonperturbative effects that invalidate perturbative approach.
However, simple analyticity properties of the correlation functions
allow us to evaluate the integral in (\ref{eq:IntM}). Indeed, the function
$\Pi$ is analytic in the complex $s$ plane everywhere except the positive
real axis. According to the Cauchy integral theorem, an integral over $s$
 along
the closed contour $C_1+C_2$ (Fig.\ 11) of the product of $\Pi(s)$ with
any nonsingular function $f(s)$ is zero.

\vspace{63mm}

\begin{center}
FIG.\ 11. \hspace{2mm} Integration contour
\end{center}

On the other hand, the imaginary
part of the correlation function is proportional to its
 discontinuity across the
positive real axis. So, the following relation holds
\begin{equation}
\int_{0}^{M_\tau^2} \hspace{1mm} ds \hspace{1mm} f(s)\mbox{Im}\Pi(s)
 = \frac{1}{2i} \int_{C_1} \hspace{1mm} ds \hspace{1mm} f(s)\Pi(s)
 = -\frac{1}{2i} \int_{C_2} \hspace{1mm} ds \hspace{1mm} f(s)\Pi(s),
\label{eq:disc}
\end{equation}
where the $C_2$ is the circle of radius $\mid s\mid =M_{\tau}^2$ (Fig.\ 11).
The benefit of the above relation is that
in the r.h.s. one needs to calculate the correlation function
for $\mid s\mid$ at $M_{\tau}^2$. Hopefully, $M_{\tau}$ is large enough
to use the operator product expansion in powers of $1/M_{\tau}^2$
and the $\alpha_s(M_{\tau})$ is small enough to use perturbative
expansion in $\alpha_s$.
Then the perturbative method can, in principle, be used to calculate
the leading term in the operator product expansion and the higher twist terms
can be estimated semi-phenomenologically.

Using eq.\ (\ref{eq:disc}), the perturbative part of the ratio
 $R_{\tau}$
can be expressed by an
integral over the invariant mass $s$ of the final state hadrons
along the contour $C_2$ in the complex $s$-plane (Fig.\ 11).
In the chiral limit, $m_{u}=m_{d}=m_{s}=0$, the currents are conserved
 and the longitudinal part of the $\Pi(s)$ is absent.
In the axial channel $\Pi^L(s)=O(m_f^2/s)$
(see section 4). For the $R_{\tau}^{\mbox{\scriptsize{pert}}}$ we get
\begin{equation}
R_{\tau}^{\mbox{\scriptsize{pert}}}=\frac{3i}{8\pi}\int_{C_2}\frac{ds}
{M_{\tau}^2}
\biggl(1-\frac{s}{M_{\tau}^2}\biggr)^2
\biggl[\biggl(1+2\frac{s}{M_{\tau}^2}\biggr)\Pi^{T}(s)\biggr].
\label{eq:Contourint}
\end{equation}
Note that the factor $(1-s/M_{\tau}^2)^2$ suppresses the contribution
from the region near the positive real axis where the $\Pi(s)$ has a
branch cut (Braaten, 1988). To simplify the description, we use
the chiral limit which is a perfect approximation for $R_{\tau}$.
On the other hand, the mass corrections can be included with the calculation
very similar to that in section 4. The actual calculations
show (Chetyrkin and Kwiatkowski, 1993; see also recent analyses in Pich, 1994a)
that the effects of quark mass corrections on $R_{\tau}$ are
well below 1\% and can be neglected. Note also that, in the massless
quark limit the contributions from vector and axial channels to $\Pi$
coincide at any given order of perturbation theory and evidently
 the results are flavor independent. So, in this case, for evaluation of
$\Pi^T(s)$ in eq.\ (\ref{eq:Contourint})
 we use our earlier
results for the electromagnetic two-point correlation function that
contributes to R(s) in electron-positron annihilation (section 6).

The function $\Pi^T(s)$ can be related to the
 $D(s)$ function defined in section 2 as follows
\begin{equation}
-\frac{3}{4}s\frac{d}{ds}\Pi^T(s)=\frac{\sum_{f=d,s} \mid V_{uf}\mid^2}
{\sum_{f}Q_f^2}D(s),
\label{eq:Corrfunct}
\end{equation}
where $V_{ud}$ and $V_{us}$ are the  Kobayashi-Maskawa matrix elements.
$\mid V_{ud}\mid^2+\mid V_{us}\mid^2=0.998\pm 0.002$ (see, e.g., Pich, 1994b).
The factor in the r.h.s of eq.\ (\ref{eq:Corrfunct}) is due to the replacement
of the electromagnetic currents by charged weak currents in the correlation
function. Note also that evidently the ``light-by-light'' type  graphs
(Fig.\ 8) do not contribute to the decay width of the $\tau$ lepton.
Thus, the term $\sim (\sum_{f}Q_f)^2$ drops out in the $D$ function.
The perturbative coefficients of $D(s)$ have been
given in the previous section up to the four-loop level
in the vector channel
(see eqs.\ (\ref{eq:RGDsolut}) and (\ref{eq:Ranalytic1}) ).

Performing the contour integration in eq.\ (\ref{eq:Contourint})
using the relations (\ref{eq:Corrfunct}) and (\ref{eq:RGDsolut}),
and replacing $\alpha_s(s)$ by $\alpha_s(M_{\tau})$ using the
evolution equation (\ref{eq:Astransform}), we obtain in terms of
perturbative coefficients of $R(s)$
\begin{eqnarray}
\lefteqn{R_{\tau}^{\mbox{\scriptsize{pert}}}=
    \frac{\mid V_{ud}\mid^2+\mid V_{us}\mid^2}{\sum_{f}Q_{f}^{2}}
        \biggl\{R_{0}
     +\frac{\alpha_s(M_{\tau}^2)}{\pi}R_1
     +\biggl(\frac{\alpha_s(M_{\tau}^2)}{\pi}\biggr)^2
     \biggl(R_{2}+\frac{19}{12}\beta_0 R_1\biggr)}
                                                     \nonumber\\
 && \quad
     +\biggl(\frac{\alpha_s(M_{\tau}^2)}{\pi}\biggr)^3
     \biggl[R_3+\frac{19}{6}R_2\beta_0+\frac{19}{12}R_1\beta_1
     +\biggl(\frac{265}{72}-\frac{\pi^2}{3}\biggr)R_1\beta_0^2\biggr]
     +O(\alpha_s^4)\biggr\},
\label{eq:Rtaumain}
\end{eqnarray}
where, as we have already mentioned, the term $\sim (\sum_{f}Q_f)^2$
should be omitted in $R_3$.

Substituting the relevant expressions for $R_{i}$ and $\beta_{i}$
from the previous sections, we obtain the $O(\alpha_s^3)$ analytical result
in the $\overline{\mbox{MS}}$ scheme
\begin{eqnarray}
\lefteqn{R_{\tau}^{\mbox{\scriptsize{pert}}}(M_{\tau}^2)=
     N_F(\mid V_{ud}\mid^2+\mid V_{us}\mid^2)\biggl\{ 1
        +\frac{\alpha_s(M_{\tau}^2)}{\pi}
                \biggl(\frac{3}{4}C_F\biggr)}
                                                         \nonumber\\
 && \quad
            +\biggl(\frac{\alpha_s(M_{\tau}^2)}{\pi}\biggr)^2
          \biggl[C_F^2\biggl(-\frac{3}{32}\biggr)
             +C_FC_A\biggl(\frac{947}{192}-\frac{11}{4}\zeta(3)\biggr)
               +NTC_F\biggl(-\frac{85}{48}+\zeta(3)\biggr)\biggr]
                                                          \nonumber\\
  && \quad
        +\biggl(\frac{\alpha_s(M_{\tau}^2)}{\pi}\biggr)^3
          \biggl[C_F^3\biggl(-\frac{69}{128}\biggr)
    +C_F^2C_A\biggl(-\frac{1733}{768}-\frac{143}{16}\zeta(3)
                                       +\frac{55}{4}\zeta(5)\biggr)
                                                        \nonumber\\
  && \quad \hspace{34mm}
    +C_FC_A^2\biggl(\frac{559715}{13824}-\frac{2591}{96}\zeta(3)
                       -\frac{55}{24}\zeta(5)\biggr)
                                                       \nonumber\\
  && \quad \hspace{34mm}
     +NTC_F^2\biggl(-\frac{125}{192}+\frac{19}{4}\zeta(3)-5\zeta(5)\biggr)
                                                        \nonumber\\
  && \quad \hspace{34mm}
           +NTC_FC_A\biggl(-\frac{24359}{864}+\frac{73}{4}\zeta(3)
                                 +\frac{5}{6}\zeta(5)\biggr)
                                                           \nonumber\\
  && \quad \hspace{5mm}
           +N^2T^2C_F\biggl(\frac{3935}{864}
                                 -\frac{19}{6}\zeta(3)\biggr)
              -\frac{\pi^2}{64}
                       C_F\biggl(\frac{11}{3}C_A
              -\frac{4}{3}NT\biggr)^2\biggr]+O(\alpha_s^4)\biggr\}.
\label{eq:Rtauanalytic0}
\end{eqnarray}
Within the standard QCD with the SU$_{c}$(3) gauge group we obtain
\begin{eqnarray}
\lefteqn{\hspace{-9mm} R_{\tau}^{\mbox{\scriptsize{pert}}}(M_{\tau}^2)=
   3(0.998\pm 0.002)\biggl\{ 1
        +\frac{\alpha_s(M_{\tau}^2)}{\pi}}
                                                          \nonumber\\
 && \quad
        +\biggl(\frac{\alpha_s(M_{\tau}^2)}{\pi}\biggr)^2
          \biggl[\frac{313}{16}-11\zeta(3)-N\left(\frac{85}{72}
                -\frac{2}{3}\zeta(3)\right)\biggr]
                                                            \nonumber\\
  && \quad
           +\biggl(\frac{\alpha_s(M_{\tau}^2)}{\pi}\biggr)^3
   \biggl[\frac{544379}{1152}-\frac{121}{8}\zeta(2)-\frac{8917}{24}\zeta(3)
        +\frac{275}{6}\zeta(5)
                                                           \nonumber\\
  && \quad \hspace{17mm}
            +N\biggl(-\frac{8203}{144}+\frac{11}{6}\zeta(2)
             +\frac{733}{18}\zeta(3)-\frac{25}{9}\zeta(5)\biggr)
                                                       \nonumber\\
  && \quad \hspace{17mm}
        +N^2\biggl(\frac{3935}{2592}-\frac{1}{18}\zeta(2)
             -\frac{19}{18}\zeta(3)\biggr) \biggr]
       +O(\alpha_s^4) \biggr\}_{N=3},
\label{eq:Rtauanalytic1}
\end{eqnarray}
and a numerical form reads
\begin{eqnarray}
\lefteqn{\hspace{-9mm}R_{\tau}^{\mbox{\scriptsize{pert}}}(M_{\tau}^2)=}
                                                          \nonumber\\
 && \quad \hspace{-21mm}
    3(0.998\pm 0.002)\biggl[ 1
        +\frac{\alpha_s(M_{\tau})}{\pi}
        +5.2023
         \left(\frac{\alpha_s(M_{\tau})}{\pi}\right)^2
        +26.366
         \left(\frac{\alpha_s(M_{\tau})}{\pi}\right)^3
          +O(\alpha_s^4) \biggr]
\label{eq:Rtaunumer}
\end{eqnarray}

\vspace{5mm}

\renewcommand{\thesection}{\arabic{section}}
\subsection{\tenbf On the Nonperturbative and Electroweak contributions}

\indent
The nonperturbative contributions to $R_{\tau}$ can be expressed as
a power series of corrections in $1/M_{\tau}^2$
\begin{equation}
R_{\tau}^{\mbox{\scriptsize{nonpert}}}
     \sim \frac{C_2{\cal f}(m_f^2(M_{\tau})
                ,\theta_c)}{M_{\tau}^2}
    +\sum_{i\geq2}\frac{C_{2i}<O_{2i}>_0}{M_{\tau}^{2i}},
\label{eq:Rtaunonp}
\end{equation}
where the $m_f$ are $u, d, s$ running quark masses, $<O_{2i}>_{0}$ are
the so-called vacuum condensates, which can be obtained
 phenomenologically
and the $C_i$ are their coefficient functions describing
short distance effects. Note that, in eq.\ (\ref{eq:Rtaunonp})
we formally include part of the pure perturbative corrections
(the first term) which is due to the nonvanishing $u, d, s$ quark
masses. These corrections for the $u$ and $d$ quarks are completely
negligible. The contribution coming from the $s$ quark is suppressed by
$\sin^2\theta_{C}$ and is also below 1\% (Pich, 1990).
Presently, the only way to estimate the
strong interaction effects in the condensate contributions
 is by perturbation
theory. The coefficient functions $C_{2i}$ are asymptotic
perturbative series in terms of $\alpha_s$. In order to estimate
the nonperturbative contributions, one needs to sum up the
power series of the QCD perturbative series.
 In the previous section we have described the calculation of the
high-order perturbative QCD contributions to the coefficient functions
of the dimension 4 power corrections (gluon, $<\alpha_sG^2>_0$
and quark, $<m_f\overline{q}_fq_f>_0$ condensates).
It was shown (Loladze, Surguladze and Tkachov, 1985; Surguladze and Tkachov,
1989b, 1990) that the high-order perturbative
corrections to some of the coefficient functions are too large.
For instance, for the coefficient function of the condensate
$<m_s\overline{s}s>_0$ in the vector channel (see eq.\ (\ref{eq:CqqnewV}))
$\Lambda_{\mbox{\scriptsize eff}}
\approx 30\Lambda_{\overline{\mbox{\scriptsize MS}}}$.
This indicates that the renormalization group invariant criteria to the
perturbative calculability of the QCD contributions to the coefficient
function is not fulfilled. The coefficient functions of the
dimension 6 condensates are calculated up to $O(\alpha_s)$
(Lanin, Spiridonov and Chetyrkin, 1986) and to analyze the corresponding
series one needs at least the next to leading correction.
The above uncertainty in coefficient functions $C_{2i}$ allows one
to estimate the condensate contributions probably not better than their
order of magnitude. There is another source of theoretical uncertainties
in the evaluation of condensate contributions of dimension 6
and higher, where the operator basis of expansion includes a large
number of operators. Presently, there are no precise methods to estimate
their matrix elements. For the matrix elements of four quark operators
(dimension 6) the vacuum saturation approximation (Shifman,
 Vainshtein and Zakharov, 1979) is used to express them as
 the square of the two-quark matrix elements. However, the
 vacuum saturation approximation is not expected to be precise
 enough in order to use it in the analyses of the tiny
 nonperturbative contributions (see, e.g., analysis by Altarelli, 1992;
see also a brief discussion in Surguladze and Samuel, 1992b ).
Indeed, as it was found by Braaten (1988) and Pich (1990, 1992a,b, 1994a),
the nonperturbative corrections are below the $1\%$ level with large
theoretical error. The contributions of dim=4 condensates  start
at $O(\alpha_s^2)$ and thus are suppressed by two powers of $\alpha_s$.
The dim=6 and dim=8 corrections are suppressed by the inverse powers
of $M_{\tau}$ ($M_{\tau}^6$ and $M_{\tau}^8$ respectively) and are small.
On the other hand, the corrections in vector and axial channels have
opposite signs and they largely cancel each other, so the total relative
error is even larger.
In the works by Pumplin (1989, 1990) it was shown that the uncertainty
due to threshold effects makes a significant contribution in the
theoretical error for $R_{\tau}$. In the works by Altarelli (1992) and
Altarelli, Nason and Ridolfi (1994) an ambiguity $\sim \Lambda^2/M_{\tau}^2$
is discussed. Earlier, Zakharov (1992) has argued that such dim=2 terms
in eq.\ (\ref{eq:Rtaunonp}) can be generated by ultraviolet renormalons.
For an alternative point of view on the effects of possible dim=2 terms,
see Narison (1994). However, this issue is still a subject of intensive
discussions and likely is far from being settled.

  Summarizing, we note that the above mentioned major
sources of theoretical uncertainties in the evaluation of small power
corrections makes certain restriction on the precision theoretical
prediction
of $R_{\tau}$ and consequently on $\alpha_{s}(M_{\tau})$.
Fortunately, the nonperturbative corrections are suppressed
and the hadronic decay of the $\tau$  still remains as a good
source to extract the low energy $\alpha_s$.

Finally, we note that the electroweak contributions
$R_{\tau}^{\mbox{\scriptsize{weak}}}$ were calculated
by Marciano and Sirlin (1988), and Braaten and Li (1990).
Those corrections contain logarithms of $M_{\tau}/M_{Z}$ and
are not negligible. The leading order electroweak corrections
give roughly $+2\%$ contributions to $R_{\tau}$
(see, e.g., Pich, 1994a).

\vspace{2cm}

\renewcommand{\thesection}{\Roman{section}}
\section{\bf \ \ \  Four-loop QED Renormalization Group
                 Functions}
\renewcommand{\thesection}{\arabic{section}}
\setcounter{equation}{0}

In this section we outline the calculation of the standard
QED renormalization group functions at the four-loop level in the
minimal and momentum subtraction schemes.
These quantities can  be obtained as an intermediate result
of the calculations of $R(s)$, described in the previous sections,
by replacing the SU$_{\mbox{\scriptsize c}}$(3) gauge group invariants for the
corresponding diagrams in a proper way. The results of two
independent calculations of the four-loop QED $\beta$-function
by Gorishny, Kataev and  Larin (1990), and by Surguladze (1990)
have been reported in the joint publications by
Gorishny, Kataev, Larin and Surguladze (1991a,c).

\renewcommand{\thesection}{\arabic{section}}
\subsection{\tenbf General formulae}

\indent
The Lagrangian density of standard QED is
\begin{eqnarray}
\lefteqn{L_{QED}=} \nonumber\\
  && -\frac{1}{4}F_{\mu\nu}F^{\mu\nu}
      +i\sum_{j}\overline{\psi}_j\gamma^{\mu}D_{\mu}\psi_{j}
      -\sum_{j}m_{j}\overline{\psi}_j\psi_j
      -\frac{1}{2\alpha_{\mbox{\tiny{G}}}}
           \partial_{\mu}A^{\mu}\partial_{\mu}A^{\mu},
\label{eq:QEDlagr}
\end{eqnarray}
where $F_{\mu\nu}=\partial_{\mu}A_{\nu}-\partial_{\nu}A_{\mu}$ and
$D_{\mu}=\partial_{\mu}-ieA_{\mu}$. $\alpha_{\mbox{\tiny{G}}}$ is the gauge
 parameter,
$m_j$ are the fermion masses, $\psi$ and $A_{\mu}$ are the fermion and
photon fields and  $e$ is the electric charge.

  Renormalization constants are defined by the relations
\begin{displaymath}
  \psi_{\mbox{\tiny{B}}} = \mu^{-\varepsilon}\sqrt{Z_{\mbox{\tiny{F}}}}\psi,
\end{displaymath}
\begin{equation}
A^{\mu}_{\mbox{\tiny{B}}} =
    \mu^{-\varepsilon}\sqrt{Z_{\mbox{\scriptsize{ph}}}}A^{\mu},
\label{eq:QEDrenormal}
\end{equation}
\begin{displaymath}
  \alpha_{\mbox{\tiny{B}}} =\mu^{2\varepsilon}
              Z_{\alpha} \alpha \hspace{1cm} (\alpha=e^2/4\pi),
\end{displaymath}
\begin{displaymath}
  \alpha_{\mbox{\tiny{G}}}^{\mbox{\tiny{B}}} =
              Z_{\mbox{\tiny{G}}} \alpha_{\mbox{\tiny{G}}}.
\end{displaymath}
For the fermion-fermion-photon vertex renormalization one has
\begin{equation}
\mu^{-2\varepsilon}Z_{\mbox{\scriptsize{vert}}}
e\overline{\psi}\gamma_{\mu}A^{\mu}\psi
=\mu^{-2\varepsilon}\sqrt{Z_{\alpha}Z_{\mbox{\scriptsize{ph}}}}
Z_{\mbox{\tiny{F}}}e\overline{\psi}\gamma_{\mu}A^{\mu}\psi.
\label{eq:FFPH}
\end{equation}
According to Ward identity in QED (Ward, 1950)
$Z_{\mbox{\scriptsize{vert}}}=Z_{\mbox{\tiny{F}}}$, which implies
from eq.\ (\ref{eq:FFPH}) the identity
\begin{equation}
Z_{\alpha}Z_{\mbox{\scriptsize{ph}}}=1.
\label{eq:QEDWI}
\end{equation}
{}From eqs.\ (\ref{eq:QEDrenormal}) and (\ref{eq:QEDWI}) we get
\begin{equation}
\alpha_{\mbox{\tiny{B}}}
=\mu^{2\varepsilon}Z_{\mbox{\scriptsize{ph}}}^{-1}\alpha.
\label{eq:QEDcouplren}
\end{equation}
The gauge invariance of the QED lagrangian implies the absence of the
conterterm for the gauge fixing term in (\ref{eq:QEDlagr}) and, thus,
$Z_{\mbox{\tiny{G}}}=Z_{\mbox{\scriptsize{ph}}}$.

   Using the  relation (\ref{eq:QEDcouplren}) and the renormalization
group invariance of ``bare'' coupling
$\mu^2 d\alpha_{\mbox{\tiny{B}}}/d\mu^2=0$, taking into account that
$Z_{\mbox{\scriptsize{ph}}}$ depends on $\mu$ only via $\alpha$ and also
the standard definition of the QED MS $\beta$-function
\begin{equation}
\beta_{\mbox{\tiny{QED}}}^{\mbox{\tiny{MS}}}(\alpha)
=\frac{1}{4\pi}\mu^2\frac{d\alpha}{d\mu^2}
  \biggr|_{\alpha_{\mbox{\tiny{B}}}\mbox{\scriptsize{ fixed}}},
\label{eq:QEDbeta}
\end{equation}
we obtain a convenient expression for the further evaluation of the
$\beta$ function
\begin{equation}
\beta_{\mbox{\tiny{QED}}}^{\mbox{\tiny{MS}}}(\alpha)=
    -\frac{1}{4\pi}\lim_{\varepsilon \rightarrow 0}
  \frac{\varepsilon \alpha}{1-\alpha\frac{\partial}{\partial\alpha}
  \log Z_{\mbox{\scriptsize{ph}}}}.
\label{eq:QEDmain}
\end{equation}

\renewcommand{\thesection}{\arabic{section}}
\subsection{\tenbf Four-loop results}

\noindent
The photon field renormalization constant $Z_{\mbox{\scriptsize{ph}}}$
can be found from the
QED relation, analogous to eq.\ (\ref{eq:Zanalres}), where only 58
QED four-loop diagrams contribute to $\Pi(\mu^2/Q^2,\alpha)$.
The prescription for the evaluation of the diagram contributions to the
$\Pi_{\mbox{\tiny{B}}}$ is analogous to the one described in section 2.
The total {\small CPU} time on the three {\small IBM} compatible mainframes was
approximately
400 hours.
Setting $C_F=1$, $C_A=0$, $T=1$ and $\alpha_s=\alpha$ in
eq.\ (\ref{eq:Zanalres}), we obtain the four-loop photon
renormalization constant in QED, corresponding to the {\em minimal
subtraction} prescription
\begin{eqnarray}
\lefteqn{\hspace{-1mm}Z_{\mbox{\scriptsize{ph}}}=
   N-\frac{\alpha}{4\pi}\frac{4}{3\varepsilon}N
              -\biggl(\frac{\alpha}{4\pi}\biggr)^2
               \frac{2}{\varepsilon}N
              -\biggl(\frac{\alpha}{4\pi}\biggr)^3
        \biggl[\frac{8}{9\varepsilon^2}N
   -\frac{1}{\varepsilon}\biggl(\frac{2}{3}+\frac{44}{27}N\biggr) \biggr]N}
                                                          \nonumber\\
 && \hspace{-4mm}
   -\biggl(\frac{\alpha}{4\pi}\biggr)^4
   \biggl\{\frac{16}{27\varepsilon^3}N^2+\frac{1}{\varepsilon^2}
    \biggl(\frac{14}{9}N-\frac{88}{81}N^2\biggr)
    -\frac{1}{\varepsilon}
    \biggl[\frac{23}{2}-\biggl(\frac{190}{9}-\frac{208}{9}\zeta(3)\biggr)N
     +\frac{308}{243}N^2 \biggr]\biggr\}N \nonumber\\
\label{eq:ZQED}
\end{eqnarray}
Substituting the expression for $Z_{\mbox{\scriptsize{ph}}}$ into eq.\
(\ref{eq:QEDmain}),
we obtain the following result for the four-loop QED $\beta$-function
in the MS type schemes
\begin{eqnarray}
\lefteqn{\hspace{-23mm}\beta_{\mbox{\tiny{QED}}}^{\mbox{\tiny{MS}}}(\alpha)=
      \frac{4}{3}N\left(\frac{\alpha}{4\pi}\right)^2
     +4N\left(\frac{\alpha}{4\pi}\right)^3
     -N\left(2+\frac{44}{9}N\right)\left(\frac{\alpha}{4\pi}\right)^4}
                                                                   \nonumber\\
  && \hspace{17mm}
      -N\biggl[46-\biggl(\frac{760}{27}-\frac{832}{9}\zeta(3)\biggr)N
     +\frac{1232}{243}N^2\biggr]\biggl(\frac{\alpha}{4\pi}\biggr)^5.
\label{eq:betaQED}
\end{eqnarray}
It is useful for further applications to present the result for the
Johnson-Willey-Baker $F_1$
function (Johnson, Willey and Baker, 1967; Baker and Johnson, 1971;
Johnson and Baker, 1973). This function can be obtained from the result for
$\beta_{\mbox{\tiny{QED}}}^{\mbox{\tiny{MS}}}$
by subtracting the contributions of the diagrams with fermion loop
insertions into the photon lines and reducing the power in $\alpha/4\pi$
by one. We obtain
\begin{equation}
F_1(\alpha)= \frac{4}{3}\left(\frac{\alpha}{4\pi}\right)
            +4\left(\frac{\alpha}{4\pi}\right)^2
            -2\left(\frac{\alpha}{4\pi}\right)^3
            -46\left(\frac{\alpha}{4\pi}\right)^4.
\label{eq:QEDF1}
\end{equation}
Note that all coefficients up to four-loop level are rational numbers.
The results for most of the individual graphs do contain
transcendental $\zeta(3)$, $\zeta(4)$ and $\zeta(5)$. The
$\zeta(4)$ and $\zeta(5)$ cancel within each gauge-invariant set of
diagrams.
The three-loop results agree with the ones obtained by de~Rafael and Rosner
(1974).
It is possible to recalculate the MS QED $\beta$ function in the form of
the Gell-Man-Low $\Psi(\alpha)$ function - the QED $\beta$ function in the
MOM scheme. See details in Gorishny, Kataev, Larin and Surguladze (1991a)
(see also Adler, 1972; de~Rafael and Rosner, 1974).
We obtain the Gell-Mann-Low $\Psi$ function at the four-loop level
\begin{eqnarray}
\lefteqn{\hspace{-9mm}\Psi(\alpha)=
       \frac{4}{3}N\biggl(\frac{\alpha}{4\pi}\biggr)^2
     +4N\biggl(\frac{\alpha}{4\pi}\biggr)^3
     -N\biggl[2
     +\biggl(\frac{184}{9}-\frac{64}{3}\zeta(3)\biggr)N\biggr]
         \biggl(\frac{\alpha}{4\pi}\biggr)^4}
                                                      \nonumber\\
  &&
     -N\biggl[46-\biggl(104+\frac{512}{3}\zeta(3)
      -\frac{1280}{3}\zeta(5)\biggr)N
     -\biggl(128-\frac{256}{3}\zeta(3)\biggr)N^2\biggr]
     \biggl(\frac{\alpha}{4\pi}\biggr)^5. \nonumber\\
\label{eq:QEDpsi}
\end{eqnarray}
The $O(\alpha^4)$ result agrees with the one obtained
by Baker and Johnson (1969) and Acharya and Nigam (1978, 1985).

Recently, Broadhurst, Kataev and Tarasov (1993) have carried out
 an additional
calculation necessary to convert the four-loop MS QED $\beta$ function
to the four-loop QED on-shell $\beta$ function, usually called
the Callan-Symanzik function $\beta_{\mbox{\tiny{QED}}}^{\mbox{\tiny{CS}}}$
(Callan, 1970; Symanzik, 1970, 1971). This function is defined as
follows
\begin{equation}
\beta_{\mbox{\tiny{QED}}}^{\mbox{\tiny{CS}}}(\alpha)
=\frac{m_{e}}{\alpha}\frac{d\alpha}{d m_{e}}
  \biggr|_{\alpha_{\mbox{\tiny{B}}}\mbox{\scriptsize{ fixed}}},
\label{eq:QEDbetaCS}
\end{equation}
where $m_{e}$ is the electron pole mass. The subtraction prescription
 in this case requires all subtractions to be on-shell. The three-loop
$\beta_{\mbox{\tiny{QED}}}^{\mbox{\tiny{CS}}}$ was calculated long ago
by de Rafael and Rosner (1974). The four-loop result has the following
form  (Broadhurst, Kataev and Tarasov, 1993)
\begin{eqnarray}
\lefteqn{\beta_{\mbox{\tiny{QED}}}^{\mbox{\tiny{CS}}}(\alpha)=
      \frac{2}{3}N\left(\frac{\alpha}{\pi}\right)
     +\frac{1}{2}N\left(\frac{\alpha}{\pi}\right)^2
     -N\left(\frac{1}{16}+\frac{7}{9}N\right)
                 \left(\frac{\alpha}{\pi}\right)^3}
                                                              \nonumber\\
 && \hspace{-7mm}
    -N\biggl[\frac{23}{64}-\biggl(\frac{1}{24}-\frac{5}{3}\zeta(2)
    +\frac{8}{3}\zeta(2)\ln 2-\frac{35}{48}\zeta(3)\biggr)N
   -\biggl(\frac{901}{648}-\frac{8}{9}\zeta(2)-\frac{7}{48}\zeta(3)\biggr)N^2
                                                                      \biggr]
            \biggl(\frac{\alpha}{\pi}\biggr)^4 \nonumber\\
\label{eq:betaQEDCS}
\end{eqnarray}

\vspace{2cm}

\renewcommand{\thesection}{\Roman{section}}
\section{\bf         Renormalization Group
                     Ambiguity of Perturbative QCD Predictions}
\renewcommand{\thesection}{\arabic{section}}
\setcounter{equation}{0}

In the previous sections we have demonstrated the calculation of
some of the important observables within the framework of
perturbative QCD. This involves calculation of a large number of
Feynman diagrams and requires a
a very large amount of computer and human resources.
For example, to $O(\alpha_{s}^{3})$ we have calculated 98
(effectively 250) four-loop Feynman diagrams. The next
order requires calculation of approximately 600-700 five-loop
diagrams. Calculations of such a scale are extremely difficult.
On the other hand, perturbative QCD series are asymptotic ones
and the question of how many orders need to be calculated, can be
answered only from estimates of remainders (see, e.g., the textbook by
Collins, 1984). Moreover, perturbative
coefficients beyond the two-loop level, as well as the expansion
 parameter, are scheme-scale dependent. The scheme-scale ambiguity
 - a fundamental property of the renormalization group calculations
in QCD, does not allow one to obtain reliable  estimates
from the first few calculated terms without involving additional
criteria.

In this section we discuss the extraction of reliable
estimates for observable quantities within perturbation theory.
The problem of scheme-scale dependence of perturbative QCD predictions
will be considered first within the MS prescription and then we
outline a scheme invariant approach along the lines of Stevenson (1981a,b).
We apply the three known approaches for resolving the scheme-scale
ambiguity. As a result, we fix the scheme-scale parameter,
within the framework of MS prescription, for which all of the criteria tested
are satisfied for the quantity $R(s)$ at the four-loop level
(Surguladze and Samuel, 1993). On the other hand, we estimate the
theoretical error by using the scheme-scale dependence as a measure
of the theoretical uncertainty (Surguladze and Samuel, 1993;
Surguladze, 1994b). We also mention the recent discovery of
commensurate scale relations by Brodsky and Lu (1994, 1995).
These relations allow one to connect several physical observables,
providing important tests of QCD without scheme-scale ambiguity.

\vspace{1cm}

\renewcommand{\thesection}{\arabic{section}}
\subsection{\tenbf Perturbative QCD series: How many loops should be
 evaluated?}

  The R-ratio in electron-positron annihilation is given within
perturbation theory in the following form
\begin{equation}
R(s)=r_0\left(1+r_1(\frac{s}{\mu^2})\frac{\alpha_s(\mu)}{\pi}
+r_2(\frac{s}{\mu^2}) (\frac{\alpha_s(\mu)}{\pi})^2
+r_3(\frac{s}{\mu^2}) (\frac{\alpha_s(\mu)}{\pi})^3 + ...\right).
\label{eq:Rgeneral}
\end{equation}
Our further discussion is quite general and can be applied to other
observables like R$_{\tau}$ or Higgs decay rates.
We consider high enough energies, where R is a function of a single
variable - the center-of-mass energy squared. Our aim is to
evaluate pure QCD effects in R, which start with the term $O(\alpha_s)$,
within the minimal subtraction prescription ('t Hooft, 1973).
We should stress here that the calculational methods allowing one to
evaluate perturbative corrections
up to the four-loop order (up to the five-loop in some cases)
is essentially based on some of the unique features of the MS prescription
and our choice seems to be well justified.
There is an ambiguity in the choice of renormalization scale parameter $\mu$.
 Usually we set $\mu^2=s$ and absorb the large logarithms in the definition
of the running coupling. On the other hand, the choice $\mu^2=\chi s$
($\chi \equiv e^{t}$) for all $\chi$ gives equivalent expansions.
Evidently, the sum  of ``all'' terms in eq.\ (\ref{eq:Rgeneral}) does not
depend on the choice of $\mu$. However, in practice,
we deal with truncated series, where the sum has a nontrivial
dependence on the
choice of renormalization parameter. Here we keep the ``natural''
choice $\mu^2=s$ and the ambiguity is transferred to the prescription
$ \int d^4p  \longrightarrow \int d^{4-2\varepsilon}p
(\mu^2 e^{(t+O(\varepsilon))})^{\varepsilon}$.
By changing $t$ one gets different MS type
schemes. One can always reexpand (\ref{eq:Rgeneral}) in a new scheme (with a
new $\Lambda$ in (\ref{eq:Asparametr}) ) and so redistribute the values of
$r_{i}$ ($i > 1$). All these schemes are equivalent. On the other hand,
a new scheme may be ``better'', but one can conclude this only based on the
knowledge of remainders. The problem of scheme-scale ambiguity which,
in fact, is a problem of remainders can be formulated as follows.
{\em How does one choose (``optimize'') the scheme (or $\Lambda$)
in order to make the remainder minimal in the series of the type
(\ref{eq:Rgeneral}) for the given range of energy and what is
the numerical uncertainty of the approximation (\ref{eq:Rgeneral})?}
Here one should also distinguish the following two questions.
First, what is the best accuracy to which the given quantity is
calculable via perturbation theory? Second, what is the accuracy of
the given approximation?
 A few notes are in order. It is known that
perturbative QCD series are asymptotic ones.
No reliable estimates of the remainders are
known at present. However, it is known from the theory of asymptotic
series (see, e.g., Dingle, 1973) that
\begin{equation}
\mid \sum_{i=1}^{\cal N}r_i\alpha^i(s)-R(s)\mid=R_{\cal N}\rightarrow
\Delta R_{\mbox{\scriptsize{min}}},
\mbox{\hspace{2mm} when \hspace{2mm}} {\cal N}\rightarrow
{\cal N}_{\mbox{\scriptsize{opt}}}.
\label{eq:asestimate}
\end{equation}
This means that, the remainder $R_{\cal N}$ goes to  its minimal
value $\Delta R_{\mbox{\scriptsize{min}}}$ when the number
of orders goes to its optimal
value ${\cal N}_{\mbox{\scriptsize{opt}}}$.
Inclusion of the next to ${\cal N}_{\mbox{\scriptsize{opt}}}$ orders
will lead away from the correct value. It is known (see, e.g., Dingle, 1973)
that for a sign-alternating asymptotic series the remainder can be estimated
by the first neglected term (or by the last included term).
However, it is still unknown if the QCD perturbative series
has this character. We assume as a hypothesis that within QCD one can
estimate the remainder by the first neglected or last included term.
Now, the minimal possible error, which defines the best accuracy of the
perturbation theory for the given quantity has an order of
$\Delta R_{\mbox{\scriptsize{min}}}\sim r_{{\cal N}+1}\alpha^{{\cal N}+1}(s)$,
${\cal N}\rightarrow {\cal N}_{\mbox{\scriptsize{opt}}}$.
Note that, both the number ${\cal N}_{\mbox{\scriptsize{opt}}}$
and the value of the $\Delta R_{\mbox{\scriptsize{min}}}$ depend on
the range of energy for the given process. We once again
emphasize that {\em the remainder depends on the choice of particular
scheme and scale parameters and its estimate makes sense only for
the ``optimized'' renormalization scheme which is unique for the
given physical observable}. In fact, it was argued (Stevenson, 1984, 1994)
that, the ``optimized'' series can still converge even when the series
in any fixed renormalization scheme is factorially divergent,
if the ``optimized'' couplant shrinks in higher orders
(see also Buckley, Duncan and Jones, 1993). However, whether this
applies to QCD is unknown.

\vspace{1cm}

\renewcommand{\thesection}{\arabic{section}}
\subsection{\tenbf $R(s)$ within the one parametric family of the
 MS type schemes and scale ambiguity problem}

Using  the results of our four-loop calculations,
we obtain the analytical result for $R(s)$ with perturbative coefficients
explicitly depending on the scheme-scale parameter
(Surguladze and Samuel, 1993)
\begin{eqnarray}
\lefteqn{\hspace{-7mm}R(s,t)=R_0+\frac{\alpha_s(s,t)}{\pi}R_1
         +\biggl(\frac{\alpha_s(s,t)}{\pi}\biggr)^2
                                    (R_2+\beta_0 R_1 t) } \nonumber\\
 && \hspace{33mm}
         +\biggl(\frac{\alpha_s(s,t)}{\pi}\biggr)^3
      [R_3-\frac{\pi^2}{3}\beta_0^2R_1
       +(2\beta_0 R_2 + \beta_1 R_1)t+\beta_0^2 R_1 t^2]. \nonumber\\
\label{eq:Rstgen}
\end{eqnarray}
Recalling the values of the $\overline{\mbox{MS}}$
perturbative coefficients $R_i$ from
eqs.\ (\ref{eq:Rexpans}) and (\ref{eq:Ranalytic0}) and the $\beta_i$
coefficients from eq.\ (\ref{eq:beta}), we obtain numerically
\begin{eqnarray}
\lefteqn{\hspace{-14mm}R(s,t)=
   3\sum_{f}Q_{f}^2\biggl\{ 1+
         \frac{\alpha_s(s,t)}{\pi}
        +\biggl(\frac{\alpha_s(s,t)}{\pi}\biggr)^2
        [(1.9857+2.75t)-N(0.1153+0.1667t)]}                 \nonumber\\
 && \quad  \hspace{3cm} +\biggl(\frac{\alpha_s(s,t)}{\pi}\biggr)^3
                  [(-6.6369+17.2964t+7.5625t^2)               \nonumber\\
 && \quad \hspace{50mm} -N(1.2001+2.0877t+0.9167t^2)
                                                               \nonumber\\
 && \quad \hspace{50mm} +N^{2}(-0.0052+0.0384t+0.0278t^2)]\biggr\}
                                                               \nonumber\\
 && \quad \hspace{12mm}  -\biggl(\sum_{f}Q_f\biggr)^2
          \biggl(\frac{\alpha_s(s,t)}{\pi}\biggr)^3 1.2395
          +O(\alpha_s^4),
\label{eq:Rstnumer}
\end{eqnarray}
where $\alpha_{s}(s,t)$ can be parametrized in the form of
(\ref{eq:Asparametr}) with $\mu=s$ and $\Lambda\rightarrow
\Lambda_t=e^{-t/2}\Lambda_{\overline{\mbox{\scriptsize{MS}}}}$.
Obviously, $t=0$ corresponds to the $\overline{\mbox{MS}}$ scheme
( eq.\ (\ref{eq:Rnumerical}) ).
$t=\ln 4\pi-\gamma$
will transform the result to the original MS scheme ('t Hooft, 1973).
( eq.\ (\ref{eq:RMSnumerical}) ).
t=-2 corresponds to the G scheme introduced by Chetyrkin and Tkachov
(1979, 1981) ( eq.\ (\ref{eq:RGnumerical}) ).
Note that because of a one-parametric nature of the MS prescription,
the $t$-dependent terms in eq.\ (\ref{eq:Rstnumer}) would represent also
the scale dependence of the perturbative coefficients within the
$\overline{\mbox{MS}}$ if one changes $t\rightarrow
\log\mu^2/s$ and takes $\alpha_s(s,t)$ with $s$ replaced by $\mu^2$ and $t=0$.

    Several approaches were suggested to deal with the
scheme-scale-remainder problem. Among them we consider the following ones.
{\it Fastest Apparent Convergence} (FAC) (Grunberg, 1980, 1982, 1984),
where the next to leading perturbative correction is absorbed in
the definition of the ``effective'' running coupling and the scheme-scale
parameter is fixed accordingly.
{\it Principle of Minimal Sensitivity} (PMS) of the approximant
to the variation of nonphysical parameters
(Stevenson, 1981a,b, 1982, 1984;
see also Mattingly and Stevenson, 1992, 1994).
{\it Brodsky-Lepage-Mackenzie} (BLM) approach (Brodsky, Lepage
 and Mackenzie, 1983), which suggests one
fix the scale by the size of the quark vacuum polarization
effects resulting in the independence of the next to leading order
perturbative correction of the number of quark flavors $N$.
For discussions of the above scheme-scale setting methods
see Celmaster and Stevenson (1983), Brodsky and Lu (1992), and
Stevenson (1992). The optimization of perturbation theory has previously
been studied by Kramer and Lampe (1988), and Bethke (1989) for jet cross
sections in electron positron annihilation. The optimized perturbation
theory is tested for different physical quantities in QED and QCD by
Field (1993). The scale ambiguity problem has been considered by Lu and
de Melo (1991) for the $\phi^{3}$ model. The scheme-scale ambiguity
problem for the quantities $R(s)$ and $R_{\tau}$ has been discussed by
Maxwell and Nicholls (1990), Chyla, Kataev and Larin (1991), and
Grunberg and Kataev (1992). Further study of the PMS method has been done
in Raczka (1995).

    We apply the above methods to eq.\ (\ref{eq:Rgeneral})
and we find a scale which gives good results for all criteria
considered (Surguladze and Samuel, 1993).
We start by noting
that, in general, the renormalizatin scheme-scale dependence
of perturbative results are parametrized by
the scale parameter, say, $\mu$ and the renormalization
prescription dependent coefficients of $\beta$ function
(Stevenson, 1981a,b). We should stress however, that
the $\beta$ function is independent of any modification
of the MS type prescriptions, but starting from $\beta_2$,
the coefficients of $\beta$ function do depend on the
particular choice of subtraction prescription other than
MS.
In order to better visualize our discussion, we consider
first the optimization procedures within the MS prescription.
In other words, we fix the scheme dependent perturbative
coefficients of $\beta$ function to their MS values
and consider only the scale variation.

In Fig.\ 12 we have plotted $r_{3}(t)$ for different $N$
(see eqs.\ (\ref{eq:Rgeneral}) and (\ref{eq:Rstnumer})).
As one can see, within the region $t\sim(-1.5,-0.5)$ $r_3$ has a very
weak dependence on the number of flavors $N$ as well as on the
parameter $t$.

\vspace{9cm}
\begin{center}
FIG.\ 12. $r_{3}(t)$ for different $N$
\end{center}
\vspace{2mm}
Corresponding to the three-loop coefficient $r_2(t)$, straight
lines intersect in one point for $t\approx -0.7 $, which
is obvious from eq.\ (\ref{eq:Rstnumer}).
This value corresponds to the BLM result
(Brodsky, Lepage and Mackenzie, 1983)
$\mu^2=\mu^{2}_{\overline{\mbox{\scriptsize MS}}}e^{0.710}$,
and at this scale
the flavor dependence is absorbed into the definition
of the coupling.

   In Fig.\ 13 we have plotted the dependence of the partial sums
\begin{displaymath}
R_{n}(t)=\sum_{m=1}^{n}r_{m}(t)(\alpha_s/\pi)^m, \hspace{5mm} n=1,2,3
\end{displaymath}
on the parameter $t$.
Here the parametrisation (\ref{eq:Asparametr}) was used,
$\log s/\Lambda_{\overline{\mbox{\scriptsize MS}}}^2=9$ and
$N=5$. The general picture does not change for other
reasonable values of $\log$ and $N$. One can see that
PMS
(Stevenson, 1981) works perfectly for a wide range of
the logarithmic scale parameter $t\sim(-1,+3)$
for the four-loop approximant and $t\sim(-2,0)$ for the three-loop
approximant. A similar analysis at the three-loop level was done
by Radyushkin (1983). According to the above analysis we found that
the BLM scale $t=-0.710$ is good at the four-loop level as well (Fig.\ 12)
and this value is
within minimal sensitivity region (Fig.~13).

\vspace{77mm}
\begin{center}
FIG.\ 13. The approximants $R_{n}$ vs the scale parameter $t$
\end{center}
\vspace{2mm}

\noindent
Moreover, we found that
if the $t$-parameter is chosen in the following analytical form
$t=4\zeta(3)-11/2+O(\varepsilon)$,
which is equivalent to the definition of a new, say,
$\widetilde{\mbox{MS}}$ modification of the MS scheme
\begin{equation}
\Lambda_{\widetilde{\mbox{\scriptsize MS}}}=\mbox{exp}
[-2\zeta(3)+11/4+O(\varepsilon)]
\Lambda_{\overline{\mbox{\scriptsize MS}}},
\label{eq:mstilda}
\end{equation}
then the $N$ dependence and the $\zeta(3)$ terms cancel exactly
at the 3-loop level. As a result, $r_2=1/12$.
Within this scheme the four-loop correction is almost independent of the
number of flavors.  The full result for the R-ratio for the {\em arbitrary}
number of flavors can be written in the following simple form
\begin{equation}
R(s)=3\sum_{f}Q_{f}^2\biggl[ 1+\frac{\alpha_s}{\pi}
       +\frac{1}{12}\biggl(\frac{\alpha_s}{\pi}\biggr)^2
    -\biggl(\frac{\alpha_s}{\pi}\biggr)^3 (16.2 \pm 0.5)\biggr]
     -\biggl(\sum_{f}Q_f\biggr)^2
          \biggl(\frac{\alpha_s}{\pi}\biggr)^3 1.2
          +O(\alpha_s^4)
\label{eq:Rsmstilda}
\end{equation}
where the small uncertainty $\pm 0.5$ stands for the remainder
dependence on the number of flavors at $O(\alpha_s^3)$
for all physically reasonable $N$ and is completely negligible
for phenomenology. The last term is also very small
$\sim 0.4(\alpha_{s}/\pi)^3$. The running coupling
can be parametrized in the standard form (\ref{eq:Asparametr}) with
$\Lambda_{\widetilde{\mbox{\scriptsize MS}}}
=1.41\Lambda_{\overline{\mbox{\scriptsize MS}}}$.

   Using the FAC approach
(Grunberg, 1980, 1982, 1984), we rewrite eq.\ (\ref{eq:Rsmstilda})
as follows.
\begin{equation}
R(s)= 3\sum_{f}Q_{f}^2\left[ 1+\frac{\alpha_{s}^{\mbox{\scriptsize eff}}}{\pi}
          +O(\alpha_s^3)\right],
\label{eq:Rseffect}
\end{equation}
where the 3-loop correction is absorbed into the definition of the
effective coupling given by eq.\ (\ref{eq:Asparametr})
with the $\Lambda$ replaced by
\begin{displaymath}
\Lambda_{\mbox{\scriptsize eff}}
 \approx \Lambda_{\widetilde{\mbox{\scriptsize MS}}}
                            \mbox{exp}\left(\frac{1}{2\beta_0}
      \frac{r_2}{r_1}\right)
         \approx 1.02\Lambda_{\widetilde{\mbox{\scriptsize MS}}}.
\end{displaymath}
As one can see, the new scheme $\widetilde{\mbox{MS}}$ almost
coincides with the effective one and the  fastest convergence is
guaranteed within the wide range of energy defined by the
renormalization group invariant criteria
\begin{displaymath}
\frac{s}{\Lambda_{\mbox{\scriptsize eff}}^{2}} \sim
\frac{s}{\Lambda_{\widetilde{\mbox{\scriptsize MS}}}^{2}} \gg 1.
\end{displaymath}

The similar analyses can be done for the semi-hadronic decay rates
of the $\tau$ lepton calculated to $O(\alpha_s^3)$ in section 7.
The result for the ratio $R_{\tau}$ in the $\widetilde{\mbox{MS}}$
scheme reads
\begin{equation}
R^{\tau}= 3(0.998\pm 0.002)\left[ 1+\frac{\alpha_s(M_{\tau}^2)}{\pi}
     +3.65\left(\frac{\alpha_s(M_{\tau}^2)}{\pi}\right)^2
    +9.83\left(\frac{\alpha_s(M_{\tau}^2)}{\pi}\right)^3 \right]
          +O(\alpha_s^4)
\label{eq:Rtaumstilde}
\end{equation}
and to be compared to eq.\ (\ref{eq:Rtaunumer}). Note that
the $\alpha_s(M_Z)$ is parametrized with the
$\Lambda_{\widetilde{\mbox{\scriptsize MS}}}
=1.41\Lambda_{\overline{\mbox{\scriptsize MS}}}$.

In Fig.\ 14 we plot one-, two- and three-loop approximants to the
$\Gamma_{H\rightarrow b\overline{b}}$ in terms of the running quark mass
(eqs.\ (\ref{eq:Danalyt})-(\ref{eq:GHtot}),
with $N=5$ and $m_f=m_b$) vs. the scale parameter $t$
(Surguladze, 1994b).

\newpage
\mbox{}

\vspace{75mm}
\begin{center}
FIG.\ 14. The approximants of the
              $\Gamma_{H\rightarrow b\overline{b}}$ vs the scale parameter $t$
\end{center}
\vspace{2mm}

One can see that the higher order corrections diminish the scale dependence
from 40\% to nearly 5\%. The solid curve, corresponding to the three-loop
result, became flat
in the wide range of the logarithmic scale parameter $t$.
Moreover, the choice $t=0$
($\overline{\mbox{MS}}$-scheme) satisfies Stevenson's
{\it Principle of Minimal Sensitivity} (Stevenson, 1981).

Let us now try to estimate the theoretical uncertainty in calculations
of $R$ by the last included term in the corresponding perturbative
expansion.  We get for the QCD contribution within the
$\widetilde{\mbox{MS}}$ scheme the following result.
\begin{equation}
\delta^{\widetilde{\mbox{\scriptsize MS}}}_{\mbox{\scriptsize QCD}}
\equiv \frac{R(s)-r_0}{r_0}=\frac{\alpha_s}{\pi}
+\frac{1}{12}\left(\frac{\alpha_s}{\pi}\right)^2
-(16.2 \pm 0.5)
 \left(\frac{\alpha_{s}}{\pi}\right)^3
\pm (\delta_{\mbox{\scriptsize QCD}}^{\mbox{\scriptsize err}}=4\%).
\label{eq:Rserrorestincl}
\end{equation}
The analysis of Fig.\ 13 shows that the deviation of the four-loop approximant
from the constant is also about
$4\%$ within a reasonably wide range of the t-parameter. This is consistent
with Stevenson's principle. One should note that the above error estimate
is only for the massless quark limit. There are several different types
of additional contributions, including those due to nonvanishing quark
masses. This may change the above error estimate.
All of the necessary information on the status of the
additional corrections can be found in Kniehl (1994b, 1995b).

As we have already mentioned, recently Brodsky and Lu (1994, 1995)
have found the relations between the effective couplings
$\alpha_{\mbox{\tiny A}}$ and
$\alpha_{\mbox{\tiny B}}$ for the physical observables
A and B in the following form.
\begin{equation}
\alpha_{\mbox{\tiny A}}(\mu_{\mbox{\tiny A}})
=\alpha_{\mbox{\tiny B}}(\mu_{\mbox{\tiny B}})
\biggl(1+r_{\mbox{\tiny A/B}}\frac{\alpha_{\mbox{\tiny B}}}{\pi}
+\cdots\biggr).
 \label{eq:CSR}
\end{equation}
The ratio of the scales of the corresponding processes
$\mu_{\mbox{\tiny A}}/\mu_{\mbox{\tiny B}}$ is chosen according
to the BLM scale setting prescription so that
$r_{\mbox{\tiny A/B}}$ is independent of the number of flavors.
Thus, evolving $\alpha_{\mbox{\tiny A}}$ and
$\alpha_{\mbox{\tiny B}}$, they pass the quark thresholds
at the same scale. It is shown that the relative scales
satisfy the transitivity rule
\begin{displaymath}
\frac{\mu_{\mbox{\tiny A}}}{\mu_{\mbox{\tiny B}}}
=\frac{\mu_{\mbox{\tiny A}}}{\mu_{\mbox{\tiny C}}}
\times\frac{\mu_{\mbox{\tiny C}}}{\mu_{\mbox{\tiny B}}}.
\end{displaymath}
So, C may correspond to any intermediate theoretical scheme
such as MS, $\overline{\mbox{MS}}$, etc. and the perturbative
results can be tested without a reference to them.
One of the impressive results of this method is
a surprisingly simple relation between the effective
couplings for the quantities $R$ and $R_{\tau}$
to the next-to-next leading order (Brodsky and Lu, 1994, 1995)
\begin{displaymath}
\frac{\alpha_{\tau}(M_{\tau})}{\pi}=
\frac{\alpha_{R}(\mu)}{\pi}, \hspace{9mm}
\mu=M_{\tau}\mbox{exp}\biggl[-\frac{19}{24}-\frac{169}{128}
\frac{\alpha_R(M_{\tau})}{\pi}\biggr].
\end{displaymath}
For more details and the relations between various other
observables we refer to the original works
by Brodsky and Lu (1994, 1995).

\vspace{1cm}

\renewcommand{\thesection}{\arabic{section}}
\subsection{\tenbf On scheme invariant analyses}

Let us now outline the original method of
scheme-invariant analyses for the perturbation theory
results by Stevenson (1981a,b, 1982, 1984).
We note first, that our analyses of perturbation
series for $R(s)$ and $R_{\tau}$ has been done
in the previous subsection within
the one parametric family of the MS type schemes,
where all $\beta$ function coefficients are the same
for any modification of MS.
In the PMS method, renormalizarion scale and scheme
dependence is parametrized by the scale parameter
$\mu/\Lambda$ and the scheme dependent coefficients
of the $\beta$ function $\beta_2$, $\beta_3,\cdots$.
Then the {\it Principle of Minimal Sensitivity}
is applied to the variation of the above parameters
and to $O(\alpha_s^3)$ the ``optimized'' scheme
corresponds to a flat two dimensional surface.
Our curve for $R_3$ in Fig.\ 13 is just a one-dimensional slice
at the particular MS value of the $\beta_2$.
The main points of the PMS formalism is as follows.
(For the scheme invariant analyses of $R(s)$ to $O(\alpha_s^3)$
see Mattingly and Stevenson, 1994).
To use familiar standard notation, we rewrite eq.\ (\ref{eq:RGfunctions})
for the couplant $a\equiv\alpha_s(\mu)/\pi$
\begin{equation}
b\frac{\partial a}{\partial \tau}
=-b a^2(1+ca+c_2a^2+\cdots),
\label{eq:ab}
\end{equation}
where
\begin{equation}
\tau=b\ln\frac{\mu}{\Lambda}, \hspace{4mm}
b=2\beta_0, \hspace{4mm} c=\frac{\beta_1}{\beta_0}
\label{eq:ta}
\end{equation}
and for any modification of the minimal subtraction prescription, the
scheme dependent coefficient $c_2=\beta_2/\beta_0$.
The scheme and scale can now be parametrized by the quantities
$RS\equiv (\tau,c_2,c_3,...)$. The
{\it Principle of Minimal Sensitivity} can be written as
\begin{equation}
\frac{dR_n}{d(\tau;c_2,c_3,...)}=0.
 \label{eq:PMS}
\end{equation}
The number of scheme-scale parameters in the above equation
is strongly correlated with $n$. Indeed,
it is not difficult to show that the following self-consistency
condition should hold for the $n$th approximant
\begin{equation}
\frac{\partial R_n}{\partial(RS)}=O(a^{n+1}).
 \label{eq:SCC}
\end{equation}
This shows that the perturbative coefficients $r_i$ can depend on
renormalization scheme only through parameters $\tau;c_2,...,c_{i-1}$.
Applying the {\it Principle of Minimal Sensitivity} in a form
(\ref{eq:PMS}) to the approximants $R_2$ and $R_3$ and taking
into account (\ref{eq:SCC}), one finds that the quantities
\begin{displaymath}
\rho_1\equiv \tau-r_2,
\end{displaymath}
\begin{equation}
\rho_2\equiv r_3+c_2-\biggl(r_2+\frac{c}{2}\biggr)^2
\label{eq:RSinv}
\end{equation}
are renormalization scheme independent. Similar invariants can
be constructed at each order of perturbation theory.
 The choice of $\tau$ as a
function of the ratio $\mu/\Lambda$ emphasizes that the renormalization
scheme dependence involves only the ratio of these quantities
and the optimization deals with $\tau$ but not $\mu$.
The ``optimal'' values of renormalization scheme parameters
$\overline{\tau}$ and $\overline{c_2}$ are defined by the following equations.
To $O(\alpha_s^2)$,
\begin{equation}
\frac{d R_2(\tau)}{d\tau}\biggr|_{\tau=\overline{\tau} }=0.
\label{eq:tauopt}
\end{equation}
To $O(\alpha_s^3)$,
\begin{equation}
\frac{\partial R_3(\tau,c_2)}{\partial\tau}\biggr|_{\tau=\overline{\tau}}=0,
\end{equation}
\begin{equation}
\frac{\partial R_3(\tau,c_2)}{\partial c_2}\biggr|_{c_2=\overline{c}_2}=0.
\label{eq:taucopt}
\end{equation}
Solving the above equations along with eqs.\ (\ref{eq:RSinv}) for the
renormalization scheme invariants and eq.\ (\ref{eq:ab}) for the couplant
with the truncated MS $\beta$ function, using the $\overline{\mbox{MS}}$
values of $r_2$ and $r_3$, one finds the ``optimized'' values of
$\overline{\tau}$,
$\overline{c}_2$ and corresponding ``optimized'' approximants
to $O(\alpha_s^3)$. The theoretical error can be estimated, as in the
previous subsection, by the last calculated term.
One obtains the following ``optimized'' result for the QCD contribution
in $R(34\mbox{ GeV})$ in the massless quark limit
(Mattingly and Stevenson, 1994; Stevenson, 1994).
\begin{equation}
\delta_{\mbox{\tiny QCD}}^{\mbox{\tiny PMS}} = 0.051 \pm 0.001.
\label{eq:PMSres}
\end{equation}

It is important to note that the above optimization procedure yields
a negative value for the $\rho_2$ invariant. This results in the
existence of a solution of equation
\begin{equation}
\frac{7}{4}+c\overline{a}^{\ast}+3\rho_2(\overline{a}^{\ast})^2=0
\label{eq:IRfix}
\end{equation}
with respect to $\overline{a}^{\ast}$ - the value of the couplant
for which the optimized third order $\beta$ function vanishes.
This allows, in principle, to do some analyses for $R(s)$ at the
low energies $\sqrt{s}\rightarrow 0$ (Mattingly and Stevenson, 1992).

Finally, we also mention that the
FAC approach (Grunberg, 1980, 1982, 1984) is a special case of the
PMS (Stevenson, 1981a,b, 1982, 1984) method. Indeed, in the FAC approach
all higher order approximants are equal to the effective couplant
(compare to eqs.\ (\ref{eq:tauopt}) and (\ref{eq:taucopt}) ).
{}From eqs.\ (\ref{eq:RSinv}) one gets $\rho_1=\tau$ and $\rho_2=c_2$
in the FAC approach.

\vspace{7mm}

\section*{\bf Conclusions}
\addtocontents{toc}{{\bf Conclusions} \hspace{124mm}  {\bf 72}}

\indent

In the present article we reviewed the
current development of calculational methods, algorithms and
computer programs which allow one to evaluate the characteristics
of the phenomenologically important physical processes
to higher orders of perturbative QCD.
We have considered
$Z\rightarrow \mbox{hadrons}$,
$\tau^{-} \rightarrow \nu_{\tau} +\mbox{hadrons}$,
$H \rightarrow \mbox{hadrons}$. The described methods are
applicable to a wide class of calculational problems of
modern high energy physics. We outlined the analytical three- and
four-loop calculations for the above mentioned processes.

The methods of analytical perturbative calculations available at present
allow, in principle, one to evaluate various decay rates, cross-sections,
coefficient functions in the operator product expansion,
renormalization group functions etc.  up to and including five-loop level.
This would correspond, for instance, the decay rate in the process
$Z \rightarrow \mbox{hadrons}$ to $O(\alpha_s^4)$. It seems that
such a high order
will completely fit the experimental state of the problem in the
observable future. Indeed, for example, the 4\% estimate of the
theoretical error for the decay rate of $Z$-boson  is based on
the $O(\alpha_s^3)$ calculation. The present
experimental error at LEP is about 5\%.

The involvement of the heavier quarks in the physical processes makes it
necessary to develop methods for calculation of the Feynman graphs
with the propagators of massive particles.
The expansion in terms of large or small masses
may not always give satisfactory results.

The problem of the renormalization group ambiguity of the perturbation
theory results and various methods for resummation of higher order
corrections is a subject of growing interest and discussions
in the literature.

The future development of analytical programming tools towards
the full automation of high order calculations would be
welcome. This would greatly reduce the chance of
errors in the calculations.
On the other hand, the computer package  with full
implementation of the algorithm of high order analytical perturbative
calculations would  make it realistic to step up by one more order.

 We recognize that it is unavoidable that some of the relevant
references have not been mentioned. We assure the reader that this is
only due to our unintentional ignorance.

\vspace{1cm}
\noindent
{\bf ACKNOWLEDGMENTS}

\indent
It is a pleasure to thank D.\ Soper for numerous discussions and his support.
We are grateful to  N.\ Deshpande and R.\ Hwa for discussions on
the present status of the Standard Model.
We thank E.\ Braaten, S.\ Brodsky and B.\ Kniehl for their
comments at various stages of this work. We would especially like to
thank P.\ Stevenson for reading the manuscript, illuminating discussions,
suggestions and correcting the errors.
L.R.S. would like to thank members of the experimental high energy physics
laboratory at the University of Oregon, especially J.\ Brau, R.\ Fray and
D.\ Strom for encouraging discussions.
L.R.S. would  like to thank N.\ S.\ Amaglobeli, V.\ A.\ Matveev,
V.\ A.\ Rubakov and A.~N.~Tavkhelidze
for their interest in our work and their support, the
members of the Theory Division of the
Moscow Institute for Nuclear Research for collaboration on various
problems which further became topics of the present article, and
the members of the Department of High Energy Physics, Tbilisi State
University for discussions.
We are grateful to C.\ Quigg for encouraging us to write this review.

\vspace{3mm}

\noindent
This work was supported by the U.S. Department of Energy under grant
No. DE-FG06-85ER-40224 and under grant No. DE-FG05-84ER40215.

\newpage
\noindent
{\bf REFERENCES}

\vspace{6mm}

\noindent
Abers, E.\ S., and B.\ W.\ Lee, 1973, Phys.\ Rep.\ {\bf 9,} 1.\\
Abbot, L., 1980, Phys.\ Rev.\ Lett.\ {\bf 44,} 1569.\\
Acharya, A., and B.\ P.\ Nigam, 1978, Nucl.\ Phys.\ {\bf B 141,} 178.\\
Acharya, A., and B.\ P.\ Nigam, 1985, Nuovo Cim.\ {\bf A 88,} 293.\\
Adler, S.\ L., 1972, Phys.\ Rev.\ {\bf D 5,} 3021.\\
Adler, S.\ L., 1974, Phys.\ Rev.\ {\bf D 10,} 3714.\\
Altarelli, G., 1982, Phys.\ Rep.\ {\bf 81,} 1.\\
Altarelli, G., 1989, Annu.\ Rev.\ Nucl.\ Sci.\ {\bf 39,} 357.\\
Altarelli, G., 1992, ``QCD and experiment: status of $\alpha_s$,''
                                            CERN preprint No.\ TH.6623/92.\\
Altarelli, G., P.\ Nason, and G.\ Ridolfi, 1994,
           A study of ultraviolet renormalon ambiguities in the determination
           of $\alpha_s$ from $\tau$ decay,''
                                            CERN preprint No.\ TH.7537/94.\\
Appelquist, T., and J.\ Carazzone, 1975, Phys.\ Rev.\ {\bf D 11,} 2856.\\
Appelquist, T., and D.\ Politzer, 1975, Phys.\ Rev.\ Lett.\ {\bf 34,} 43.\\
Ashmore, J.\ F., 1972,  Nuov.\ Cimm.\ Lett.\ {\bf 4,} 289.\\
Baker, M., and K.\ Johnson, 1969, Phys.\ Rev.\ {\bf 183,} 1292.\\
Baker, M., and K.\ Johnson, 1971, Phys.\ Rev.\ {\bf D 3,} 2541.\\
Banyai, L., S.\ Marculescu, and T.\ Vescan, 1974, Lett.\ Nuov.\ Cim.\ {\bf 11,}

151.\\
Barclay, D.\ T., and C.\ J.\ Maxwell, 1992a, Phys.\ Rev.\ Lett.\ {\bf 69,}
3417.\\
Barclay, D.\ T., and C.\ J.\ Maxwell, 1992b, Phys.\ Rev.\ {\bf D 45,} 1760.\\
Bardeen, W., A.\ Buras, D.\ Duke, and T.\ Muta, 1978, Phys.\ Rev.\ {\bf 18,}

3998.\\
Barger, V.\ D., and R.\ J.\ N.\ Phillips, 1987, {\it Collider Physics,}
                          Frontiers in Physics Series {\bf 71}
(Addison-Wesley).\\
Barnett, M.\ R., M.\ Dine, and L.\ McLerran, 1980, Phys.\ Rev.\ {\bf D 22,}
594.\\
Barnett, M.\ R., H.\ E.\ Haber, and D.\ E.\ Soper, 1988, Nucl.\ Phys.\ {\bf B
306,}
                                                                     697.\\
Bechi, C., S.\ Narison, E.\ de Rafael, and F.\ Yndurain, 1981, Z.\ Phys.\
                                                               {\bf 1981,}
335.\\
Bernreuter, W., and W.\ Wetzel, 1982, Nucl.\ Phys.\ {\bf B 197,} 228.\\
Bethke, S., 1989, Z.\ Phys.\ {\bf C 43,} 331.\\
Bethke, S., 1992, in {\sl Proceedings of the 26th International Conference on
                          High Energy Physics} (Dallas, USA),  p.\ 81.\\
Bethke, S., and J.\ E.\ Pilcher, 1992, Annu.\ Rev.\ Nucl.\ Sci. {\bf 42,}
251.\\
Bjorken, J.\ D., 1968, in {sl Proceedings of 1967 Int. School of Physics,
                              Enrico Fermi, Course 41,} Varenna, Italy
                              (Academic Press, New York), p.\ 55.\\
Bjorken, J.\ D., 1969, Phys.\ Rev.\ {\bf 179,} 1547.\\
Bloch, F., and A.\ Nordsieck, 1937, Phys.\ Rev.\ {\bf 52,} 54.\\
Bogolyubov, N.\ N., and O.\ S.\ Parasyuk, 1955a,
              Dokl. Akad. Nauk SSSR [Sov.\ Phys.\ Dokl.]  {\bf 100,} 25.\\
Bogolyubov, N.\ N., and O.\ S.\ Parasyuk, 1955b,
              Dokl.\ Akad.\ Nauk SSSR [Sov.\ Phys.\ Dokl.]  {\bf 100,} 429.\\
Bogolyubov, N.\ N., and O.\ S.\ Parasyuk, 1956,
                  Izv.\ Akad.\ Nauk SSSR, ser.\ matem.\ {\bf 20,} 585.\\
Bogolyubov, N.\ N., and O.\ S.\ Parasyuk, 1957, Acta Mathem.\ {\bf 97,} 227.\\
Bogolyubov, N.\ N., and D.\ V.\ Shirkov, 1955, Dokl.\ Akad.\ Nauk SSSR
                                     [Sov.\ Phys.\ Dokl.]  {\bf 103,} 203.\\
Bogolyubov, N.\ N., and D.\ V.\ Shirkov, 1956a, Sov.\ Phys.-JETP
                         (translation of Zh.\ Eksp.\ Teor.\ Fiz.) {\bf 30,}
77.\\
Bogolyubov, N.\ N., and D.\ V.\ Shirkov, 1956b, Nuov.\ Cim. {\bf 3,} 845.\\
Bogolyubov, N.\ N., and D.\ V.\ Shirkov, 1980, {\sl Introduction to
             the Theory of Quantized Fields} (John Wiley \& Sons, Inc.).\\
Bogolyubov, N.\ N., B.\ V.\ Struminsky, and A.\ N.\ Tavkhelidze, 1965,
                                  JINR report No.\ JINR-D-1968.\\
Bollini, C.\ G., and J.\ J.\ Giambiagi, 1972, Phys.\ Lett.\ {\bf 40 B,} 566.\\
Bonneau, G., 1980, Phys.\ Lett.\ {\bf 96 B,} 147.\\
Boos, E.\ E., and A.\ I.\ Davydychev, 1992, Theor.\ Math.\ Phys.\ {\bf 89,}
1052.\\
Braaten, E., 1988, Phys.\ Rev.\ Lett.\ {\bf 60,} 1606.\\
Braaten, E., and J.\ P.\ Leveille, 1980, Phys.\ Rev.\ {\bf D 22,} 715.\\
Braaten, E., C.\ S.\ Li, 1990, Phys.\ Rev.\ {\bf D 42,} 3888.\\
Braaten, E., S.\ Narison, and A.\ Pich, 1992, Nucl.\ Phys.\ {\bf B 373,} 581.\\
Broadhurst, D.\ J., and S.\ G.\ Generalis, 1982, ``Pseudoscalar QCD sum
rules,''
                                Open University preprint No.\ {\bf
OUT-4102-8}.\\
Broadhurst, D.\ J., and S.\ G.\ Generalis, 1985, Phys.\ Lett.\ {\bf 165 B,}
175.\\
Broadhurst, D.\ J., N.\ Gray and K.\ Schilcher, 1991, Z.\ Phys.\ {\bf C 52,}
111.\\
Broadhurst, D.\ J., A.\ L.\ Kataev, and O.\ V.\ Tarasov, 1993, Phys.\ Lett.\
                                                     {\bf B 298,} 445.\\
Broadhurst, D.\ J., {\it et al.}, 1994, Phys.\ Lett.\ {\bf B 329,} 103.\\
Brock, R., {\it et al.}, CTEQ Collaboration, 1993, {\sl Handbook of
Perturbative
                   QCD,} Edited by G.\ Sterman.\\
Brodsky, S.\ J., 1993, ``New perspective in Quantum Chromodynamics,''
                         SLAC preprint No.\ SLAC-PUB-6304.\\
Brodsky, S.\ J., and H.\ J.\ Lu, 1992,
           ``On the selfconsistency of scale setting methods,''
                                           SLAC preprint No.\ SLAC-PUB-6000.\\
Brodsky, S.\ J., and H.\ J.\ Lu, 1994, ``Commensurate scale relations:
      precise tests of Quantum Chromodynamics without scale or scheme
ambiguity,''
                                               SLAC preprint No.\
SLAC-PUB-6683.\\
Brodsky, S.\ J., and H.\ J.\ Lu, 1995, Phys.\ Rev.\ {\bf D 51,} 3652.\\
Brodsky, S.\ J., and G.\ R.\ Ferrar, 1973, Phys.\ Rev.\ Lett.\ {\bf 31,}
1153.\\
Brodsky, S.\ J., G.\ P.\ Lepage, and P.\ B.\ Mackenzie, 1983,
                                                 Phys.\ Rev.\ {\bf D 28,}
228.\\
Brown, L.\ S., and L.\ G.\ Yaffe, 1992, Phys.\ Rev.\ {\bf D 45,} 398.\\
Brown, L.\ S., L.\ G.\ Yaffe, and C.\ X.\ Zhai, 1992,
       ``Large order perturbation theory for the electromagnetic
current-current
         correlation function,''
         Washington University preprint No.\ UW-PT-92-07.\\
Buckley, I.\ R.\ C., A.\ H.\ Duncan, and H.\ F.\ Jones, 1993,
                                       Phys.\ Rev.\ {\bf D 47,} 2554.\\
Buras, A.\ J., E.\ G.\ Floratos, D.\ A.\ Ross, and C.\ T.\ Sachrajda, 1977,
                                         Nucl.\ Phys.\ {\bf B 131,} 308.\\
Callan, C., 1970, Phys.\ Rev.\ {\bf D 2,} 1541.\\
Caswell, W.\ E., and F.\ Wilczek, 1974, Phys.\ Lett.\ {\bf B 49,} 291.\\
Celmaster, W., and R.\ G.\ Gonsalves, 1979, Phys.\ Rev.\ {\bf D 20,} 1420.\\
Celmaster, W., and R.\ G.\ Gonsalves, 1980, Phys.\ Rev.\ Lett.\ {\bf 44,}
560.\\
Celmaster, W., and P.\ M.\ Stevenson, 1983, Phys.\ Lett.\ {\bf B 125,} 493.\\
Chetyrkin, K.\ G., 1988, Teor.\ Mat.\ Fiz.\ {\bf 76,} 207
                             [Theor.\ Math.\ Phys.\ {\bf 76,} 809 (1988)].\\
Chetyrkin, K.\ G., 1991,
``Combinatorics of R, R$^{-1}$, and R$^{\ast}$ operations and asymptotic
expansions of Feynman integrals in the limit of large momenta and masses,''
                        Max Planck Institute preprint No.\ MPI-PAE/PTh 13/91.\\
Chetyrkin, K.\ G., 1992, Phys.\ Lett.\ {\bf B 282,} 221.\\
Chetyrkin, K.\ G., 1993a, Phys.\ Lett.\ {\bf B 307,} 169.\\
Chetyrkin, K.\ G., 1993b, ``Possible and impossible in multiloop
renormalization
                    group,'' Karlsruhe University preprint No.\ TTP93-37.\\
Chetyrkin, K.\ G., S.\ G.\ Gorishny, and V.\ P.\ Spiridonov, 1985, Phys.\
Lett.\
                                                             {\bf B 160,}
149.\\
Chetyrkin, K.\ G., S.\ G.\ Gorishny ,and F.\ V.\ Tkachov, 1982, Phys.\ Lett.\
                                                              {\bf B 119,}
407.\\
Chetyrkin, K.\ G., A.\ L.\ Kataev, and F.\ V.\ Tkachov, 1979, Phys.\ Lett.\
                                                                 {\bf 85,}
277.\\
Chetyrkin, K.\ G., A.\ L.\ Kataev, and F.\ V.\ Tkachov, 1980, Nucl.\ Phys.\
                                                              {\bf B 174,}
345.\\
Chetyrkin, K.\ G., and A.\ Kwiatkowski, 1993, Z.\ Phys.\ {\bf C 59,} 525.\\
Chetyrkin, K.\ G., and A.\ Kwiatkowski, 1995, ``Second order QCD corrections to
                scalar and pseudoscalar Higgs decays into massive bottom
quarks,''
                LBL preprint No.\ LBL-37269.\\
Chetyrkin, K.\ G., and J.\ H.\ K\"{u}hn, 1990, Phys.\ Lett.\ {\bf B 248,}
359.\\
Chetyrkin, K.\ G., and J.\ H.\ K\"{u}hn, 1992, Phys.\ Lett.\ {\bf B 282,}
359.\\
Chetyrkin, K.\ G., J.\ H.\ K\"{u}hn, and A.\ Kwiatkowski, 1992, Phys.\ Lett.\
                                                          {\bf B 282,} 221.\\
Chetyrkin, K.\ G., and F.\ V.\ Tkachov, 1979, ``New approach to evaluations
                                                of multiloop Feynman
diagrams,''
                     Moscow Institute for Nuclear Research preprint No.
P-0018.\\
Chetyrkin, K.\ G., and F.\ V.\ Tkachov, 1981, Nucl.\ Phys.\ {\bf B 192,} 159.\\
Chetyrkin, K.\ G., and F.\ V.\ Tkachov, 1982, Phys.\ Lett.\ {\bf B 114,} 340.\\
Chyla, J., A.\ L.\ Kataev, and S.\ A.\ Larin, 1991, Phys.\ Lett.\
                                                            {\bf B 267,} 269.\\
Cicuta, G.\ M., and E.\ Montaldi, 1972, Nuov.\ Cimm.\ Lett.\ {\bf 4,} 329.\\
Collins, J.\ C., 1974, Nucl.\ Phys.\ {\bf B 80,} 341.\\
Collins, J.\ C., 1984, {\sl Renormalization}
                                  (Cambridge University Press, Cambridge,
UK).\\
Collins, J.\ C., A.\ Duncan, and S.\ D.\ Joglekar, 1977, Phys.\ Rev.\
                                                              {\bf D 16,}
438.\\
Collins, J.\ C., A.\ J.\ Macfarlane, 1974, Phys.\ Rev.\ {\bf D 10,} 1201.\\
Collins, J.\ C., and D.\ E.\ Soper, 1987, Annu.\ Rev.\ Nucl.\ Sci.\
                                                                {\bf 37,}
383.\\
Collins, J.\ C., D.\ E.\ Soper, and G.\ Sterman, 1983, in
     {\sl Proceedings of the 18th Rencontres de Moriond,} Edited
                                             by J.~Tran Thanh Van, p.\ 157.\\
Collins, J.\ C., D.\ E.\ Soper, and G.\ Sterman, 1984,
                                              Phys.\ Lett.\ {\bf B 134,} 263.\\
Collins, J.\ C., D.\ E.\ Soper, and G.\ Sterman, 1985,
                                              Nucl.\ Phys.\ {\bf B 261,} 104.\\
Collins, J.\ C., D.\ E.\ Soper, and G.\ Sterman, 1989, in
                       {\sl Perturbative Quantum Chromodinamics,}
                        Edited by A.\ H.\ Muller (World Scientific), p.\ 1.\\
Collins, J.\ C., F.\ Wilczek, and A.\ Zee, 1978, Phys.\ Rev.\ {\bf D 18,}
242.\\
Davydychev A.\ I., 1991, J.\ Math.\ Phys.\ {\bf 32,} 1052.\\
de~Rafael, E., and J.\ L.\ Rosner, 1974, Annals of Phys.\ {\bf 82,} 369.\\
de~R\'{u}jula, A., and H.\ Georgi, 1976, Phys.\ Rev.\ {\bf D 13,} 1296.\\
Delbourgo, R., and D.\ A.\ Akyeampong, 1974, Nuov.\ Cim.\ {\bf A 19,} 219.\\
de~Witt, B., 1967, Phys.\ Rev.\ {\bf 162,} 1195.\\
Diberder, F.\ L., and A.\ Pich, 1992a, Phys.\ Lett.\ {\bf B 286,} 147.\\
Diberder, F.\ L., and A.\ Pich, 1992b, Phys.\ Lett.\ {\bf B 289,} 165.\\
Dine, M., and J.\ Sapirstein, 1979, Phys.\ Rev.\ Lett.\ {\bf 43,} 668.\\
Dingle, R.\ B., 1973, {\it Asymptotic Expansions: Their derivation and
                              Interpretation} (Academic Press, New York).\\
Drees, M., and K.\ Hikasa, 1990, Phys.\ Rev.\ {\bf D 41,} 1547.\\
Drell, S.\ D., and T.\ M.\ Yan, 1971, Ann.\ Phys.\  {\bf 66,} 578.\\
Duncan, A.\ H., {\it et al.}, 1993,  Phys.\ Rev.\ Lett.\ {\bf 70,} 4159.\\
Efremov, A.\ V., and A.\ V.\ Radyushkin, 1980a, Teor.\ Mat.\ Fiz.\ {\bf 44,} 17
                             [Theor.\ Math.\ Phys.\ {\bf 44,} 573 (1980)].\\
Efremov, A. V., and A. V. Radyushkin, 1980b, Teor.\ Mat.\ Fiz.\ {\bf 44,} 157
                             [Theor.\ Math.\ Phys.\ {\bf 44,} 664 (1981)].\\
Ellis, J., M.\ Karliner, and M.\ Samuel, 1995, Phys.\ Rev.\ Lett.\\
Ellis, R.\ K., 1993, in {\sl Proceedings of the 7th 1992 Fermilab
                             Meeting of the American Physical Society,}
                             edited by C.\ H.\ Albright {\it et al.},
                             (World Scientific) p.\ 167.\\
Ellis, R.\ K., and W.\ J.\ Stirling, 1990, ``QCD AND COLLIDER PHYSICS,''
                                 Fermilab preprint No.\
FERMILAB-Conf-90/164-T.\\
Faddeev, L.\ D, and U.\ N.\ Popov, 1967, Phys.\ Lett.\ {\bf B 25,} 29.\\
Faddeev, L.\ D, and A.\ A.\ Slavnov, 1980, {\it Gauge Fields: Introduction
                                     to Quantum Theory} (Benjamin, New York).\\
Feynman, R., 1963, Acta Phys.\ Polonica {\bf 26,} 697.\\
Feynman, R., 1969, Phys.\ Rev.\ Lett.\ {\bf 23,} 1415.\\
Feynman, R., 1972, {\it Photon Hadron Interactions} (Benjamin, New York).\\
Field, J.\ H., 1993, Ann.\ Phys.\ {\bf 226,} 209.\\
Fritzsch, H., M.\ Gell-Mann, and H.\ Leutwyler, 1973, Phys.\ Lett.\
                                                             {\bf B 47,} 365.\\
Furry, W., 1937, Phys.\ Rev.\ {\bf 51,} 125.\\
Gell-Mann, M., 1964, Phys.\ Lett., {\bf 8,} 214.\\
Gell-Mann, M., and F.\ Low, 1954, Phys.\ Rev.\ {\bf 95,} 1300.\\
Georgi, H., H.\ D.\ Politzer, 1976, Phys.\ Rev.\ {\bf D 14,} 1829.\\
Glashow, S.\ L., J.\ Iliopoulos, and L.\ Maiani, 1970, Phys.\ Rev.\
                                                            {\bf D 2,} 1285.\\
Gorishny, S.\ G., A.\ L.\ Kataev, S.\ A.\ Larin, 1986, Nuov.\ Cim.\ {\bf A 92,}

119.\\
Gorishny, S.\ G., A.\ L.\ Kataev, S.\ A.\ Larin, 1988, Phys.\ Lett.\ {\bf B
212,}
                                                                         238.\\
Gorishny, S.\ G., A.\ L.\ Kataev, S.\ A.\ Larin, 1990, ``Four-loop QED
                                  $\beta$ function'' (private
communications).\\
Gorishny, S.\ G., A.\ L.\ Kataev, S.\ A.\ Larin, 1991, Phys. Lett. {\bf B 259,}

144.\\
Gorishny, S.\ G., A.\ L.\ Kataev, S.\ A.\ Larin, and L.\ R.\ Surguladze, 1990,
                                          Mod.\ Phys.\ Lett.\ {\bf A 5,}
2703.\\
Gorishny, S.\ G., A.\ L.\ Kataev, S.\ A.\ Larin, and L.\ R.\ Surguladze, 1991a,
                                                Phys.\ Lett.\ {\bf B 256,}
81.\\
Gorishny, S.\ G., A.\ L.\ Kataev, S.\ A.\ Larin, and L.\ R.\ Surguladze, 1991b,
                                                 Phys.\ Rev.\ {\bf D 43,}
1633.\\
Gorishny, S.\ G., A.\ L.\ Kataev, S.\ A.\ Larin, and L.\ R.\ Surguladze, 1991c,
                 in {\sl Proceedings of the  International Seminar
``QUARKS-90''}
         (Telavi, Georgia, USSR, May 1990) edited by V.~A.~Matveev {\it et
al.},
                                                    (World Scientific) p.\
194.\\
Gorishny, S.\ G., and  S.\ A.\ Larin, 1987, Nucl.\ Phys.\ {\bf B 283,} 452.\\
Gorishny, S.\ G., S.\ A.\ Larin, L.\ R.\ Surguladze, and F.\ V.\ Tkachov, 1989,
                                        Comput.\ Phys.\ Commun.\ {\bf 55,}
381.\\
Gorishny, S.\ G., S.\ A.\ Larin, and F.\ V.\ Tkachov, 1983,
                                               Phys.\ Lett.\ {\bf B 124,}
217.\\
Greenberg, O.\ W., 1964, Phys.\ Rev.\ Lett.\ {\bf 13,} 598.\\
Gross, D., and  F.\ Wilczek, 1973, Phys.\ Rev.\ Lett.\ {\bf 30,} 1343.\\
Grunberg, G., 1980, Phys.\ Lett.\ {\bf B 95,} 70.\\
Grunberg, G., 1982, Phys.\ Lett.\ {\bf B 110,} 501.\\
Grunberg, G., 1984, Phys.\ Rev.\ {\bf D 29,} 2315.\\
Grunberg, G., and A.\ L.\ Kataev, 1992, Phys.\ Lett.\ {\bf B 279,} 352.\\
Han, M.\ Y., and Y.\ Nambu, 1965, Phys.\ Rev.\ {\bf 139,} 1005.\\
Hearn, A.\ C., 1973, ``{\small REDUCE}, {\sl User's Manual}
                                     (University of Utah), Report No.\
UCP-19.\\
Hoang, A.\ H., M.\ Jezabek, J.\ H.\ K\"{u}hn, and T.\ Teubner, 1994,
                                           Phys.\ Lett.\ {\bf B 338,} 330.\\
Inami, T., and T.\ Kubota, 1981, Nucl.\ Phys.\ {\bf B 179,} 171.\\
Johnson, K., R.\ Willey, and M.\ Baker, 1967, Phys.\ Rev.\ {\bf 163,} 1699.\\
Johnson, K., and M.\ Baker, 1973, Phys.\ Rev.\ {\bf D 8,} 1110.\\
Kartvelishvili, V., and M.\ Margvelashvili, 1995, Phys.\ Lett.\
                                                    {\bf B 345,} 161.\\
Kataev, A.\ L., 1990, ``Next-next-to-leading perturbative QCD corrections:
                        the current status of investigations,''
                        Montpellier Preprint No.\ PM/90-41.\\
Kataev, A.\ L., 1991, Nucl.\ Phys.\ B (Proc.\ Suppl.) {\bf  A 23,} 72.\\
Kataev, A.\ L., and V.\ V.\ Starshenko, 1994, CERN Preprint No.\ TH.7400/94.\\
Kleinert, H., {\it et al.}, 1991, Phys.\ Lett.\ {\bf B 272,} 39.\\
Kniehl, B.\ A., 1990, Phys.\ Lett.\ {\bf B 237,} 127.\\
Kniehl, B.\ A., 1994a, Phys.\ Rep.\ {\bf 240,} 211.\\
Kniehl, B.\ A., 1994b,  in {\sl Proceedings of the 1994
                        Tennessee International Symposium on Radiative
                        Corrections: Status and Outlook,} (to be published);
                        Bulletin Board: hep-ph/9410391.\\
Kniehl, B.\ A., 1995a, Phys.\ Lett.\ {\bf B 343,} 299.\\
Kniehl, B.\ A., 1995b, Int.\ J.\ Mod.\ Phys.\  {\bf A 10,} 443.\\
Kniehl, B.\ A., and J.\ H.\ K\"{u}hn, 1989, Phys.\ Lett.\ {\bf B 224,} 229.\\
Kniehl, B.\ A., and J.\ H.\ K\"{u}hn, 1990, Nucl.\ Phys.\ {\bf B 329,} 547.\\
Kotikov, A.\ V., 1991, Mod.\ Phys.\ Lett.\ {\bf A 6,} 677.\\
Kramer, G., and B.\ Lampe, 1988, Z.\ Phys.\ {\bf C 39,} 101.\\
Krasnikov, N.\ V., and A.\ A.\ Pivovarov, 1982, Phys.\ Lett.\ {\bf B 116,}
168.\\
Krasnikov, N.\ V., A.\ A.\ Pivovarov, and N.\ N.\ Tavkhelidze, 1983,
                                                 Z.\ Phys.\ {\bf C 19,} 301.\\
Krasnikov, N.\ V., and N.\ N.\ Tavkhelidze, 1982,
       ``The contribution of instantons into cross-section of the $e^{+}e^{-}$
          annihilation into hadrons'' Moscow Institute for Nuclear Research
                                                       Preprint No.\ P-227.\\
Lam, C.\ S., and T.\ M.\ Yan, 1977, Phys.\ Rev.\ {\bf D 16,} 703.\\
Langacker, P., and L.\ Mingxing, and A.\ K.\ Mann, 1992, Rev.\ Mod.\ Phys.\
                                                               {\bf 64,} 87.\\
Lanin L.\ V., V.\ P.\ Spiridonov, and K.\ G.\ Chetyrkin, 1986,
                                                  Yad.\ Fiz.\ {\bf 44,} 1374.\\
Larin S.\ A., 1993,  Phys.\ Lett.\ {\bf B 303,} 113.\\
Lee, B.\ W., and J.\ Zinn-Justin, 1972, Phys.\ Rev.\ {\bf D 5,} 3121.\\
Lee, B.\ W., and J.\ Zinn-Justin, 1973, Phys.\ Rev.\ {\bf D 7,} 1049.\\
Le Guillou, J.\ C., and J.\ Zinn-Justin, 1990, Eds., {\sl Large-Order
              Behaviour of Perturbation Theory} (Elsevier Science Publishers
                                          B.\ V., North-Holland, Amsterdam).\\
Leibbrandt, G., 1975, Rev.\ Mod.\ Phys.\ {\bf 47,} 849.\\
Libby, S.\ B., and G.\ Sterman, 1978, Phys.\ Rev.\ {\bf D 18,} 3252.\\
Logunov, A.\ A., L.\ D.\ Soloviov, and A.\ N.\ Tavkhelidze, 1967,
                                      Phys.\ Lett.\ {\bf B 24,} 181.\\
Loladze, G.\ T., L.\ R.\ Surguladze, and F.\ V.\ Tkachov, 1984,
                     Bull.\ Acad.\ Sci.\ Georgian SSR {\bf 116,} 509.\\
Loladze, G.\ T., L.\ R.\ Surguladze, and F.\ V.\ Tkachov, 1985, Phys.\ Lett.\
                                                            {\bf B 162,} 363.\\
Lovett-Turner, C.\ N., and C.\ J.\ Maxwell, 1994,
                                                Nucl.\ Phys.\ {\bf B 432,}
147.\\
Lu, H.\ J., and C.\ A.\ R.\ de Melo, 1991,  Phys.\ Lett.\ {\bf B 273,} 260.\\
Mandelstam, S., 1968, Phys.\ Rev.\ {\bf 175,} 1580.\\
Marciano, W.\ J., 1975, Phys.\ Rev.\ {\bf D 12,} 3861.\\
Marciano, W.\ J., 1984, Phys.\ Rev.\ {\bf D 29,} 580.\\
Marciano, W.\ J., and H.\ Pagels, 1978, Phys.\ Rep.\ {\bf C 36,} 137.\\
Marciano, W.\ J., and A.\ Sirlin, 1988, Phys.\ Rev.\ Lett.\ {\bf 61,} 1815.\\
Marciano, W.\ J., 1991, {\sl Annu.\ Rev.\ Nucl.\ Sci.\ } {\bf 41,} 469.\\
Marciano, W.\ J., 1992, ``$\tau$ decays: a theoretical perspective,''
                       Brookhaven National Laboratory Preprint No.\
BNL-48179.\\
Marciano, W.\ J., 1993a, in {\sl Proceedings of the 7th 1992 Fermilab Meeting
                                of the American Physical Society,} edited by
                                C.\ H.\ Albright {\it et al.}
                                (World Scientific), p.\ 185.\\
Marciano, W.\ J., 1993b,  ``Standard Model Status'',
           Brookhaven National Laboratory Preprint No.\ BNL-48760.\\
Matveev, V.\ A., R.\ M.\ Muradyan, and A.\ N.\ Tavkhelidze, 1970,
     Fiz.\ Elem.\ Chastits At Yadra {\bf 1,} 91 [Sov.\ J.\ Part.\ Nucl.].\\
Matveev, V.\ A., R.\ M.\ Muradyan, and A.\ N.\ Tavkhelidze, 1972,
                 Lett.\ Nuov.\ Cim.\ {\bf 5,} 907.\\
Matveev, V.\ A., R.\ M.\ Muradyan, and A.\ N.\ Tavkhelidze, 1973,
                 Lett.\ Nuov.\ Cim.\ {\bf 7,} 719.\\
Mattingly, A.\ C., and P.\ M.\ Stevenson, 1992, Phys.\ Rev.\ Lett.\
                                                          {\bf 69,} 1320.\\
Mattingly, A.\ C., and P.\ M.\ Stevenson, 1994, Phys.\ Rev.\ {\bf D 49,} 437.\\
Maxwell, C.\ J., and J.\ A.\ Nicholls, 1990, Phys.\ Lett.\ {\bf B 236,} 63.\\
Miamoto, Y., 1965, Prog.\ Theor.\ Phys.\ Suppl.\ Extra {\bf 187}.\\
Monsay, E.\ and C.\ Rosenzweig, 1981, Phys.\ Rev.\ {\bf D 23,} 1217.\\
Mueller, A.\ H., 1978, Phys.\ Rev.\ {\bf D 18,} 3705.\\
Mueller, A.\ H., 1981, Phys.\ Rep.\ {\bf 73,} 237.\\
Mueller, A.\ H., 1992, in
                {\sl Proceedings of the Workshop QCD-Twenty Years Later,}
                edited by P.\ M.\ Zerwas, and H.\ A.\ Kastrup,
                (World Scientific) {\bf 1,} p.\ 162.\\
Muta, T., 1987, {\sl Foundations of Quantum Chromodinamics,}
                     Lecture Notes in Physics Vol.\ 5 (World Scientific).\\
Nason, P., and M.\ Porrati, 1994, Nucl.\ Phys.\ {\bf B 421,} 518.\\
Narison, S., 1981a, Phys.\ Lett.\ {\bf B 104,} 485.\\
Narison, S., 1981b, Nucl.\ Phys.\ {\bf B 182,} 59.\\
Narison, S., 1982, Phys.\ Rep.\ {\bf 84,} 263.\\
Narison, S., 1986, ``QCD duality sum rules: introduction and some recent
                     developments,'' CERN Preprint No.\ TH.4624/86.\\
Narison, S., 1994, ``$\alpha_s$ from tau decays'',
                                      CERN Preprint No.\ TH.7506/94.\\
Narison, S., and E.\ de Rafael, 1980, Nucl.\ Phys.\ {\bf B 169,} 253.\\
Narison, S., and E.\ de Rafael, 1981, Phys.\ Lett.\ {\bf B 103,} 57.\\
Narison, S., A.\ Pich, 1988, Phys.\ Lett.\ {\bf B 211,} 183.\\
Narison, S., and R.\ Tarrach, 1983, Phys.\ Lett.\ {\bf B 125,} 217.\\
Nielsen, N.\ K., 1977, Nucl.\ Phys.\ {\bf B 120,} 212.\\
Novikov, V.\ A.,  {\it et al.}, 1978, Phys.\ Rep.\  {\bf 41,} 1.\\
Novikov, V.\ A., M.\ A.\ Shifman, A.\ I.\ Vainshtein, and V.\ I.\ Zakharov,
                                         1985, Nucl.\ Phys.\ {\bf B 249,}
445.\\
Ovsyannikov, L.\ V., 1956, Dok.\ Akad.\ Nauk SSSR {\bf 109,} 112
                                                    [Sov.\ Phys.\ Dokl.].\\
Pauli, W., and F.\ Villars, 1949, Rev.\ Mod.\ Phys.\ {\bf 21,} 433.\\
Pennington M.\ R., and G.\ G.\ Ross, 1982, Phys.\ Lett.\ {\bf B 102,} 167.\\
Peterman, A., 1979, Phys.\ Rep.\ {\bf 53,} 159.\\
Pich, A., 1990, ``Hadronic tau decays and QCD,'' CERN Preprint No.\
TH.5940/90.\\
Pich, A., 1991, in {\sl Heavy flavours,} edited by A.~J.~Buras and
                                          M.~Lindner (CERN, Geneva), p.\ 375.\\
Pich, A., 1992a, ``Tau physics and tau charm factories,''
                  CERN Preprint No.\ TH.6672/92.\\
Pich, A., 1992b, ``QCD predictions for the tau hadronic width and determination
                   of $\alpha_s(M_{\tau}^2)$,'' CERN Preprint No.\
TH.6738/92.\\
Pich, A., 1994a, ``QCD predictions for the tau hadronic width: determination
                   of $\alpha_s(M_{\tau}^2)$,''
                   Val\`{e}ncia University Preprint No.\ FTUV/94-71.\\
Pich, A., 1994b, ``The Standard Model of electroweak interactions,''
                   Val\`{e}ncia University Preprint No.\ FTUV/94-62.\\
Pivovarov, A.\ A., and L.\ R.\ Surguladze, 1991, Nucl.\ Phys.\ {\bf B 360,}
97.\\
Pivovarov, A.\ A., 1992a, Nuovo Cim.\ {\bf A 105,} 813.\\
Pivovarov, A.\ A., 1992b, Z.\ Phys.\ {\bf C 53,} 461.\\
Pivovarov, G.\ B., and F.\ V.\ Tkachov, 1988, Teor.\ Mat.\ Fiz.\ {\bf 77,} 51
                                   [Theor Math. Phys. {\bf 77,} 1038 (1988)].\\
Pivovarov, G.\ B., and F.\ V.\ Tkachov, 1993, Int.\ J.\ Mod.\ Phys.\
                                                          {\bf A 8,} 2241.\\
Poggio, E., H.\ Quinn, and S.\ Weinberg, 1976, Phys.\ Rev.\ {\bf D 13,} 1958.\\
Politzer, H.\ D., 1973, Phys.\ Rev.\ Lett.\ {\bf 30,} 1346.\\
Pumplin, J., 1989, Phys.\ Rev.\ Lett.\ {\bf 63,} 576.\\
Pumplin, J., 1990, Phys.\ Rev.\ {\bf D 41,} 900.\\
Quigg C., 1983, {\sl Gauge Theories of the Strong, Weak and Electromagnetic
                 interactions,} Frontiers In Physics 56
                 (Benjamin).\\
Raczka, P.\ A., 1995, Z.\ Phys.\ {\bf C 65,} 481.\\
Radyushkin, A.\ V., 1982, ``Optimized $\Lambda$ -parametrization for
                             the QCD running coupling constant in spacelike
                             and timelike regions,''
                             Dubna Joint Institute for Nuclear Research
Preprint
                             No.\  E2-82-159.\\
Radyushkin, A.\ V., 1983, Fiz. Elem. Chastits At Yadra {\bf 14,} 58
                                                   [Sov.J. Part. Nucl.].\\
Reinders, L.\ J., H.\ R.\ Rubinstein, and S.\ Yazaki, 1985,
                                               Phys.\ Rept.\ {\bf 127,} 1.\\
Reya, E., 1981, Phys.\ Rep.\ {\bf 69,} 195.\\
Rodrigo, G., and A.\ Santamaria, 1993, Phys.\ Lett.\ {\bf B 313,} 441.\\
Sakai, N., 1980, Phys.\ Rev.\ {\bf D 22,} 2220.\\
Salam, A., 1969, in {\sl Elementary particle Theory,}
           edited by N.\ Svartholm (Almqvist \& Wiksells, Stockholm), p.\
367.\\
Samuel, M.\ A., and G.\ Li, 1994a, Int.\ J.\ Theor.\ Phys.\ {\bf 33,} 1461.\\
Samuel, M.\ A., and G.\ Li, 1994b, Phys.\ Lett.\ {\bf B 331,} 114.\\
Samuel, M.\ A., and G.\ Li, 1994c, Int.\ J.\ Theor.\ Phys.\ {\bf 33,} 2207.\\
Samuel, M.\ A., G.\ Li, and E.\ Steinfelds, 1994a,
    ``On estimating perturbative coefficients in quantum field theory
      and statistical physics,''
      Oklahoma State University Preprint No.\ RN-278.\\
Samuel, M.\ A., G.\ Li, and E.\ Steinfelds, 1994b, Phys.\ Lett.\
                                                       {\bf B 323,} 188.\\
Samuel, M.\ A., G.\ Li, and E.\ Steinfelds, 1994c, Phys.\ Rev.\ {\bf D 48,}
869.\\
Samuel, M.\ A., and L.\ R.\ Surguladze, 1991, Phys.\ Rev.\ {\bf D 44,} 1602.\\
Schilcher, K., and M.\ D.\ Tran, 1984, Phys.\ Rev.\ {\bf D 29,} 570.\\
Shankar, R., 1977, Phys.\ Rev.\ {\bf D 15,} 755.\\
Shifman, M.\ A., 1992, {\sl Vacuum Structure and QCD Sum Rules,} (Elsevier
                                                       Science Publishers).\\
Shifman, M.\ A., A.\ I.\ Vainshtein, and V.\ I.\ Zakharov, 1979, Nucl.\ Phys.\
                                                            {\bf B 147,} 385.\\
Shirkov, D.\ V., 1980, ``Three loop approximation for running coupling constant
in
                         Quantum Chromodynamics,''
                         Dubna Joint Institute for Nuclear Research Preprint
                                                No.\  E2-80-609.\\
Shirkov, D.\ V., 1992, ``Historical remarks on the renormalization group,''
                    Max Planck Institute Preprint No.\ MPI-PAE/PTh 55/92.\\
Sirlin, A., 1993a, ``Universality of the weak interactions,''
                     New York University Preprint No.\ 93-0526.\\
Sirlin, A., 1993b, ``Status of the standard electroweak model,''
                     New York University Preprint No.\ NYU-TH-93-06-04.\\
Smirnov V.\ A., 1990, Commun.\ Math.\ Phys.\ {\bf 134,} 109.\\
Smirnov V.\ A., 1991, {\sl Renormalization and Asymptotic Expansions}
                                                             (Birkhauser).\\
Smirnov, V.\ A., and K.\ G.\ Chetyrkin, 1985, Teor. Mat. Fiz. {\bf 63,} 208
                             [Theor.\ Math.\ Phys.\ {\bf 63,} 462 (1985)].\\
Soper, D.\ E., 1995, in {\sl Proceedings of the XXXth Rencontres de Moriond
                             ``QCD and High Energy Interactions''}
                                                   (Les Arcs, France).\\
Soper, D.\ E., and L.\ R.\ Surguladze, 1994, Phys.\ Rev.\ Lett.\ {\bf 73,}
2958.\\
Soper, D.\ E., and L.\ R.\ Surguladze, 1995, in {\sl Proceedings of the XXXth
                                                        Rencontres de Moriond
                                          ``QCD and High Energy Interactions''}
                                                            (Les Arcs,
France).\\
Soper, D.\ E., and L.\ R.\ Surguladze, 1995 (in preparation).\\
Speer, E.\ R., 1974, J.\ Math.\ Phys.\ {\bf 15,} 1.\\
Spiridonov, V.\ P., 1984, ``Anomalous dimension of $G^2$ and $\beta$
function,''
                    Moscow Institute for Nuclear Research Preprint No.\
P-378.\\
Spiridonov, V.\ P., 1987, Yad. Fiz. {\bf 46,} 302 [Sov.\ J.\ Nucl.\ Phys.].\\
Stevenson, P.\ M., 1981a, Phys.\ Lett.\ {\bf B 100,} 61.\\
Stevenson, P.\ M., 1981b, Phys.\ Rev.\ {\bf D 23,} 2916.\\
Stevenson, P.\ M., 1982,  Nucl.\ Phys.\ {\bf B 203,} 472.\\
Stevenson, P.\ M., 1984,  Nucl.\ Phys.\ {\bf B 231,} 65.\\
Stevenson, P.\ M., 1992, ``Response to Brodsky and Lu's Letter:
                           On the selfconsistency of scale setting methods,''
                           Rice University Preprint No.\ DOE-ER-40717-2;
                           Bulletin Board: hep-ph/9211327.\\
Stevenson, P.\ M., 1994, (private communication).\\
Strubbe, H., 1974, Comput.\ Phys.\ Commun.\ {\bf 8,} 1.\\
Stueckelberg, E.\ C.\ G., and A.\ Peterman, 1953, Helv.\ Phys.\ Acta {\bf 26,}
                                                                         499.\\
Surguladze, L.\ R., 1989a, ``$O(m^2)$ contributions to correlators of quark
                             currents: three-loop approximation,''
                             Moscow Institute for Nuclear Research
                             Preprint No.\ P-639.\\
Surguladze, L.\ R., 1989b, ``Structure of the program for multiloop
                             calculations in quantum field theory
                             on the {\small SCHOONSCHIP} system,''
                             Moscow Institute for Nuclear Research
                             Preprint No.\ P-643.\\
Surguladze, L.\ R., 1989c, ``Program {\small MINCER} in Four-loop
calculations''
                                                               (unpublished).\\
Surguladze, L.\ R., 1989d, Yad.\ Fiz.\ {\bf 50,} 604
                             [Sov.\ J.\ Nucl.\ Phys.\ {bf 50,} 372 (1989)].\\
Surguladze, L.\ R., 1990,  ``Four-loop QED $\beta$ function'' (unpublished).\\
Surguladze, L.\ R., 1992,  ``A program for analytical perturbative calculations
                             in high energy physics up to four loops for the
                             {\small FORM} system,''
                             Fermilab Preprint No.\  FERMILAB-PUB 92/191-T.\\
Surguladze, L.\ R., 1994a, Phys.\ Lett.\ {\bf B 338,} 229.\\
Surguladze, L.\ R., 1994b, Phys.\ Lett.\ {\bf B 341,} 60.\\
Surguladze, L.\ R., 1994c, ``Quark mass corrections to the Z boson decay
rates,''
                             University of Oregon Preprint No.\ OITS-554.\\
Surguladze, L.\ R., 1994d, Int.\ J.\ Mod.\ Phys.\ {\bf C 5,} 1089.\\
Surguladze, L.\ R., and F.\ V.\ Tkachov, 1986, ``Three-loop coefficient
functions
                                  of gluon and quark condensates in QCD sum
rules
                                  for light mesons,''
                                Moscow Institute for
                                Nuclear Research
                                Preprint No.\ P-501.\\
Surguladze, L.\ R., and F.\ V.\ Tkachov, 1988, Teor.\ Mat.\ Fiz.\ {\bf 75,} 245
                                  [Theor.\ Math.\ Phys.\ {\bf 75,} 502
(1988)].\\
Surguladze, L.\ R., and F.\ V.\ Tkachov, 1989a,
                               Comp.\ Phys.\ Commun.\ {\bf 55,} 205.\\
Surguladze, L.\ R., and F.\ V.\ Tkachov, 1989b, Mod.\ Phys.\ Lett.\
                                                        {\bf A 4,} 765.\\
Surguladze, L.\ R., and F.\ V.\ Tkachov, 1990, Nucl.\ Phys.\ {\bf B 331,} 35.\\
Surguladze, L.\ R., and M.\ A.\ Samuel, 1991a, in {\sl Proceedings of the
                International Conference  Beyond the Standard Model II}
                (Norman, OK, USA, 1990),
                edited by K.\ Milton, R.\ Kantowski, and M.\ A.\ Samuel
                (World Scientific), p.\ 206.\\
Surguladze, L. R., and M. A. Samuel, 1991b, Phys. Rev. Lett. {\bf 66,} 560.\\
Surguladze, L.\ R., and M.\ A.\ Samuel, 1992a, ``On West's asymptotic estimate
          of perturbative coefficients of R(s) in $e^{+}e^{-}$ annihilation,''
          Oklahoma State University Preprint No.\ RN-268A.\\
Surguladze, L.\ R., and M.\ A.\ Samuel, 1992b,
       ``Four-loop perturbative calculations of
       $\sigma_{\mbox{\scriptsize tot}}(e^{+}e^{-}\rightarrow \mbox{hadrons})$,
       $\Gamma(\tau\rightarrow \nu_{\tau}+\mbox{hadrons})$ and
       QED $\beta$ function,'' Fermilab Preprint No.\  FERMILAB-PUB 92/192-T.\\
Surguladze, L.\ R., and M.\ A.\ Samuel, 1993, Phys.\ Lett.\ {\bf B 309,} 157.\\
Symanzik, K., 1970, Commun.\ Math.\ Phys.\ {\bf 18,} 227.\\
Symanzik, K., 1971, Commun.\ Math.\ Phys.\ {\bf 23,} 49.\\
Tarasov, O.\ V., 1982, ``Anomalous dimensions of quark masses in three-loop
                        approximation,''
                        Dubna Joint Institute for Nuclear Research Preprint
                        No.\ JINR-P2-82-900.\\
Tarasov, O.\ V., A.\ A.\ Vladimirov, and A.\ Yu.\ Zharkov, 1980, Phys.\ Lett.\
                                                          {\bf B 93,} 429.\\
Tarrach, R., 1982, Nucl.\ Phys.\ {\bf B 196,} 45.\\
Tavkhelidze, A.\ N., 1965, Lect.\ High Energy Phys.\ Elem.\ Particles
(Vienna).\\
Tavkhelidze, A.\ N., 1994, ``Color, colored quarks, Quantum Chromodynamics,''
                         Dubna Joint Institute for Nuclear Research Preprint
                             No.\ JINR-E2-94-372.\\
t 'Hooft, G., 1971, Nucl.\ Phys.\  {\bf B 33,} 173.\\
t 'Hooft, G., 1973, Nucl.\ Phys.\ {\bf B 61,} 455.\\
t 'Hooft, G., and M.\ Veltman, 1972, Nucl.\ Phys.\ {\bf B 44,} 189.\\
t 'Hooft, G., and M.\ Veltman, 1973,  ``Diagrammar,'' CERN report.\\
Tkachov, F.\ V., 1981, Phys.\ Lett.\ {\bf B 100,} 65.\\
Tkachov, F.\ V., 1983a, Teor.\ Mat.\ Fiz.\ {\bf 56,} 350
                         [Theor.\ Math.\ Phys.\ {\bf 56,} 866 (1983)].\\
Tkachov, F.\ V., 1983b, Phys.\ Lett.\ {\bf B 124,} 212.\\
Tkachov, F.\ V., 1983c, Phys.\ Lett.\ {\bf B 125,} 85.\\
Tkachov, F.\ V., 1991, Fermilab Preprint No.\ FERMILAB-PUB-91/347-T.\\
Tkachov, F.\ V., 1993, Int.\ Journ.\ Mod.\ Phys.\ {\bf A 8,} 2047.\\
Trueman, T.\ L., 1979, Phys.\ Lett.\ {\bf B 88,} 331.\\
Tsai, Y.\ S., 1971, Phys.\ Rev.\ {\bf D 4,} 2821.\\
Vainshtein, A.\ I., and V.\ I.\ Zakharov, 1994, Phys.\ Rev.\ Lett.\
                                                        {\bf 73,} 1207.\\
Veltman, M., 1967, {\sl {\small SCHOONSCHIP}, A CDC 6600 program
         for symbolic evaluation of algebraic expressions} (CERN).\\
Veltman, M., 1991, {\sl {\small SCHOONSCHIP}, A program for symbol handling}
                                                               (Michigan).\\
Vermaseren, J.\ A.\ M., 1989, {\sl {\small FORM}, User's Manual} (NIKHEP,
                                                                 Amsterdam).\\
Vladimirov, A.\ A., 1978, Teor.\ Mat.\ Fiz.\ {\bf 36,} 271
                            [Theor.\ Math.\ Phys.\ {\bf 36,} 732 (1979)].\\
Vladimirov, A.\ A., 1980, Teor.\ Mat.\ Fiz.\ {\bf 43,} 280
                            [Theor.\ Math.\ Phys.\ {\bf 43,} 417 (1980)].\\
Ward, J.\ C., 1950, Phys.\ Rev.\ {\bf 78,} 182.\\
Weinberg, S., 1967, Phys.\ Rev.\ Lett.\ {\bf 19,} 1264.\\
Weinberg, S., 1973, Phys.\ Rev.\ {\bf D 8,} 3497.\\
West, G.\ B., 1991, Phys.\ Rev.\ Lett.\ {\bf 67,} 1388.\\
Wetzel, W., and W.\ Bernreuther, 1981, Phys.\ Rev.\ {\bf D 24,} 2724.\\
Wilson, K.\ G., 1969, Phys.\ Rev.\ {\bf 179,} 1499.\\
Yang, C.\ N., and R.\ L.\ Mills, 1954, Phys.\ Rev.\ {\bf 96,} 191.\\
Yang, C.\ N., 1969, in {\sl High Energy Collisions} (Gordon\& Breach, NY),
                                                                p.\ 509.\\
Yennie, D.\ R., S.\ C.\ Frautschi, and H.\ Suura, 1961,
                                           Ann.\ Phys.\  {\bf 13,} 379.\\
Yndurain, F.\ J., 1983, {\sl QCD: an Introduction to the
                      Theory of Quarks an Gluons} (Springer Verlag).\\
Zakharov, V.\ I., 1992, Nucl.\ Phys.\ {\bf B 385,} 452.\\
Zweig, G., 1964, ``An SU(3) model for strong interaction symmetry and
                   its breaking,'' CERN Preprint No.\  TH.412.
\end{document}